\documentclass[twocolumn]{aastex62}
\usepackage{graphicx}
\usepackage{color}
\usepackage[T1]{fontenc}
\usepackage{amssymb,amsmath}
\usepackage{ulem}

\newcommand{\nue}{\mbox{$\nu_e$}}
\newcommand{\nueb}{\mbox{$\bar{\nu}_e$}}
\newcommand{\nux}{\mbox{$\nu_x$}}

\received{xxx, 2018}
\revised{xxx, 2018}
\accepted{xxx, 2018}
\submitjournal{ApJ}

\shorttitle{Supernova neutrino process of Li and B}
\shortauthors{Kusakabe, Cheoun, Kim et al.}

\begin{document}

\title{Supernova Neutrino Process of Li and B Revisited}

\correspondingauthor{Motohiko Kusakabe}
\email{kusakabe@buaa.edu.cn}

\author[0000-0003-3083-6565]{Motohiko Kusakabe}
\affil{School of Physics, and International Research Center for Big-Bang Cosmology and Element Genesis, Beihang University, 37, Xueyuan Rd., Haidian-qu, Beijing 100083 China}
\affiliation{National Astronomical Observatory of Japan, 2-21-1 Osawa, Mitaka, Tokyo 181-8588, Japan}

\author{Myung-Ki Cheoun}
\affil{School of Physics, and International Research Center for Big-Bang Cosmology and Element Genesis, Beihang University, 37, Xueyuan Rd., Haidian-qu, Beijing 100083 China}
\affiliation{National Astronomical Observatory of Japan, 2-21-1 Osawa, Mitaka, Tokyo 181-8588, Japan}
\affiliation{Department of Physics and Origin of Matter and Evolution of Galaxy (OMEG) Institute, Soongsil University, Seoul 156-743, Korea}

\author{K. S. Kim}
\affiliation{School of Liberal Arts and Science, Korea Aerospace University, Goyang 412-791, Korea}

\author{Masa-aki Hashimoto}
\affiliation{Kyushu University, Hakozaki, Fukuoka 812-8581, Japan}

\author{Masaomi Ono}
\affiliation{RIKEN, 2-1 Hirosawa, Wako-shi, Saitama 351-0198, Japan}

\author{Ken'ichi Nomoto}
\affiliation{Kavli Institute for the Physics and Mathematics of the Universe (WPI), The University of Tokyo, 5-1-5 Kashiwanoha, Kashiwa, Chiba 277-8583, Japan}

\author{Toshio Suzuki}
\affiliation{National Astronomical Observatory of Japan, 2-21-1 Osawa, Mitaka, Tokyo 181-8588, Japan}
\affiliation{Department of Physics, College of Humanities and Science, Nihon University, Sakurajosui 3-25-40, Setagaya-ku, Tokyo 156-8550, Japan}

\author{Toshitaka Kajino}
\affil{School of Physics, and International Research Center for Big-Bang Cosmology and Element Genesis, Beihang University, 37, Xueyuan Rd., Haidian-qu, Beijing 100083 China}
\affiliation{National Astronomical Observatory of Japan, 2-21-1 Osawa, Mitaka, Tokyo 181-8588, Japan}
\affiliation{Graduate School of Science, University of Tokyo, 7-3-1 Hongo, Bunkyo-ku, Tokyo 113-0033, Japan}

\author{Grant J. Mathews}
\affiliation{National Astronomical Observatory of Japan, 2-21-1 Osawa, Mitaka, Tokyo 181-8588, Japan}
\affiliation{Center for Astrophysics, Department of Physics, University of Notre Dame, Notre Dame, IN 46556, U.S.A.}



\begin{abstract}

  We reinvestigate effects of neutrino oscillations on the production of $^7$Li and $^{11}$B in core-collapse supernovae (SNe). During the propagation of neutrinos from the proto-neutron star, their flavors change and the neutrino reaction rates for spallation of $^{12}$C and $^4$He are affected. In this work corrected neutrino spallation cross sections for  $^4$He and $^{12}$C are adopted. Initial abundances involving heavy $s$-nuclei and other physical conditions are derived in a new calculation of the SN 1987A progenitor in which effects of the progenitor metallicity are included. A dependence of the SN nucleosynthesis and final yields of $^7$Li and $^{11}$B on the neutrino mass hierarchy are shown in several stellar locations.
  In the normal hierarchy case, the charged-current (CC) reaction rates of $\nue$ are enhanced, and yields of proton-rich nuclei, along with $^7$Be and $^{11}$C, are increased.
  In the inverted hierarchy case, the CC reaction rates of $\nueb$ are enhanced, and yields of neutron-rich nuclei, along with $^7$Li and $^{11}$B, are increased.
  We find that variation of the metallicity modifies the yields of $^7$Li, $^7$Be, $^{11}$B, and $^{11}$C. This effect is caused by changes in the neutron abundance during SN nucleosynthesis.
  Therefore, accurate calculations of Li and B production in SNe should take into account the metallicity of progenitor stars.  

\end{abstract}



\section{Introduction}\label{sec1}

Lithium, beryllium, and boron are fragile light elements with very small solar abundances \citep{2009LanB...4B..712L}. For the most part they are destroyed during the stellar evolution \citep{1997nceg.book.....P}, although some $^7$Li could be made in red giant stars, while $^7$Li and $^{11}$B can be made by the $\nu$-process in SNe. One important mechanism for the production of LiBeB isotopes is via cosmic ray nucleosynthesis. The nuclear spallation of CNO by protons and $^4$He nuclei generates all $^{6,7}$Li, $^9$Be, and $^{10,11}$B isotopes, and $\alpha+\alpha$ reactions generate $^{6,7}$Li \citep{1970Natur.226..727R,1971A&A....15..337M}. Also, the neutrino spallation process of He and C in massive stars during SN explosions contributes to the production of $^7$Li and $^{11}$B \citep{1978Ap&SS..58..273D,1990ApJ...356..272W}. In addition, the $^3$He($\alpha$,$\gamma$)$^7$Be reaction \citep{1955ApJ...121..144C,1971ApJ...164..111C} operates in asymptotic giant branch stars \citep{2010MNRAS.402L..72V}, red giants \citep{1999ApJ...510..217S}, novae \citep{1996ApJ...465L..27H} as well as in SNe. All of these processes contribute at some level to the Galactic chemical evolution of Li, Be, and B \citep{2012A&A...542A..67P}.

Because of the small yields for these light elements by astrophysical processes, observations of their primordial abundances can test cosmological models that predict changes in the abundances from that of standard big bang nucleosynthesis (SBBN). Lithium abundances on metal-poor stars have been measured by spectroscopic astronomical observations. Inferred abundances are somewhat smaller than the SBBN prediction \citep{1982A&A...115..357S,2000ApJ...530L..57R,2009ApJ...698.1803A,2010A&A...522A..26S}. Some hypothetical BBN models involving exotic long-lived particles can explain the observed lithium abundance. For example, negatively-charged and/or strongly-interacting particles can decrease the primordial Li abundance \citep{2008PhRvD..78h3010B,2011PhRvD..83e5011K} and enhance the primordial abundances of $^9$Be and/or $^{10}$B \citep{2014ApJS..214....5K,2009PhRvD..80j3501K}. Thus, the primordial LiBeB abundance is a unique probe to explore the physics of the early universe.

SN explosions are energized by neutrinos emitted from the proto-neutron star (NS), and neutrino reactions with nuclei in the stellar interior produces a number of rare stable nuclei including $^{7}$Li, $^{11}$B, $^{138}$La, and $^{180}$Ta \citep{1978Ap&SS..58..273D,1990ApJ...356..272W,2005PhLB..606..258H,2005PhRvL..94w1101Y} as well as short-lived nuclei $^{92}$Nb and $^{98}$Tc \citep{2012PhRvC..85f5807C,2013ApJ...779L...9H,2015arXiv150501082S,2016EPJWC.10906004S,2018JPhCS.940a2054S,Hayakawa2017}, and $^{26}$Al and $^{22}$Na \citep{2015arXiv150501082S,2016EPJWC.10906004S,2018JPhCS.940a2054S,2018ApJ...865..143S}. Among them, the production of the light elements Li and B occurs in the C- and He-rich layers inside of which neutrino flavor mixing can effectively occur due to the influence of the stellar electrons. Effects on the Li and B yields from the neutrino temperature and the total neutrino energy emitted from the NS have been studied previously \citep{2004ApJ...600..204Y,2005PhRvL..94w1101Y}. In addition, effects of neutrino oscillations have been clarified and Li and B yields have been calculated as functions of neutrino mixing angle $\sin^2 2 \theta_{13}$ for both cases of the normal and inverted mass hierarchy \citep{2006PhRvL..96i1101Y,2006ApJ...649..319Y}.

Detailed results of the SN neutrino process for Li and B yields have been obtained adopting neutrino-nucleus reaction cross sections for $^4$He and $^{12}$C based on shell model calculations that explain the spin properties of $p$-shell nuclei \citep{2008ApJ...686..448Y}. Larger total neutrino energy or neutrino temperature enhances the Li and B yields, and the yield ratio of $^7$Li/$^{11}$B are different between the normal and inverted mass hierarchy. The $^7$Li and $^{11}$B abundances are also affected by collective neutrino oscillation effects in the inner region of SNe \citep{Ko2019}. Based on time-dependent neutrino spectra, effects of neutrino flavor changes on SN nucleosynthesis have been studied \citep{2015PhRvD..91f5016W}. Although collective effects on $^7$Li and $^{11}$B have not yet been analyzed, it has been found that collective neutrino oscillations enhance the yields of $^{138}$La and $^{180}$Ta. In their model, the Mikheyev-Smirnov-Wolfenstein (MSW) flavor oscillations are expected to have a stronger impact on the production of Li and B than the collective oscillations. 

It has also been pointed out that by considering the ratios Li/Si and B/Si that are larger than the solar value and isotopic ratios $^7$Li/$^6$Li and $^{11}$B/$^{10}$B measured in SiC X grains, an upper limit on the yield ratio $^7$Li/$^{11}$B can be derived \citep{2012PhRvD..85j5023M}. This could be a method to constrain the neutrino mass hierarchy from meteoritic measurements.

Uncertainties in nuclear reaction rates, however, make yields of presupernova stellar nucleosynthesis ambiguous. Effects on Li and B synthesis from variations of rates of the triple-alpha reaction and $^{12}$C($\alpha$,$\gamma$)$^{16}$O have been investigated \citep{2011PhRvL.106o2501A}. It was found that the $2\sigma$ uncertainty in the reaction rate of $^{12}$C($\alpha$,$\gamma$)$^{16}$O leads to a difference in the $^{11}$B yield by a factor of up to two. However, constraints from solar system abundances of the intermediate-mass isotopes (O to Ca) and $s$-only isotopes can limit the possible variations in rates of the triple-alpha reaction and $^{12}$C($\alpha$,$\gamma$)$^{16}$O \citep{2014PhRvL.112k1101A}. \citet{2014PhRvL.112k1101A} found that the yields of $^7$Li and $^{11}$B can change by at most $\sim$10 \%.
If progenitor stars experience convective mixing of O- and C-rich layers, SN nucleosynthesis is affected via changes in the neutron and proton abundances \citep{2014MNRAS.441..733N}. Recently, based on a two-dimensional SN explosion model, it has been suggested that the very inner region of $\alpha$-rich nucleosynthesis freezeout can contribute to the Li and B yields since part of the nuclei in that region can escape from fallback onto the NS \citep{2018EPJWC.16501045S}.

Thus, to study the neutrino oscillation effects by comparing theoretical results with observations of $^7$Li and $^{11}$B abundances an effort is needed to improve knowledge on: (1) neutrino spectra and reaction cross sections; (2) nuclear reaction rates and stellar evolution; and (3) multi-dimensional effects of SN explosions. The neutrino spallation of He in the He-rich layer of low metallicity stars can induce $r$-process nucleosynthesis on a relatively long time scale and low temperature only for the inverted mass hierarchy case \citep{2016EPJWC.10906001B}. Also, $^9$Be production via $^7$Li($n$,$\gamma$)$^8$Li($n$,$\gamma$)$^9$Li(,$e^-$\nueb)$^9$Be and $^7$Li($t$,$n$)$^9$Be can operate in such a metal-poor condition \citep{2016EPJWC.10906001B}. It has been suggested that the solar system formation was triggered by a low mass supernova with a low explosion energy. This event can explain the solar abundances of stable nuclei and the abundances of short-lived nuclei when meteorites formed. It can also explain the existence of short-lived $^{10}$Be produced via the neutrino spallation of $^{12}$C \citep{2016NatCo...713639B}.

In this paper, we study effects of the initial nuclear abundances and new reaction rates on the SN nucleosynthesis of $^7$Li and $^{11}$B. Production of those light nuclei in SNe is important because of the potential for constraining conditions of stellar evolution, and neutrino emission from NSs, and SN explosions. Therefore, we also make a detailed analysis of the results of neutrino process nucleosynthesis. In Sec.~\ref{sec2} a description of our model is given for the presupernova evolution, the SN explosion, the nucleosynthesis, neutrino cross sections, and neutrino reaction rates with the flavor change taken into account. In Sec.~\ref{sec3} effects of neutrino flavor changes are shown in the time evolution of nuclear abundances. Final yields are derived as a function of the Lagrangian mass coordinate. In Sec.~\ref{sec4} effects of initial nuclear abundances are investigated. In Sec.~\ref{sec5} the total yields of $^7$Li and $^{11}$B are listed, and the possible range of Li/B ratios in SN grains is discussed. In Sec.~\ref{sec6} we summarize this study.
In Appendix \ref{app1} we add a note on the time evolution of nuclear production rates taking the JINA REACLIB functions as an example.
In Appendices \ref{app2} and \ref{app3} we identify which reactions operate in different layers of the star, along with a study of the effects of neutrino oscillations and the initial nuclear abundances on the reaction rates.

\section{Model}\label{sec2}
\subsection{Presupernova}\label{sec2a}

Neutrino oscillations occur during transport from the neutrino-sphere through the stellar envelope.
In core-collapse SNe, a shock wave develops, and the temperature and density are changed once the shock arrives at a given radius.  The changes of temperature and density can induce nucleosynthesis, and the nuclear abundances and electron fraction are changed.  The neutrino oscillations which depend on the electron density are then different between epochs before and after the arrival of the shock.  However, most of the neutrinos emitted from the proto-neutron star are produced soon after the stellar core bounce.  Most neutrinos pass through before the shock wave reaches to the radial region where the neutrino flavor changes significantly. Therefore, we neglect the effect of the shock on the neutrino oscillations \citep{2008ApJ...686..448Y}\footnote{Because of a shock, we can have a situation in which neutrinos experience the MSW H resonance more than once during their propagation. This shock effect on the neutrino process will be reported elsewhere \citep{Ko2019b}}.

We adopt the presupernova density, temperature, and abundance profiles from a new calculation using the method of \citep{2015PTEP.2015f3E01K} where the initial metallicity is taken to be $Z=Z_\sun /4$ appropriate to that of the Large Magellanic Cloud. The density and temperature profiles are very similar to that used previously in the SN explosion calculation of model 14E1 for SN1987A \citep{1990ApJ...360..242S}. This star corresponds to the last evolution stage of a progenitor star, with an initial mass of $20 M_\odot$ which has evolved to a helium core of $6 M_\odot$ just prior to collapse.

\begin{figure}[t!]
  \epsscale{1}
  \plotone{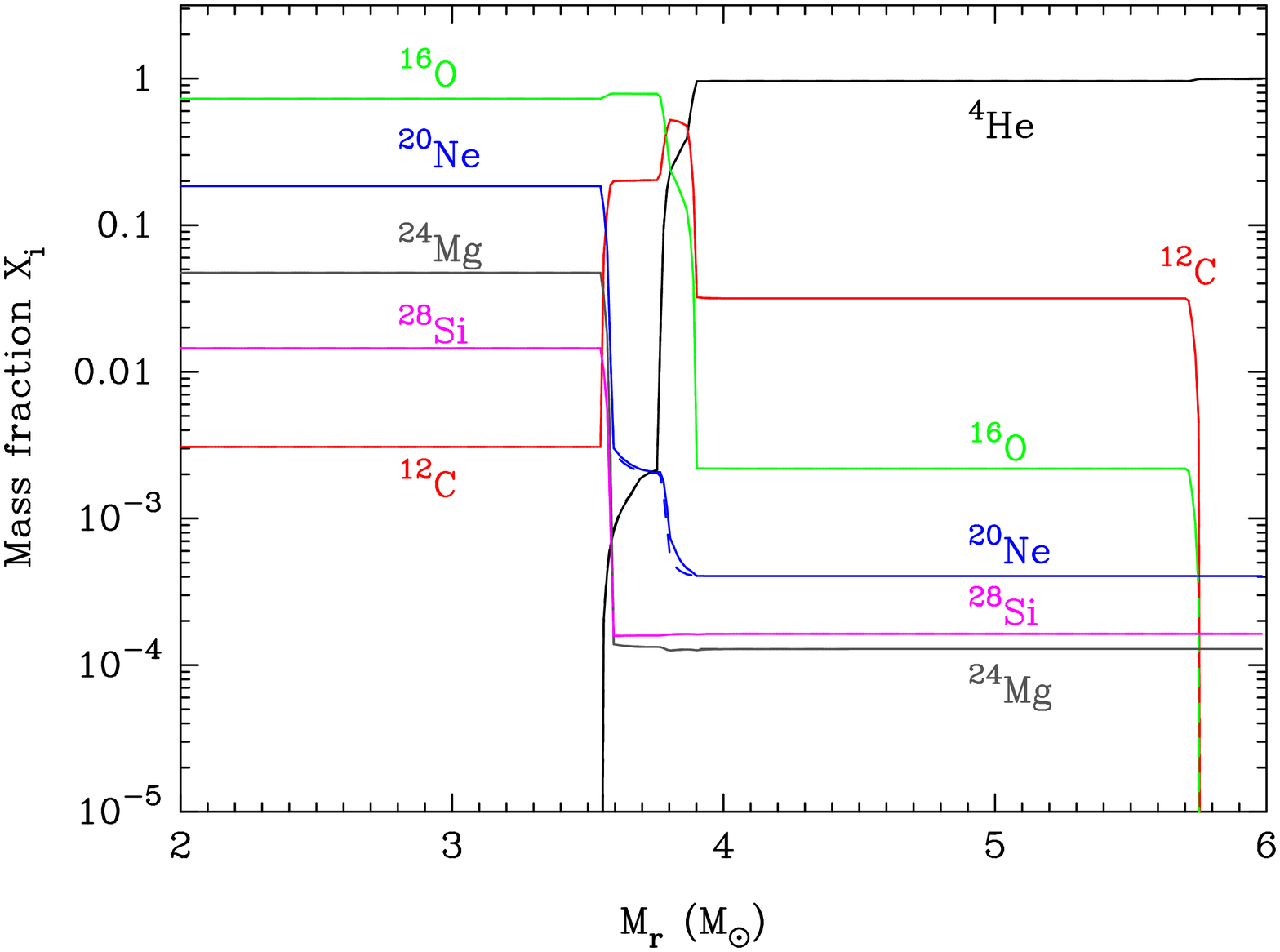}
\caption{Mass fractions of the most abundant nuclei as a function of Lagrangian mass. Dashed and solid lines correspond to before the neutrinos arrive and after the SN shock passes, respectively.}
\label{fig:composition}
\end{figure}

Figure \ref{fig:composition} shows the mass fractions of abundant nuclei as a function of the Lagrangian mass coordinate, before the neutrinos arrive (dashed lines) and after the shock passes (solid lines). The $^7$Li and $^7$Be are predominantly produced in the He-rich layer, while the $^{11}$B and $^{11}$C are produced in both the He- and C-rich layers. In these layers, the main nuclear components are not altered by the passage of the shock. The solid and dashed lines almost completely overlap except for $^{20}$Ne at the base of the $^4$He shell.

\subsection{Hydrodynamics}\label{sec2b}
Hydrodynamical data for the SN explosion are derived using the public code (blcode) (Christian Ott, Viktoriya Morozova, and Anthony L. Piro; https://stellarcollapse.org/snec). The density and temperature profiles of the presupernova stellar model are adopted for the initial state of the star, and a thermal bomb type explosion is evolved. We adopt a simple setup for the explosion, i.e., a constant luminosity for 3 s with a total explosion energy of $10^{51}$ erg. This is consistent with the neutrino luminosity described below. The calculation is stopped at 50 s after the start of the calculation, and the output is recorded every 0.02 s. An equal interval in mass of 360 zones is adopted for the stellar grid. The explosion energy is injected at the innermost five grid points above the mass cut at $M_r =1.6~M_\sun$. After $t=50$ s, we switched off two-body reactions by setting the temperature and density low enough that only one-body decay can operate.

\begin{figure}[t!]
  \epsscale{1}
\plotone{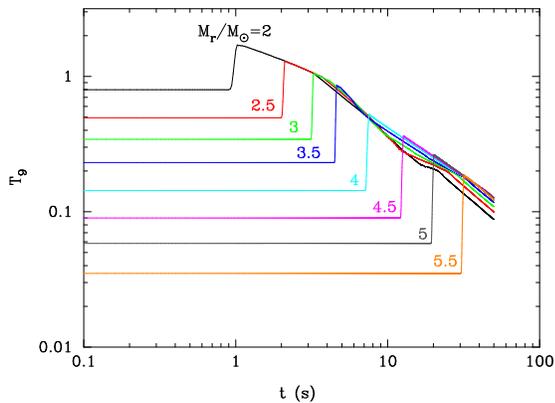}
\caption{Temperatures $T_9 \equiv T/(10^9~{\rm K})$ as a function of time at fixed Lagrangian mass points.}
\label{fig:temp}
\end{figure}

\begin{figure}[t!]
  \epsscale{1}
\plotone{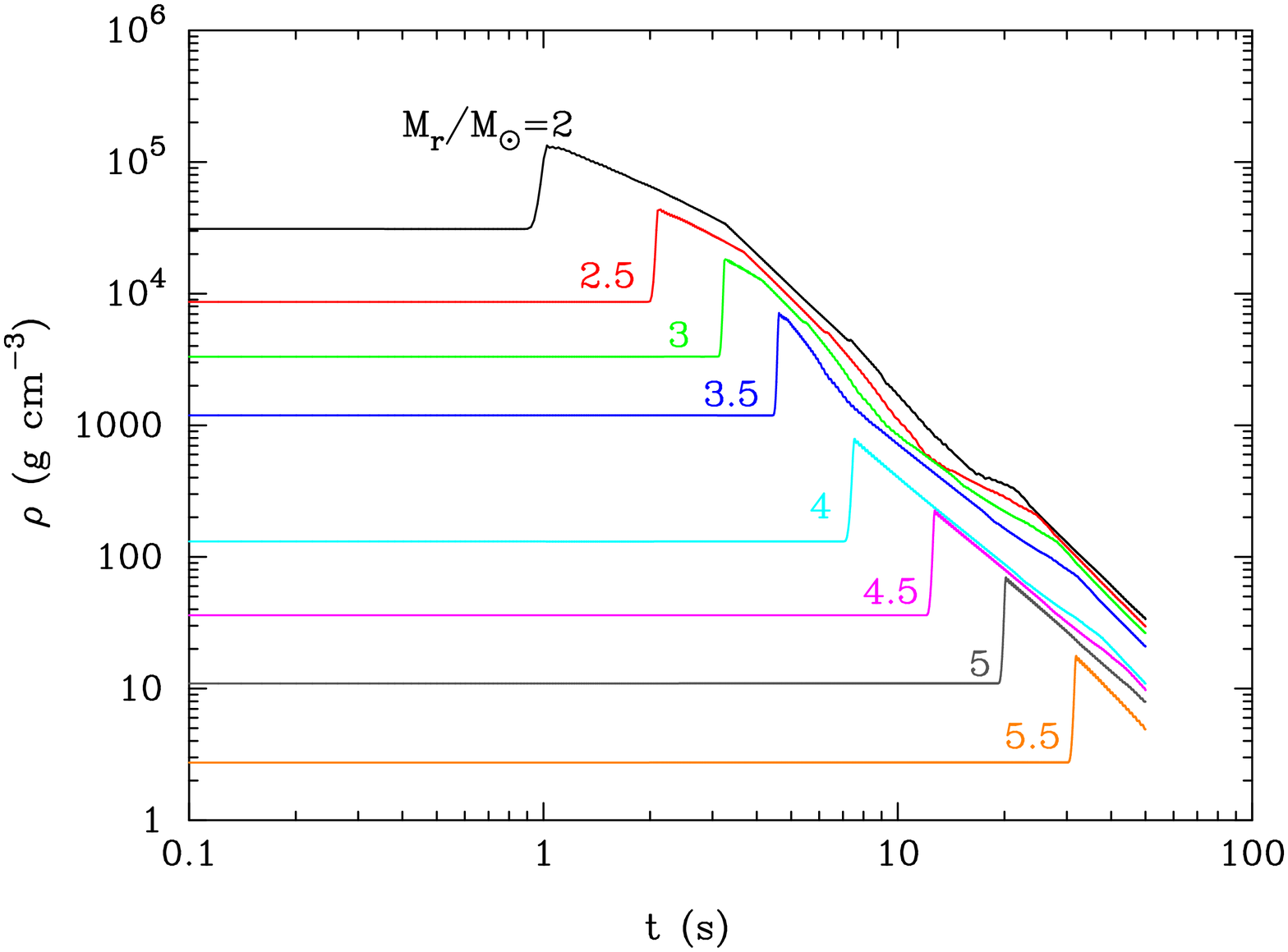}
\caption{Densities as a function of time at fixed Lagrangian mass points.}
\label{fig:density}
\end{figure}

\begin{figure}[t!]
  \epsscale{1}
\plotone{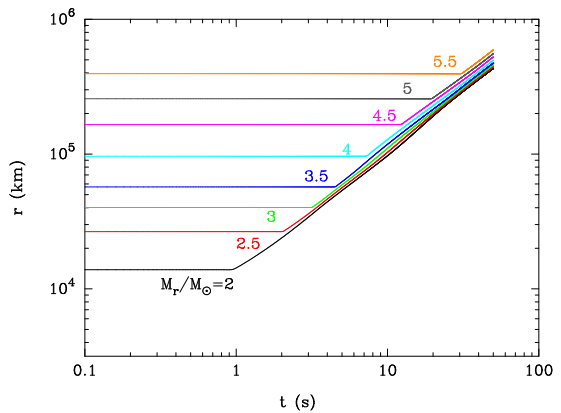}
\caption{Radii as a function of time at fixed Lagrangian mass points.}
\label{fig:radius}
\end{figure}

Figures \ref{fig:temp}, \ref{fig:density}, and \ref{fig:radius} show the time evolution of the temperature, density, and radius, respectively, for Lagrangian masses e.g. of $M_r =2$, 2.5, 3, 3.5, ..., and 5.5 $M_\sun$ as labeled.

\subsection{Nuclear reaction network}\label{sec2c}

We have constructed a code for the calculation of the SN nucleosynthesis including neutrino reactions as well as nuclear reactions, $\beta$-decay, and $\alpha$-decay.
We adopt nuclear data and reaction rates from the JINA REACLIB database \citep[JINA REACLIB V2.0, 2014]{2010ApJS..189..240C} \footnote{The data file erroneously included two different rates for the $\beta$-decay of (ga85 $\rightarrow$ ge85) and (mo111 $\rightarrow$ tc111), respectively.  We should use only one rate for each reaction, and the latest rates from wc12w were chosen for the both reactions (Richard Cyburt, Priv. Comm. April, 2014).}. Information on REACLIB rates and nuclear partition functions are available in the README file by Rauscher \footnote{http://download.nucastro.org/astro/reaclib}. The $\alpha$-decay reactions of unstable nuclides with half lives $T_{1/2} \leq 14$ Gyr are added with rates taken from National Nuclear Data Center, Brookhaven National Laboratory \footnote{http://www.nndc.bnl.gov}. Elements up to $_{87}$Fr, involving 3080 nuclides and 38197 reactions are included in the network.

The latest evaluated values \citep{2009LanB...4B..712L}\footnote{Mo, Dy, Er, Yb and Lu have updated isotopic compositions.} of solar abundances 4.56 Gyr ago, i.e., at the solar birth, are adopted. To solve for the abundance evolution the variable order Bader-Deuflhard time integration method is adopted coupled with the MA38 sparse-matrix package \citep{MA38}\footnote{http://www.hsl.rl.ac.uk/catalogue/ma38.html}. This is one of the best methods investigated in \citet{1999ApJS..124..241T}.  The subroutines METANI \citep{Bader1983} and STIFBS \citep{1992nrfa.book.....P} are used. The theoretical tolerance in final abundances of major nuclei associated with this numerical integration is set to be less than 1 \%. The method used in this paper is adopted for controlling numerical errors by a real error estimate, not by error expectation. Also, the variable order method adopted here minimizes the computational time while keeping the required numerical accuracy.

One should include all possible important reactions of the light nuclides, Li, Be, and B, to derive their yields accurately.  In addition to the JINA REACLIB database, we include the reaction $^{10}$B($n$, $p$)$^{10}$Be ($Q=0.2255$ MeV) with a rate estimated with the TALYS-1.8 code (2015) \citep{Koning2007}\footnote{http://www.talys.eu/home}. The thermally averaged reaction rate in the range of $T_9 =[0.01, 10]$ is fitted with the JINA REACLIB function [Eq. (\ref{eq_a6})] with fixed $a_1 =0$.  The result is $a_0 =-28.151$, $a_2 =-6.522$, $a_3 =50.095$, $a_4 =-4.358$, $a_5 =0.260$, and $a_6 =-12.376$.  The rate for the reverse reaction $^{10}$Be($p$, $n$)$^{10}$B is given by the same coefficients $a_i$ ($i=2$, ..., $6$) with $a_{0,{\rm rev}}=a_0+\ln(7)=-26.205$ and $a_{1,{\rm rev}}=Q/(10^9~{\rm K}) =2.617$.  It is noted, however, that this reaction has a negligible effect on the evolution of nuclear abundances in the present calculation of SN nucleosynthesis. Rates of other final states, i.e., $^{10}$B($n$, $\gamma$)$^{11}$B and $^{10}$B($n$, $\alpha$)$^7$Li, derived with the TALYS code agree with those in JINA REACLIB database to within a factor of $\sim$three in the relevant temperature region. 

\subsection{Neutrino flavor change probability}\label{sec2d}
The neutrino oscillations in matter are described by the mixing matrix
\begin{eqnarray}
i\frac{d}{dr}
\left( \begin{array}{c}
\nu_e \\
\nu_\mu \\
\nu_\tau
\end{array} \right)
&=&\left[ U \left( \begin{array}{ccc}
0 & 0 & 0 \\
0 & \frac{\Delta m^2_{21}}{2 E_\nu} & 0 \\
0 & 0 & \frac{\Delta m^2_{31}}{2 E_\nu}
  \end{array} \right) U^\dag \right. \nonumber \\
  &&+ \left. \left( \begin{array}{ccc}
\pm \sqrt{2} G_{\rm F} n_e(r) & 0 & 0 \\
0 & 0 & 0 \\
0 & 0 & 0
\end{array} \right) \right]
\left( \begin{array}{c}
\nu_e \\
\nu_\mu \\
\nu_\tau
\end{array} \right) ,
\label{eq_osc}
\end{eqnarray}
\begin{equation}
U = \left( \begin{array}{ccc}
c_{12}c_{13} & s_{12}c_{13} & s_{13} \\
-s_{12}c_{23}-c_{12}s_{23}s_{13} & c_{12}c_{23}-s_{12}s_{23}s_{13} &
s_{23}c_{13} \\
s_{12}s_{23}-c_{12}c_{23}s_{13} & -c_{12}s_{23}-s_{12}c_{23}s_{13} &
c_{23}c_{13}
\end{array} \right),
\label{matrix_osc}
\end{equation}
where
$r$ is the radius from the center of the proto-neutron star,
$\Delta m^2_{ij} = m^2_i - m^2_j$ is the difference of mass squared with $m_i$ the neutrino mass of eigenstate $i$, $G_{\rm F}$ is the Fermi constant, $n_e(r)$ is 
the electron number density, and
$s_{ij}=\sin \theta_{ij}$ while $c_{ij} = \cos \theta_{ij}$.
Plus and minus signs in the second term in the square brackets of the RHS in Eq. (\ref{eq_osc}) correspond to neutrinos and anti-neutrinos, respectively.

It has been clearly demonstrated \citep{1987PhLB..198..406K,2002PhLB..544..286Y} that the CP violation will not be seen in any observables when $\mu$- and $\tau$-neutrino energy spectra and those of anti-neutrinos are degenerate, as expected to be the case approximately for the neutrinos emitted from core-collapse supernovae.
Therefore, we set the CP phase equal to be zero.

The central values of the particle data group, 2015 \citep{2014ChPhC..38i0001O} are adopted for the parameters of this equation as follows:  
$\sin^2 2 \theta_{12}=0.846$, $\sin^2 2 \theta_{23}=0.999$, $\sin^2 2 \theta_{13}=0.093$, $\Delta m^2_{21}=7.53 \times 10^{-5}$ eV$^2$, and $|\Delta m^2_{32}|=2.4 \times 10^{-3}$ eV$^2$. Our calculation completely reproduces Figs. 1a and 2a of \citet{2006ApJ...649..319Y} when adopting the same values for neutrino parameters. 

Probabilities of neutrino flavor change are then calculated for the parameter set of neutrino energy $E_{\nu}$ and radius $r$. The probabilities do not depend on the time of neutrino emission from the neutrino sphere since we have assumed steady-state density, temperature, and abundances for the presupernova.

\subsection{$\nu$-reaction rate}\label{sec2e}

The number flux of neutrinos with flavor $\alpha$ ($\alpha$=$e$, $\mu$, $\tau$) and energy $E_\nu$ coming from the proto-NS is described \citep{2005NJPh....7...51B,2006ApJ...649..319Y} by
\begin{equation}
\frac{d \phi_{\nu_\alpha}}{d E_\nu} =
\frac{L_{\nu_\alpha}}{4 \pi r^2} \frac{1}{F_3(\eta_{\nu_\alpha}) \left(kT_{\nu_\alpha} \right)^4}
\frac{E_\nu^2}{\exp(E_\nu/kT_{\nu_\alpha} -\eta_{\nu_\alpha} ) +1}
\label{nu1}
\end{equation}
with,
\begin{equation}
F_n(\eta_{\nu_\alpha}) =
\int_0^\infty \frac{x^n}{\exp(x-\eta_{\nu_\alpha}) +1} dx,
\label{nu2}
\end{equation}
where
$L_{\nu_\alpha}$ is the luminosity of the neutrino $\alpha$,
$k$ is the Boltzmann constant,
$T_{\nu_\alpha}$ is the temperature of neutrino $\alpha$, and
$\eta_{\nu_\alpha} =\mu_{\nu_\alpha}/kT_{\nu_\alpha}$ is defined in terms of the chemical potential $\mu_{\nu_\alpha}$.

The reaction rate of neutrino ${\nu_\alpha}$ \citep{2006ApJ...649..319Y} is given by
\begin{eqnarray}
\lambda_{\nu_\alpha} (r) &=&
\int_0^\infty \sum_{\beta=e,\mu,\tau} \frac{d \phi_{\nu_\beta}}{d E_\nu}  P_{\beta\alpha}(r; E_\nu) \sigma_{\nu_\alpha}(E_\nu) dE_\nu \nonumber\\
&=&
\sum_{\beta=e,\mu,\tau} \left[
  \frac{L_{\nu_\beta}}{4 \pi r^2} \frac{1}{F_3(\eta_{\nu_\beta}) \left(kT_{\nu_\beta} \right)^4} \right. \nonumber \\
  && \left. \times
\int_0^\infty 
\frac{E_\nu^2 P_{\beta\alpha}(r; E_\nu) \sigma_{\nu_\alpha}(E_\nu) dE_\nu}{\exp(E_\nu/kT_{\nu_\beta} -\eta_{\nu_\beta} ) +1} \right] \nonumber\\
&=&
\sum_{\beta=e,\mu,\tau} \left[
\frac{L_{\nu_\beta}}{4 \pi r^2} \frac{1}{kT_{\nu_\beta}} \frac{F_2(\eta_{\nu_\beta})}{F_3(\eta_{\nu_\beta})}
\langle P_{\beta\alpha} \sigma_{\nu_\alpha} \rangle(T_{\nu_\beta}; r) \right], \nonumber \\
\label{nu3}
\end{eqnarray}
where
$P_{\beta\alpha}(r; E_\nu)$ is the probability of neutrino flavor change from $\beta$ to $\alpha$ during the propagation of a neutrino with energy $E_\nu$ from $0$ to $r$.
$\sigma_{\nu_\alpha}(E_\nu)$ is the cross section for neutrino reactions as a function of $E_\nu$, and
the thermal average value $\langle P_{\beta\alpha} \sigma_{\nu_\alpha} \rangle(T_{\nu_\beta}; r)$
of the product of the flavor-change-probability and the cross section is defined in the last equality by
\begin{eqnarray}
\langle P_{\beta\alpha} \sigma_{\nu_\alpha} \rangle(T_{\nu_\beta}; r) &=&
\frac{1}{F_2(\eta_{\nu_\beta}) \left(kT_{\nu_\beta} \right)^3} \nonumber \\
&& \times
\int_0^\infty \frac{E_\nu^2 P_{\beta\alpha}(r; E_\nu) \sigma_{\nu_\alpha}(E_\nu) dE_\nu}{\exp(E_\nu/kT_{\nu_\beta} -\eta_{\nu_\beta} ) +1} \nonumber\\
&=& \frac{1}{F_2(\eta_{\nu_\beta})}
\int_0^\infty \frac{P_{\beta\alpha}(r; E_\nu) \sigma_{\nu_\alpha}(E_\nu) x^2 dx}{\exp(x -\eta_{\nu_\beta} ) +1}. \nonumber \\
\label{nu4}
\end{eqnarray}

Specific values of the $F_n$ function include $F_2(0)=(3/2)\zeta(3)$ with $\zeta(3)=1.20205$ and $F_3(0)=7\pi^4/120$.  Reaction rates of antineutrinos can be formulated similarly to those of neutrinos. In this study, we assume $\eta_{\nu_\alpha}=0$ for all flavors $\alpha$ [see \citep{2005PhRvL..94w1101Y} for effects of chemical potentials].

\subsubsection{Neutral current reaction rate}
The total neutral current (NC) reaction rate is given by
\begin{eqnarray}
\lambda_{{\nu},{\rm tot}}^{\rm NC} (r) &=&
\sum_{\alpha=e,\mu,\tau} \lambda_{\nu_\alpha}^{\rm NC} (r) \nonumber \\
&=&
\sum_{\beta=e,\mu,\tau}
\frac{L_{\nu_\beta}}{4 \pi r^2} \frac{1}{kT_{\nu_\beta}} \frac{F_2(\eta_{\nu_\beta})}{F_3(\eta_{\nu_\beta})}
\langle \sigma_\nu^{\rm NC} \rangle(T_{\nu_\beta}),
\label{nu5}
\end{eqnarray}
where
$\sum_\alpha \langle P_{\beta\alpha} \sigma_{\nu_\alpha}^{\rm NC} \rangle =\langle \sigma_\nu^{\rm NC} \rangle$ is used in the last equality.  This rate is thus independent of the flavor-change-probabilities, and the average cross section $\langle \sigma_\nu^{\rm NC} \rangle(T_{\nu_\beta})$ is independent of radius.

\subsubsection{Charged current reaction rate}
Only the electron-type neutrinos (and antineutrinos) can react with nuclei via charged current (CC) interactions. This is because of the low neutrino energies, $E_\nu < {\mathcal O}(10)$ MeV.  The charged current reaction rates are then given by
\begin{eqnarray}
\lambda_{\nu_e}^{\rm CC} (r) &=&
\sum_{\beta=e,\mu,\tau} \left[
\frac{L_{\nu_\beta}}{4 \pi r^2} \frac{1}{kT_{\nu_\beta}} \frac{F_2(\eta_{\nu_\beta})}{F_3(\eta_{\nu_\beta})}
\langle P_{\beta e} \sigma_{\nu_e}^{\rm CC} \rangle(T_{\nu_\beta}; r) \right]. \nonumber \\
\label{nu6}
\end{eqnarray}

Note that neglecting the neutrino oscillation effect corresponds to assuming the condition $P_{\beta\alpha}(r; E_\nu) =\delta_{\beta\alpha}$. The charged current reaction rate is then given by
\begin{equation}
\lambda_{\nu_e}^{\rm CC}({\rm no-osc}) (r)=
\frac{L_{\nu_e}}{4 \pi r^2} \frac{1}{kT_{\nu_e}} \frac{F_2(\eta_{\nu_e})}{F_3(\eta_{\nu_e})}
\langle \sigma_{\nu_e}^{\rm CC} \rangle(T_{\nu_e}; r).
\label{nu7}
\end{equation}

\subsubsection{Neutrino parameters}
The time evolution of the neutrino luminosity is adopted from \citet{2004ApJ...600..204Y}
\begin{eqnarray}
  L_{\nu_\alpha}(r; t) &=& \frac{1}{6} \frac{E_\nu^{\rm SN}}{\tau_\nu} \exp\left( -\frac{t -r/c}{\tau_\nu} \right) \Theta \left( t -\frac{r}{c} \right) \nonumber\\
  &=&
  1.04025 \times 10^{58}~{\rm MeV}~{\rm s}^{-1}~\frac{E_{\nu,~10^{53}~{\rm erg}}^{\rm SN}}{\tau_{\nu,~{\rm s}^{-1}}} \nonumber \\
  && \times \exp\left( -\frac{t -r/c}{\tau_\nu} \right) \Theta \left( t -\frac{r}{c} \right),
\label{nu8}
\end{eqnarray}
where
$E_\nu^{\rm SN}$ is the total energy of the emitted neutrinos,
$c$ is the light speed, and
$\Theta(x)$ is the step function in $x$.

We assume the following values for parameters from \citet{2006ApJ...649..319Y}:   The total energy is $E_\nu^{\rm SN} =3 \times 10^{53}$ erg.  The decay time of the neutrino luminosity is $\tau_\nu =3$ s.  Neutrino temperatures are $T_{\nu_e} =3.2$, $T_{\bar{{\nu_e}}} =5.0$, and $T_{\nu_x} =6.0$ MeV (for \nux =$\nu_\mu$, $\nu_\tau$, $\bar{\nu_\mu}$, and $\bar{\nu_\tau}$), which have been constrained by consideration of the Galactic chemical evolution of B \citep{2004ApJ...600..204Y,2005PhRvL..94w1101Y,2006PhRvL..96i1101Y,2006ApJ...649..319Y,2008ApJ...686..448Y}.\footnote{Neutrino energy spectra and their evolution as well as the explosion energy affect the production of $^7$Li and $^{11}$B. Accurate calculations for realistic neutrino spectra are desired to improve the calculation of $^7$Li and $^{11}$B yields.}

The numerical values of Eqs. (\ref{nu3}) and (\ref{nu5}) for $\eta_{\nu_\beta} =0$ are given by
\begin{eqnarray}
  \lambda_{\nu_\alpha} (r) &=&
  1.263 \times 10^{-7}~{\rm s}^{-1}~\sum_{\beta=e,\mu,\tau} \left[
    \frac{L_{\nu_\beta,~10^{58}{\rm MeV/s}}}{r_{,2\times 10^{10}{\rm cm}}^2} \right. \nonumber \\
    &&\times \left. \frac{1}{(kT_{\nu_\beta})_{,5{\rm MeV}}}
\langle P_{\beta\alpha} \sigma_{\nu_\alpha} \rangle_{, 10^{-42}~{\rm cm}^2}(T_{\nu_\beta}; r) \right],~~~~~
\label{nu9}
\end{eqnarray}
and
\begin{eqnarray}
\lambda_{{\nu},{\rm tot}}^{\rm NC} (r) &=&
1.263 \times 10^{-7}~{\rm s}^{-1}
~\sum_{\beta=e,\mu,\tau} \left[
  \frac{L_{\nu_\beta,~10^{58}{\rm MeV/s}}}{r_{,2\times 10^{10}{\rm cm}}^2} \right. \nonumber \\
  &&\times \left. \frac{1}{(kT_{\nu_\beta})_{,5{\rm MeV}}}
\langle \sigma_\nu^{\rm NC} \rangle_{, 10^{-42}~{\rm cm}^2}(T_{\nu_\beta}) \right],
\label{nu10}
\end{eqnarray}
respectively,
where
the physical quantities ($G$) are normalized to typical values for the SN neutrino process in the He-rich layer ($a$), i.e., $G_{,a} \equiv G/a$.

\begin{figure}[t!]
  \epsscale{1}
\plotone{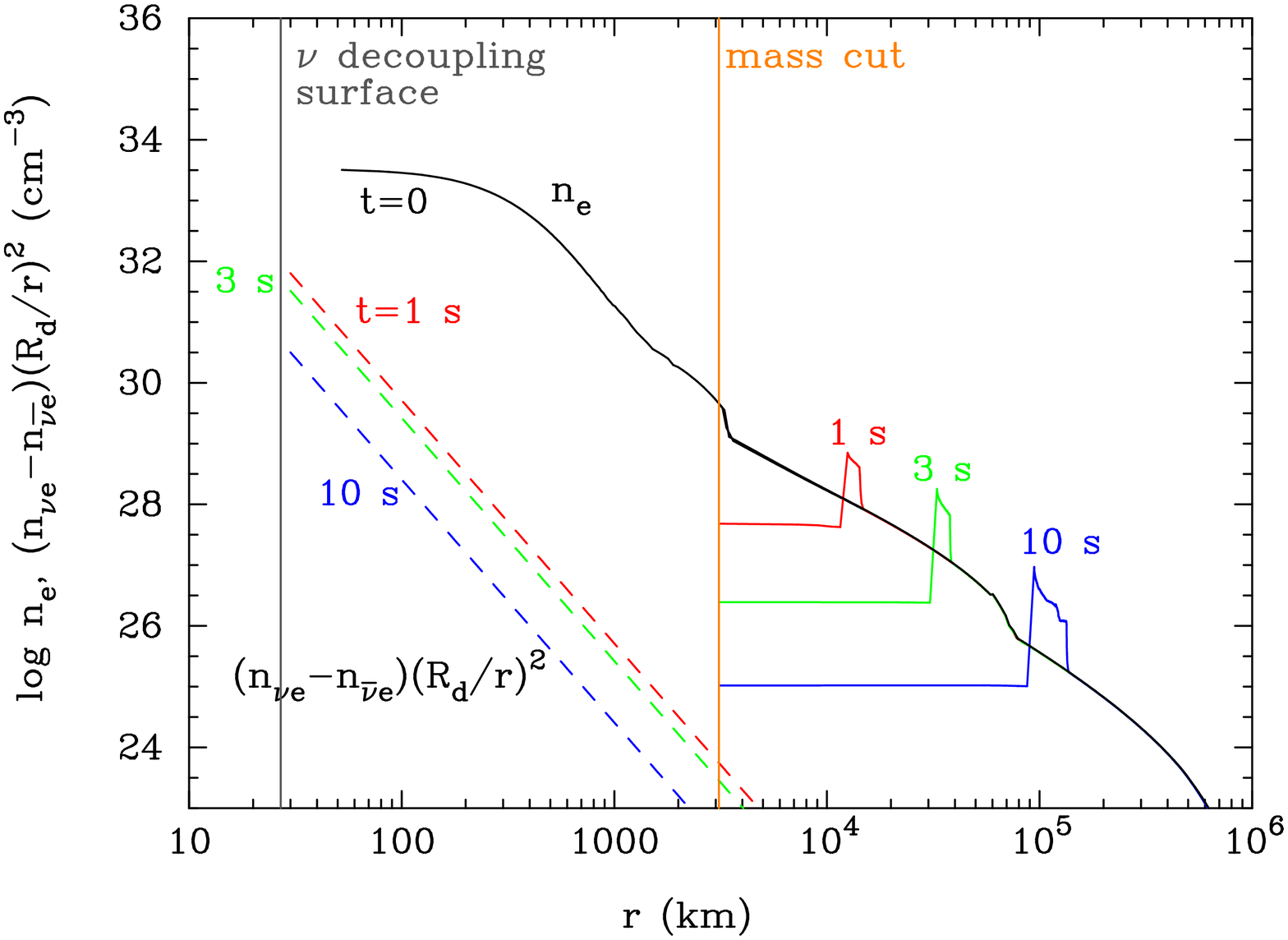}
\caption{Electron number density $n_e(r)$ and the neutrino effective number density $n^{\rm eff}_\nu(r)$ as a function of radius (see text). The inner vertical line shows the $\nu$ decoupling radius in the model of \citet{2015PhRvD..91f5016W}, while the outer vertical line is at the position of the mass cut at $t=0$. Solid lines show the $n_e$ values at $t=0$, 1, 3, and 10 s, respectively. Dashed lines show the $n^{\rm eff}_\nu(r)$ values at $t=1$, 3, and 10 s, respectively.}
\label{fig:potential}
\end{figure}

Figure \ref{fig:potential} shows the electron number density $n_e(r)$ and the quantity $n^{\rm eff}_\nu(r)\equiv [n_{\nue}(r) -n_{\nueb}(r)](R_{\rm d}/r)^2$ as a function of radius, where $n_{\nu_\alpha}$ and $n_{\bar{\nu}_\alpha}$ are the number densities of neutrinos and antineutrinos, respectively, of flavor $\alpha$, and $R_{\rm d}$ is the decoupling radius of neutrinos. We assume spherical symmetry in the electron number density and the azimuthal ($\phi$) symmetry in the neutrino number density for any fixed angle $\theta$ along a line from the stellar center to the neutrino decoupling surface. In this case, the effective Hamiltonians for neutrino-electron forward scattering and neutrino-neutrino forward scattering are, respectively, roughly proportional to $n_e(r)$ and $\sum_\alpha(n_{\nu_\alpha}(r) -n_{\nu_\alpha}(r))(1-\cos^2 \theta_{\rm max} )$. For large radii of $r \gg R_{\rm d}$, it approximately follows that $1-\cos^2 \theta_{\rm max} \sim (R_{\rm d}/r)^2$. In addition, under the present assumption, we have $n_{\nu_\alpha}=n_{\bar{\nu}_\alpha}$ for $\alpha =\mu$ and $\tau$. Then, the above two quantities provide information on the order of magnitude of respective potential terms.

The inner vertical line is located at $R_{\rm d}$ in the model of \citet{2015PhRvD..91f5016W}. The outer vertical line is at the position of the mass cut at $t=0$. Material inside this radius is assumed to collapse to the NS. Solid lines show the $n_e$ values at $t=0$, 1, 3, and 10 s, respectively. Dashed lines show the effective neutrino number density $n^{\rm eff}_\nu(r)$ at $t=1$, 3, and 10 s, respectively. It is seen that above the mass cut, the condition $n_e \gg n^{\rm eff}_\nu$ is always satisfied. Therefore, the neutrino-neutrino forward scattering term is unimportant in this region. However, it is not clear whether the opposite condition $n_e \lesssim n^{\rm eff}_\nu$ is realized in the inner region since the inner region is not treated in the present hydrodynamical calculation. The role of collective oscillations cannot be assessed with the current model and therefore is neglected. A sensitivity of the flavor change to SN explosion models should be studied in detail in the future. In order to obtain a realistic result on the neutrino forward scattering effect, one needs results of consistent calculations including the NS formation, matter heating by neutrinos and the explosion dynamics.

\subsection{$\nu$ reaction cross sections}\label{sec2f}
\subsubsection{Light nuclides}
We adopt NC cross sections for the spallation of $^4$He and $^{12}$C by neutrinos and antineutrinos separately based upon the models WBP and SFO \citep{2008ApJ...686..448Y}, respectively \citep[calculated in][but separate data have not been published]{2008ApJ...686..448Y}.  NC cross sections for neutrinos and antineutrinos differ from each other by up to a factor of 2.  Since the spectra of $\nu_e$ and $\bar{\nu}_e$ are different in general, we need to analyze the cross sections of neutrinos and antineutrinos separately.  In previous studies \citep{2008ApJ...686..448Y}, average values of the NC cross sections for neutrinos and antineutrinos were used.  However, calculations with those average values are inaccurate.

\begin{figure}[t!]
  \epsscale{1}
\plotone{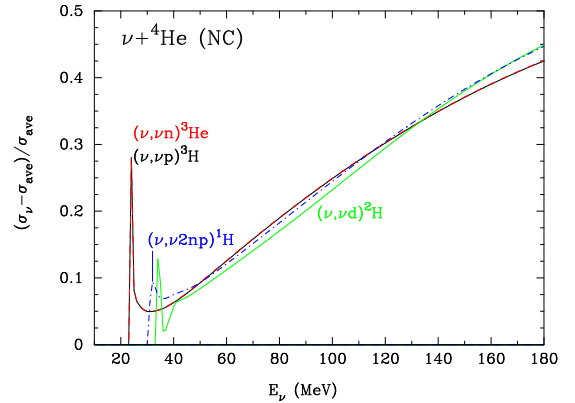}
\caption{Fractional differences of NC neutrino-induced $^{4}$He spallation cross sections from the averages for neutrinos and antineutrinos as a function of the neutrino energy $E_\nu$.}
\label{fig:cross_he4}
\end{figure}

\begin{figure*}[t!]
\plottwo{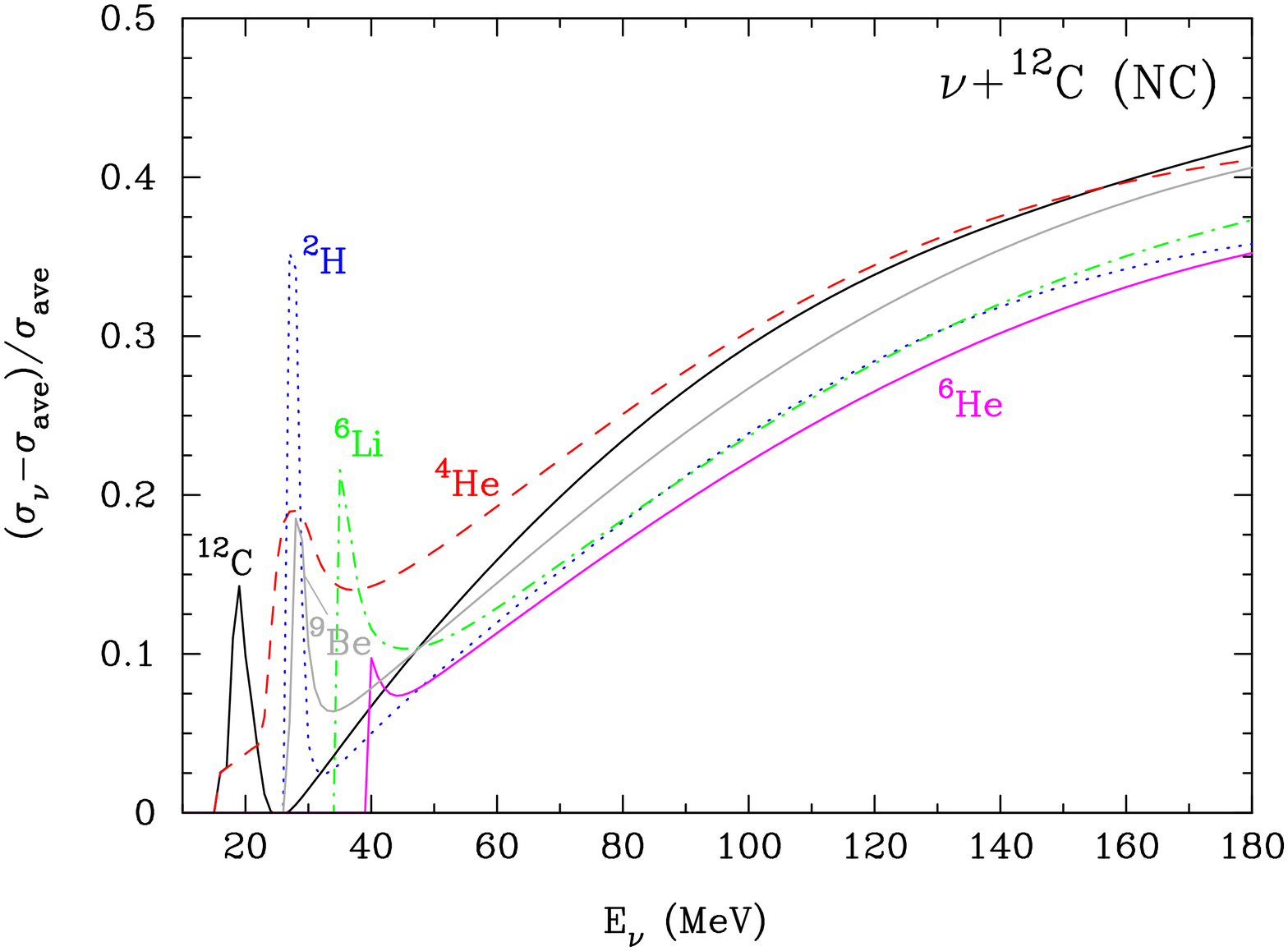}{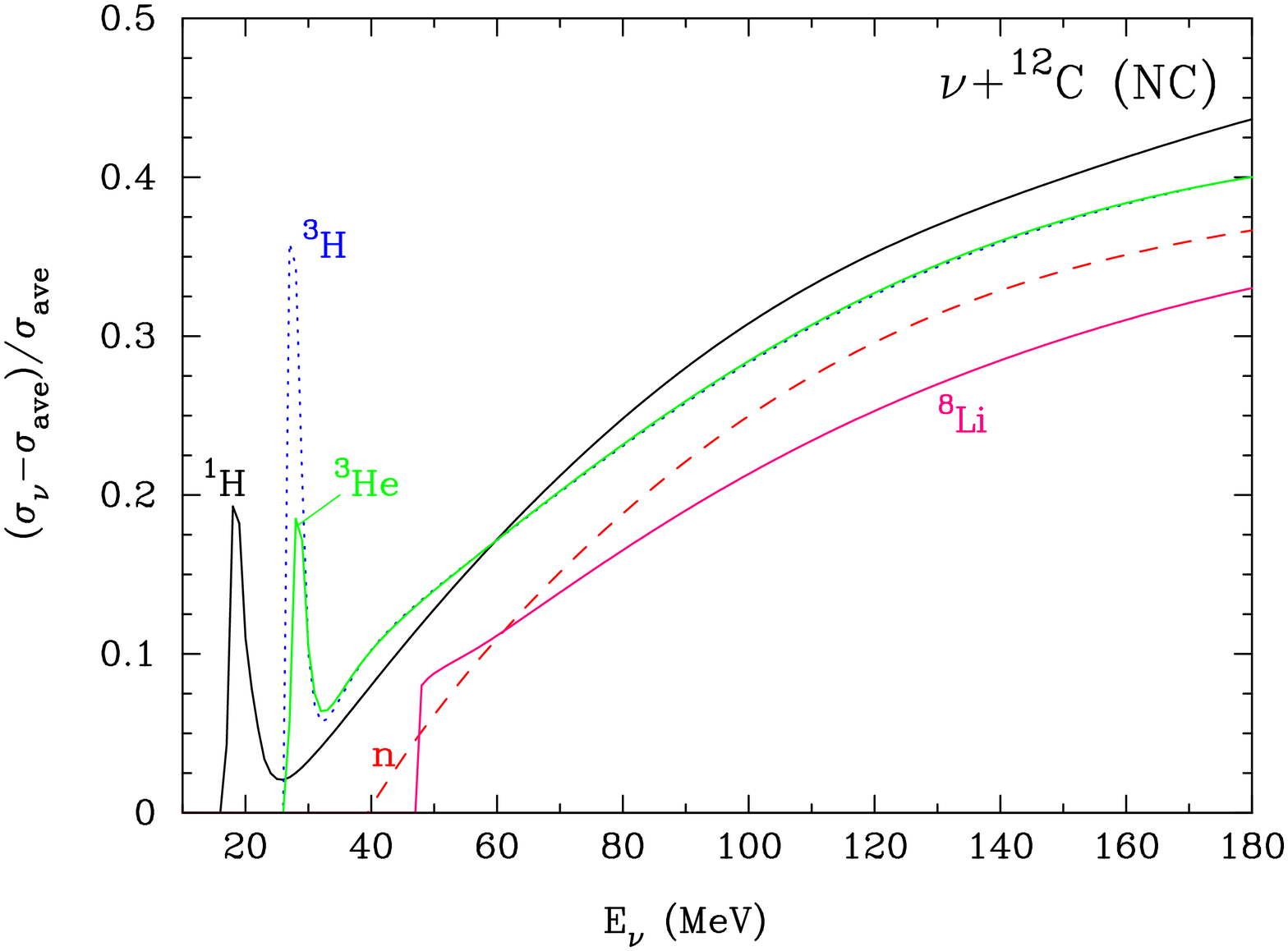}
  \epsscale{0.5}
\plotone{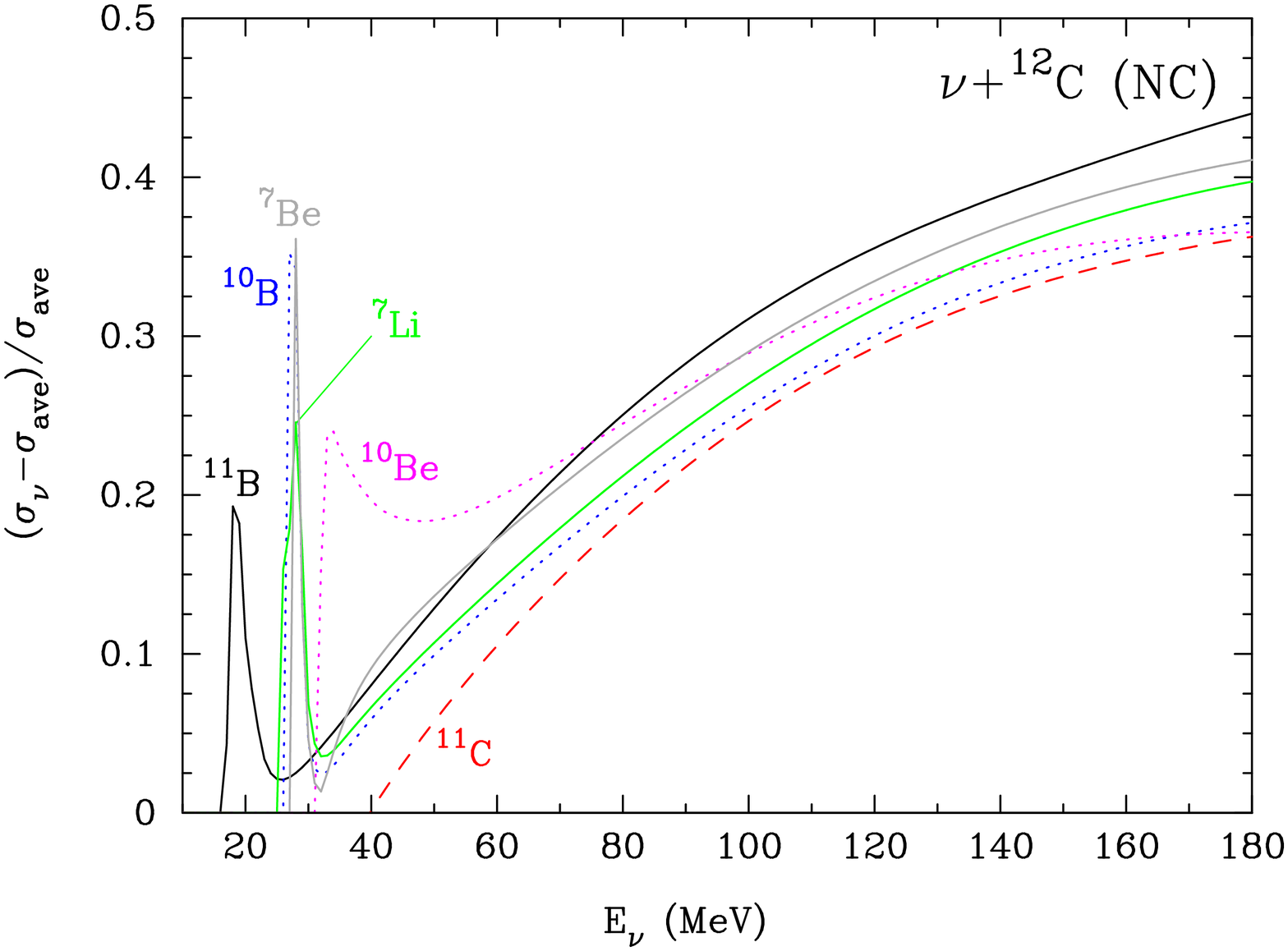}
\caption{Fractional differences of NC neutrino-induced cross sections for $^{12}$C destruction and the production of light nuclei from the average cross sections for neutrinos and antineutrinos as a function of the neutrino energy $E_\nu$.}
\label{fig:cross_c12}
\end{figure*}

  Figures \ref{fig:cross_he4} and \ref{fig:cross_c12} show the fractional differences of neutrino NC cross sections for $^{4}$He and $^{12}$C, i.e., $(\sigma_\nu -\sigma_{\rm ave})/\sigma_{\rm ave}$, respectively, as a function of the neutrino energy $E_\nu$, where $\sigma_\nu$ is the cross section of neutrinos and $\sigma_{\rm ave}$ is the average cross section of neutrinos and antineutrinos. Here we see that the difference can be as large as a factor of 0.4 particularly at the highest energies corresponding to a factor of $\sigma_\nu /\sigma_{\bar{\nu}} \approx 2$.

\subsubsection{Heavy nuclei}
We use averaged cross sections \citep{2012PhRvC..85f5807C} \footnote{We note typographical errors related to the $^{92}$Nb production in Table 1 in \citet{2012PhRvC..85f5807C}}, assuming that flavor-changes do not occur. This assumption is rather good since destruction via the $\gamma$-process operates in relatively high temperature environments realized in the inner stellar region. On the other hand, neutrino flavor changes occur for typical neutrino energies in the region outside of where the MSW resonance occurs.

Neutrino spallation of nuclides can affect the abundances of $n$ and $p$ during the SN nucleosynthesis.  The rates of neutrino reactions on nuclei with $A \leq 70$ other than $^4$He and $^{12}$C are adopted from Hoffmann and Woosley \footnote{http://dbserv.pnpi.spb.ru/elbib/tablisot/toi98/www/astro- /hw92\_1.htm}.  We take their data for $T_{\nu_e} =T_{\bar{\nu_e}} =4$ MeV and $T_x =6$ MeV for $x =\nu_\mu$, $\bar{\nu}_\mu$, $\nu_\tau$, and $\bar{\nu}_\tau$.  The branching ratios for the NC reactions of $\nu_e$ and $\bar{\nu}_e$ are not available. Hence, we take the branching ratios at $T_x =6$ MeV also for $\nu_e$ and $\bar{\nu}_e$.

There are several possible reaction pathways for $^{98}$Tc production:
$^{97}$Mo($p$, $\gamma$)$^{98}$Tc, 
$^{97}$Tc($n$, $\gamma$)$^{98}$Tc, 
$^{98}$Mo($p$, $n$)$^{98}$Tc, 
$^{98}$Mo($\nu_e$, $e^-$)$^{98}$Tc, 
$^{99}$Ru($\gamma$, $p$)$^{98}$Tc, and
$^{99}$Tc($\gamma$, $n$)$^{98}$Tc.
The reaction $^{98}$Ru($\bar{\nu}_e$, $e^+$)$^{98}$Tc is also possible. However, the abundance of $^{98}$Ru (a $p$-nuclide) is much smaller than that of $^{98}$Mo, so that the production of $^{98}$Tc via this reaction is negligible.  Also, the photodisintegration reaction $^{102}$Rh($\gamma$, $\alpha$)$^{98}$Tc is negligible since the initial abundance of the unstable isotope $^{102}$Rh is small. Results for the production of heavy nuclei including $^{98}$Tc based upon this network code are published elsewhere (See \citet{Hayakawa2017} for $^{98}$Tc production and \citet{Ko2019} for general nucleosynthesis taking into account neutrino self-interaction effects).

\subsection{Adopted models}\label{sec2g}
We adopt five models for SN nucleosynthesis. First, for the standard model of initial nuclear abundances based on a presupernova stellar model (Sec. \ref{sec2a}), the following three cases are adopted: (1) the neutrino oscillations are taken into account in the normal mass hierarchy; (2) the neutrino oscillations are taken into account in the inverted mass hierarchy; and (3) the neutrino oscillation is neglected \footnote{We also calculated for the case in which all neutrino reactions are neglected. In this case, however, yields of $^7$Li and $^{11}$B are negligibly low, and we can safely consider that their SN yields are effectively zero.}. A comparison of results for these cases shows the effect of neutrino flavor change on SN nucleosynthesis. Second, for the case of an inverted mass hierarchy, we have two additional cases of initial abundances that are different from case (2) above: (4) abundances of elements heavier than $^{40}$Ca are proportional to the metallicity, i.e., $Z =Z_\sun /4$ \citep[this setup roughly corresponds to that of the previous studies;][]{2005PhRvL..94w1101Y,2006PhRvL..96i1101Y,2006ApJ...649..319Y,2008ApJ...686..448Y,2004ApJ...600..204Y}; (5) abundances of elements heavier than $^{40}$Ca are equal to solar abundances. Thus, in Cases 4 and 5, initial abundances of the heavy nuclides are not the presupernova abundances calculated for Cases 1 to 3. We can deduce effects of a presupernova $s$-process and metallicity on the SN nucleosynthesis by comparing results of Cases 2, 4, and 5.
For each model we divide the nucleosynthesis discussion into results in five Lagrangian grid zones centered at: 1) $M_r = 2 M_\sun$; 2) $M_r = 3.7 M_\sun$; 3) $M_r =4 M_\sun$;  4) $M_r =4.5 M_\sun$; and 5) $M_r =5.9 M_\sun$.

\section{effect of neutrino oscillations}\label{sec3}
\subsection{Li and B yields}\label{sec3a}

\begin{figure*}[t!]
\plottwo{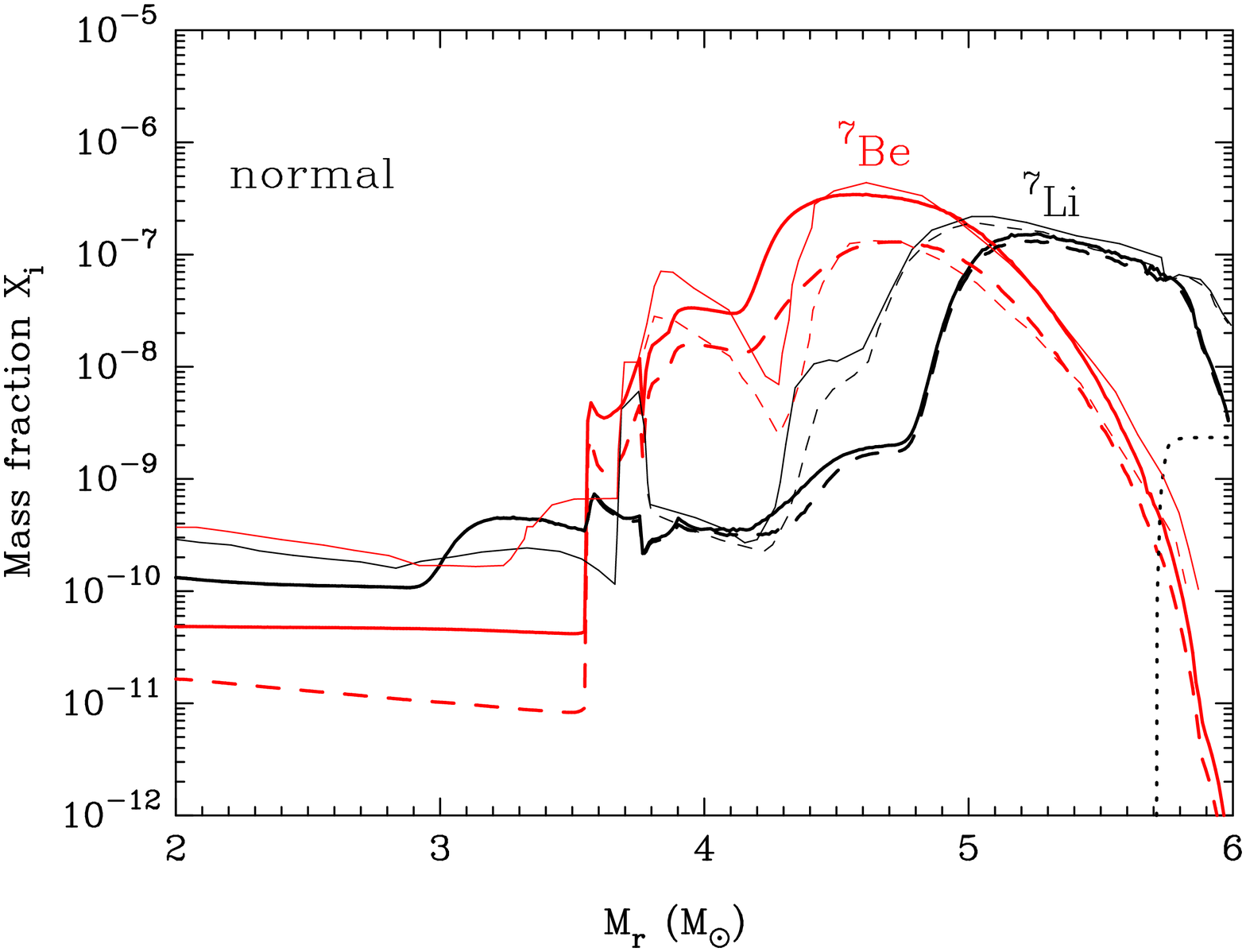}{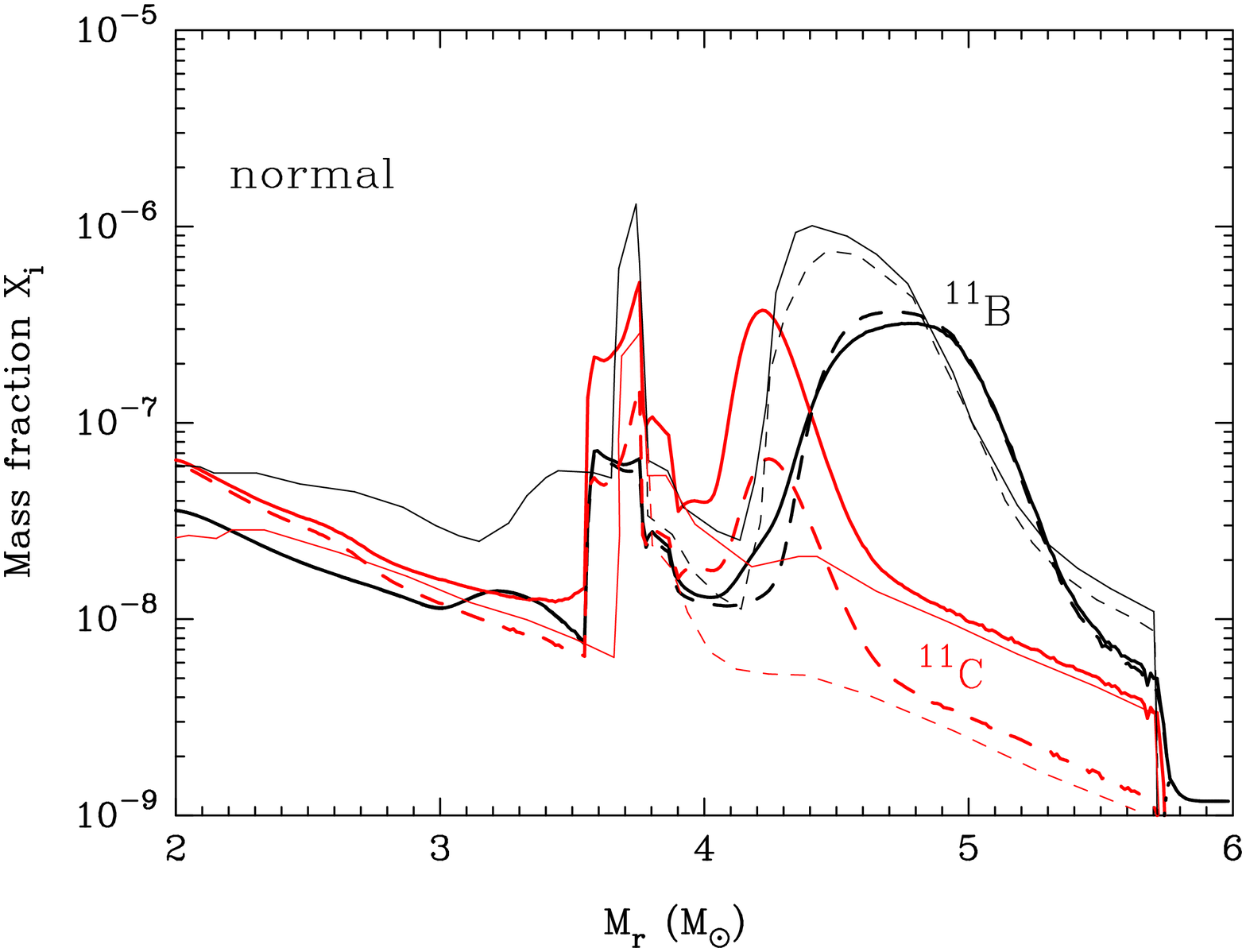}
\plottwo{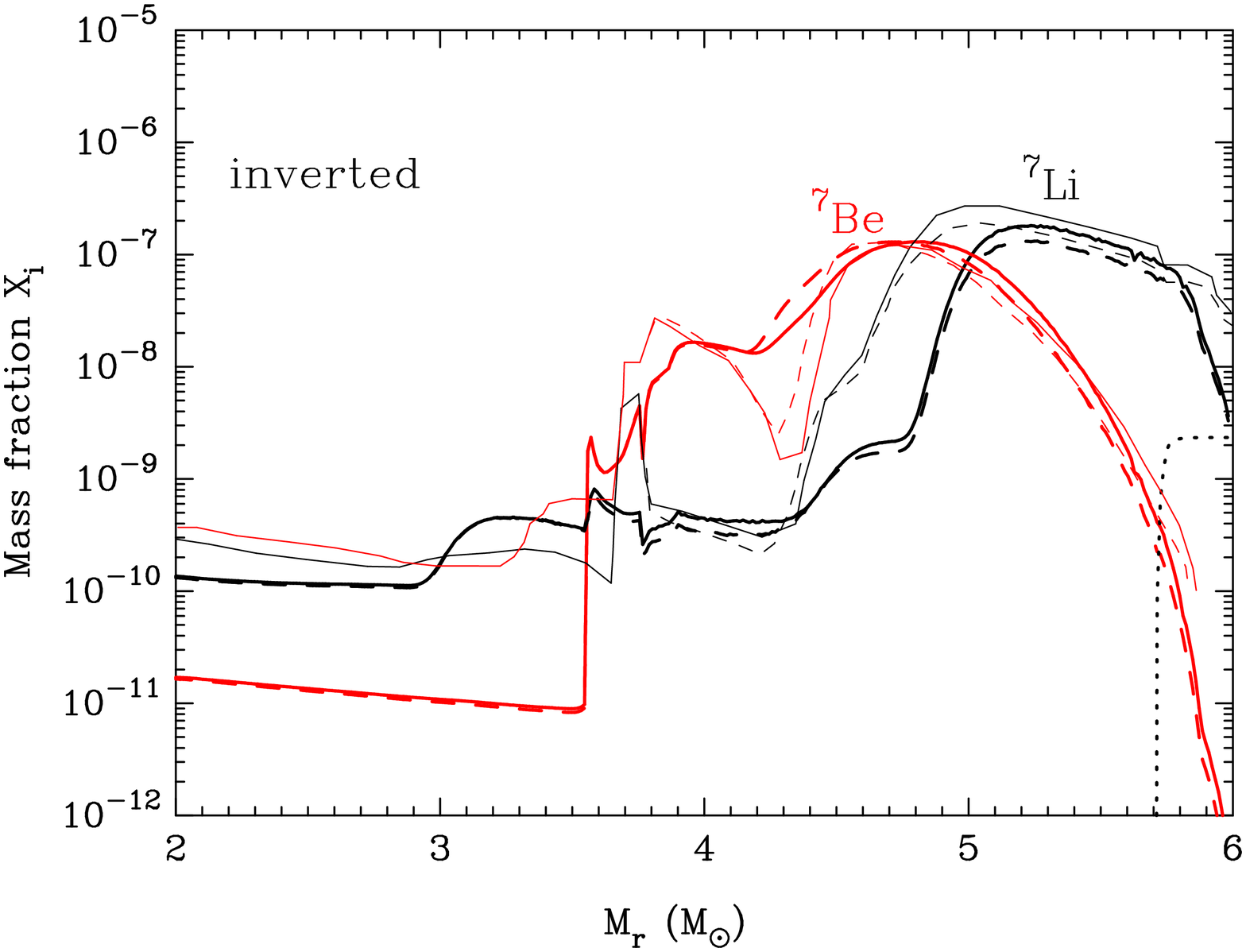}{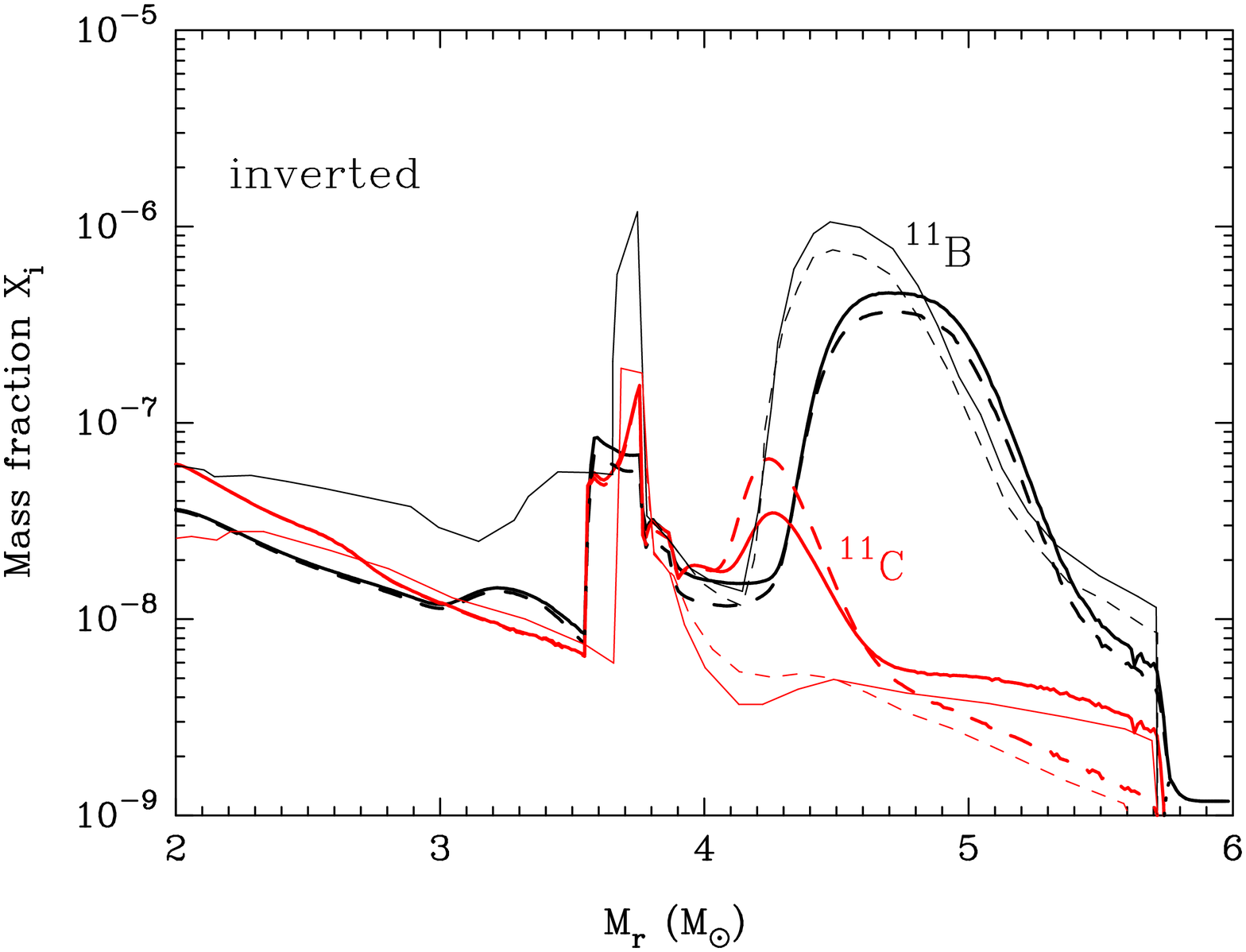}
\caption{Mass fractions of $^7$Li and $^7$Be (left panels) and $^{11}$B and $^{11}$C (right panels) versus Lagrangian mass coordinate. Thick and thin lines correspond to the present calculated results and the previous results of \citet{2006ApJ...649..319Y}, respectively. Solid lines correspond to the normal mass hierarchy (upper panels) and the inverted mass hierarchy (lower panels), respectively. Dashed lines correspond to the case in which the neutrino oscillations are neglected.}
\label{fig:li_b_nor_inv}
\end{figure*}

Figure \ref{fig:li_b_nor_inv} shows the calculated final mass fractions of $^7$Li and $^7$Be (left panels), and $^{11}$B and $^{11}$C (right panels), respectively at 50 s after the start of explosion as a function of Lagrangian mass coordinate (thick lines). Also plotted are the previous results \citep{2006ApJ...649..319Y} (thin lines) taken from figure 3 of that paper for a model with neutrino flavor change probabilities. Solid lines show results of the normal mass hierarchy (upper panels) and the inverted mass hierarchy (lower panels), respectively. Dashed lines correspond to the case in which the neutrino oscillations are neglected.

\subsubsection{Normal hierarchy}
In the He-rich layer, the global trends of $^7$Li and $^7$Be abundances are similar to the previous results. Especially, the $^7$Be yields at $M_r \gtrsim 4.5 M_\sun$ are very similar. However, there are differences. In our results, there are small peaks of $^{11}$C yields at $M_r \sim 4.2 M_\sun$.

Inside the O-rich layer at $M_r \lesssim 3.8 M_\sun$, abundances are not much different from those of the previous calculation \citep{2008ApJ...686..448Y}. However, the $^7$Be abundance differs the most. Our results for the inverted hierarchy and no oscillation cases are more than one order of magnitude smaller than the previous results.

The main region for $^7$Li and $^{11}$B production is the He-rich layer. Differences in abundances between different cases in that region are observed for all nuclides. These differences come from the different hydrodynamical results and initial nuclear abundances adopted here. Our use of realistic light- and $s$-process nuclear abundances lead to a more accurate evolution of the $n$ and $p$ abundances that affects the $\nu$-process nucleosynthesis.

One difference in the innermost region from previous results \citep{2008ApJ...686..448Y} is that the abundances of $^7$Be and $^{11}$C depend on the neutrino oscillations in our calculation. We checked the abundance evolution at $M_r =2 M_\sun$ (see Sec. \ref{sec3b}),
and found that this difference occurs at a late time of $t \gtrsim 10$ s. At $t \sim 10$ s, the radius has increased from $1 \times 10^9$ cm to $\sim 4 \times 10^{10}$ cm. When this shell expands to the MSW resonance location for typical neutrino energies, the neutrinos effectively change their flavors. Therefore, an observable effect is seen in the figure. However, since the yields of $^7$Be and $^{11}$C are very small in this innermost region, this difference is negligible in the total SN yields.

In this layer, the final abundances of $^7$Li and $^7$Be are contributed to by the late production after shock heating. A large difference between cases is seen for $^7$Be, while no large difference is seen for $^7$Li.
There are also slight differences in the abundance of $^{11}$C.

There is the following general effect of the $\nu+^4$He reactions: The MSW effect increases the $\nue$ rates. Yields of $^3$He and $p$ are then increased. As a result, the production of $^7$Be and $^{11}$C via $^3$He($\alpha$,$\gamma$)$^7$Be($\alpha$,$\gamma$)$^{11}$C becomes stronger. In addition, the destruction of $^{11}$B is stronger due to the enhanced rate of $^{11}$B($p$,2$\alpha$)$^4$He.
There is also a slight increase in the $^{11}$B abundance at $M_r \lesssim 4.4 M_\sun$ and a slight decrease at $M_r \gtrsim 4.4 M_\sun$.

$^{11}$B is burned via the $^{11}$B($p$,2$\alpha$)$^4$He reaction from protons produced by the $\nue$ reactions on $^4$He and $^{12}$C (shells 2--4 in Sec. \ref{sec3b}).
The $^1$H abundance is the highest in the normal hierarchy case (shells 2--5). This enhanced $^1$H abundance leads to effective $^{11}$B burning. An enhancement of the $^7$Li destruction via $^7$Li($p$,$\alpha$)$^4$He occurs also (shells 2--4) although the reaction $^7$Li($\alpha$,$\gamma$)$^{11}$B is the main channel for $^7$Li destruction.

\subsubsection{Inverted hierarchy}
The total abundance of $A=7$ nuclei is slightly larger in the inverted mass hierarchy case than in the case of no oscillations. The region of $^7$Be production is shifted to the outer region. On the other hand, the $^7$Li production region does not change, and yields are enhanced globally.
The total abundance of mass-11 nuclei is also slightly larger. 
The $^{11}$B abundance is larger in the whole He-rich layer. The peak $^{11}$C yield at $M_r \sim 4.2$--4.3$M_\sun$ is smaller than that of the no oscillation case.

In the range of $M_r =4.2$--4.7$M_\sun$ (shell 4), the $^7$Be abundance is smaller than in the no oscillation case, while the $^7$Li abundance is the same.
Neutrino oscillations enhance the $\bar{\nue}$ reaction rates.
The $^3$He abundance is slightly smaller than that of the no oscillation case before the shock heating.
This is because of the more effective operation of the $^3$He($n$,$p$)$^3$H reaction by the enhanced $n$ abundance.
As a result, the production of $^7$Be is smaller than in the no oscillation case, and the resultant $^7$Be abundance is smaller.
After the shock heating, $^7$Be nuclei are produced via the $^3$He($\alpha$,$\gamma$)$^7$Be reaction and destroyed by the $^7$Be($n$,$p$)$^7$Li reaction. During the temperature peak of the shock, the $^3$He abundance is smaller than that of the no oscillation case, and the $^7$Be production is slightly weaker. In addition, the destruction rate of $^7$Be($n$,$p$)$^7$Li is slightly larger. As a result, the final abundance of $^7$Be is smaller than that of the no oscillation case (cf. Fig \ref{fig:abun_238}).

In the region of $M_r \lesssim 3.5 M_\sun$, there are slight differences in the abundances of $^{11}$B and $^{11}$C. The $\bar{\nue}$ reaction rates are enhanced by the MSW effect. In general, in the region of $M_r \gtrsim 3.5 M_\sun$ where the altered $\bar{\nue}$ spectrum arrives, the $^{11}$B abundance is increased. The $^{11}$B nuclei are produced via the CC $\bar{\nue}+^{12}$C reaction only, while $^{11}$C nuclei are produced via the CC $\nue+^{12}$C reaction only. Therefore, the change of the $\bar{\nue}$ spectrum affects the $^{11}$B abundance preferentially.
The $^{11}$C abundance is smaller at $M_r \lesssim 4.6 M_\sun$ and higher for $M_r \gtrsim 4.6 M_\sun$.

\subsection{Abundance evolution}\label{sec3b}

Figures \ref{fig:abun_034}--\ref{fig:abun_353} show nuclear abundances as functions of time for shells 1--5, respectively. The left panels show abundances of $n$, $p$, $d$, $t$, $^3$He, $^7$Li, and $^7$Be while the right panels show those of $^{11}$B, $^{11}$C, and $^4$He. Solid, dotted, and dashed lines correspond to the normal, inverted hierarchy, and no neutrino oscillation cases, respectively. Long dashed lines correspond to results in the case of no neutrino flux.

Table \ref{tab:rank1} shows the highest yield cases for respective nuclides and shells: NH, IH, and ``no'' refer to the normal hierarchy, inverted hierarchy, and the no oscillation cases, respectively. They are evaluated before the decay of unstable $^7$Be and $^{11}$C. If there are significant differences in yields between the second highest and lowest yield cases, the second highest yield case is indicated in brackets. Blanks indicate that yields are not much different between the three cases.

\begin{deluxetable*}{llllll}
\caption{\label{tab:rank1}
Highest yield cases for respective nuclides and shells}
\tablehead{
  \colhead{nuclide \textbackslash shell} & \colhead{1 ($M_r =2 M_\sun$)} & \colhead{2 ($M_r =3.7 M_\sun$)} & \colhead{3 ($M_r =4 M_\sun$)} & \colhead{4 ($M_r =4.5 M_\sun$)} & \colhead{5 ($M_r =5.9 M_\sun$)}
}
\startdata
 $^3$H   & IH & NH (IH) & IH (NH) & IH (no) & IH (NH) \\
 $^3$He  & NH & NH      & NH (IH) & NH (IH) & NH (IH) \\
 $^7$Li  &    & IH (NH) & IH      & NH (IH) & IH (NH) \\
 $^7$Be  & NH & NH      & NH      & NH (no) & NH (IH) \\
 $^{11}$B &    & IH      & IH      & IH &  \\
 $^{11}$C &    & NH      & NH      & NH & \\
\enddata
\end{deluxetable*}

\subsubsection{Shell 1}
\begin{figure*}[t!]
\plottwo{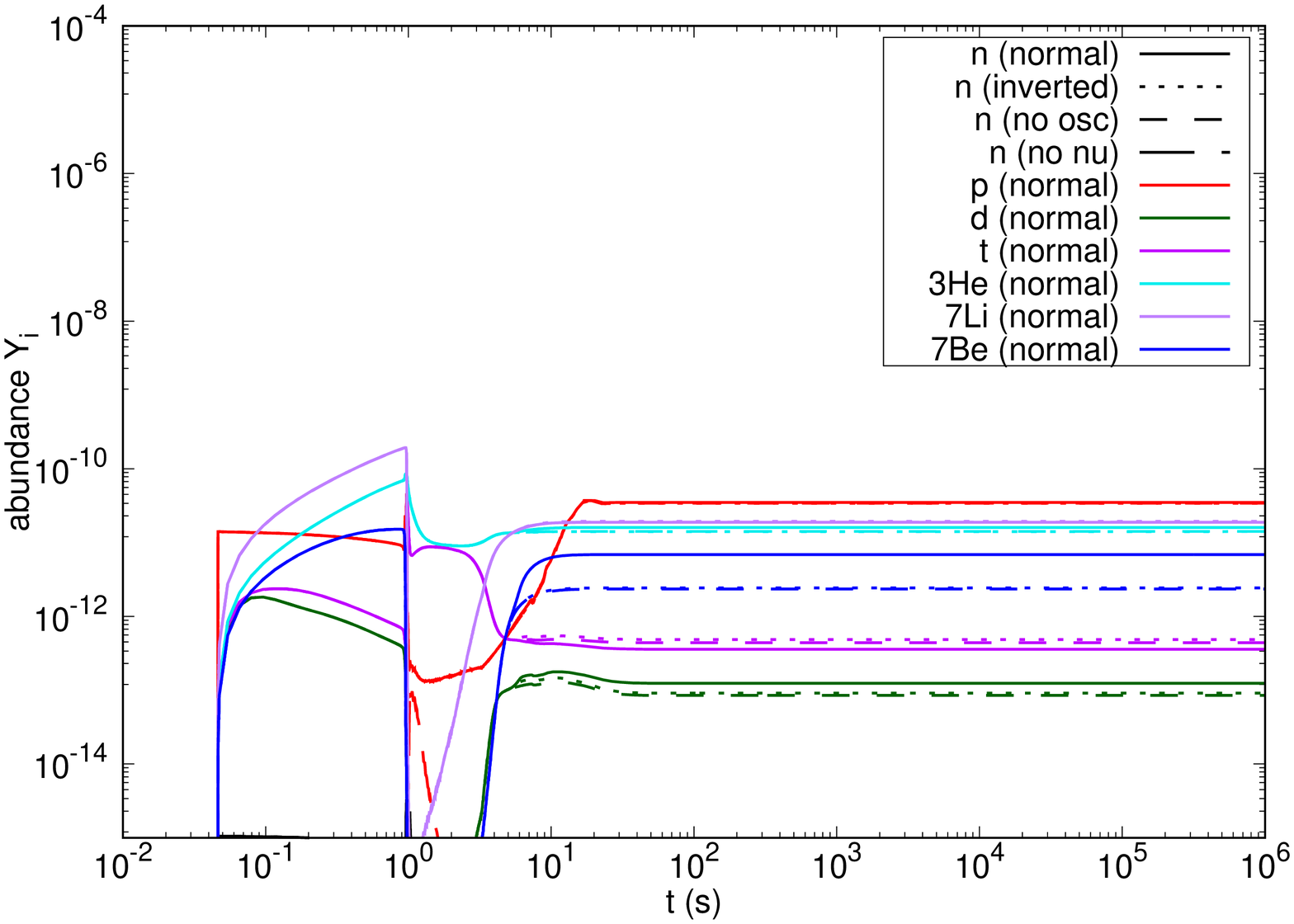}{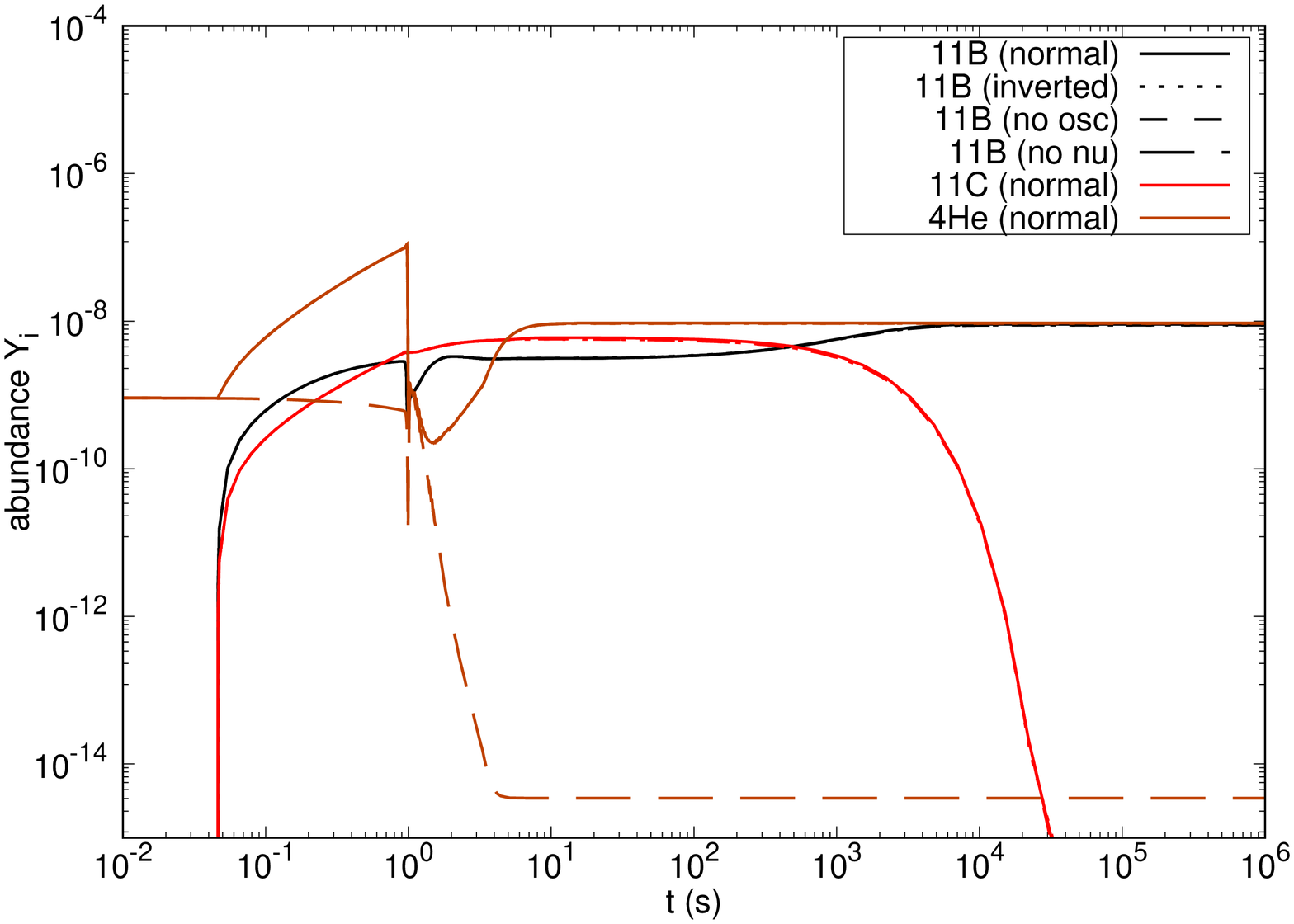}
\caption{Nuclear abundances as a function of time for shell 1 ($M_r=2 M_\sun$). Solid, dotted, and dashed lines correspond to results of the normal, inverted hierarchy, and no neutrino oscillation cases, respectively. Long dashed lines (not always discernable in the range plotted on this figure) correspond to results from the case of no neutrino flux.}
\label{fig:abun_034}
\end{figure*}

\begin{figure*}[t!]
\plottwo{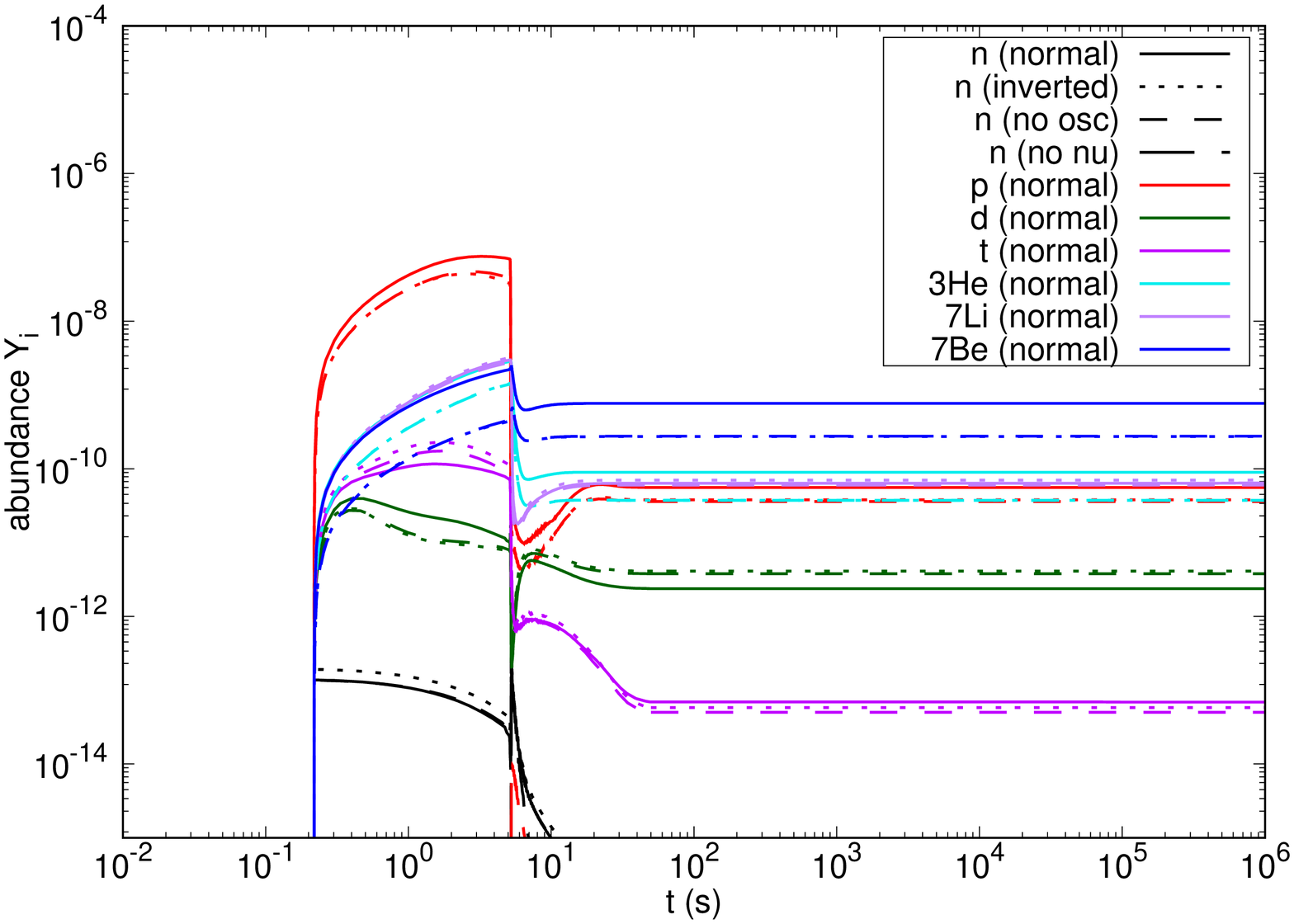}{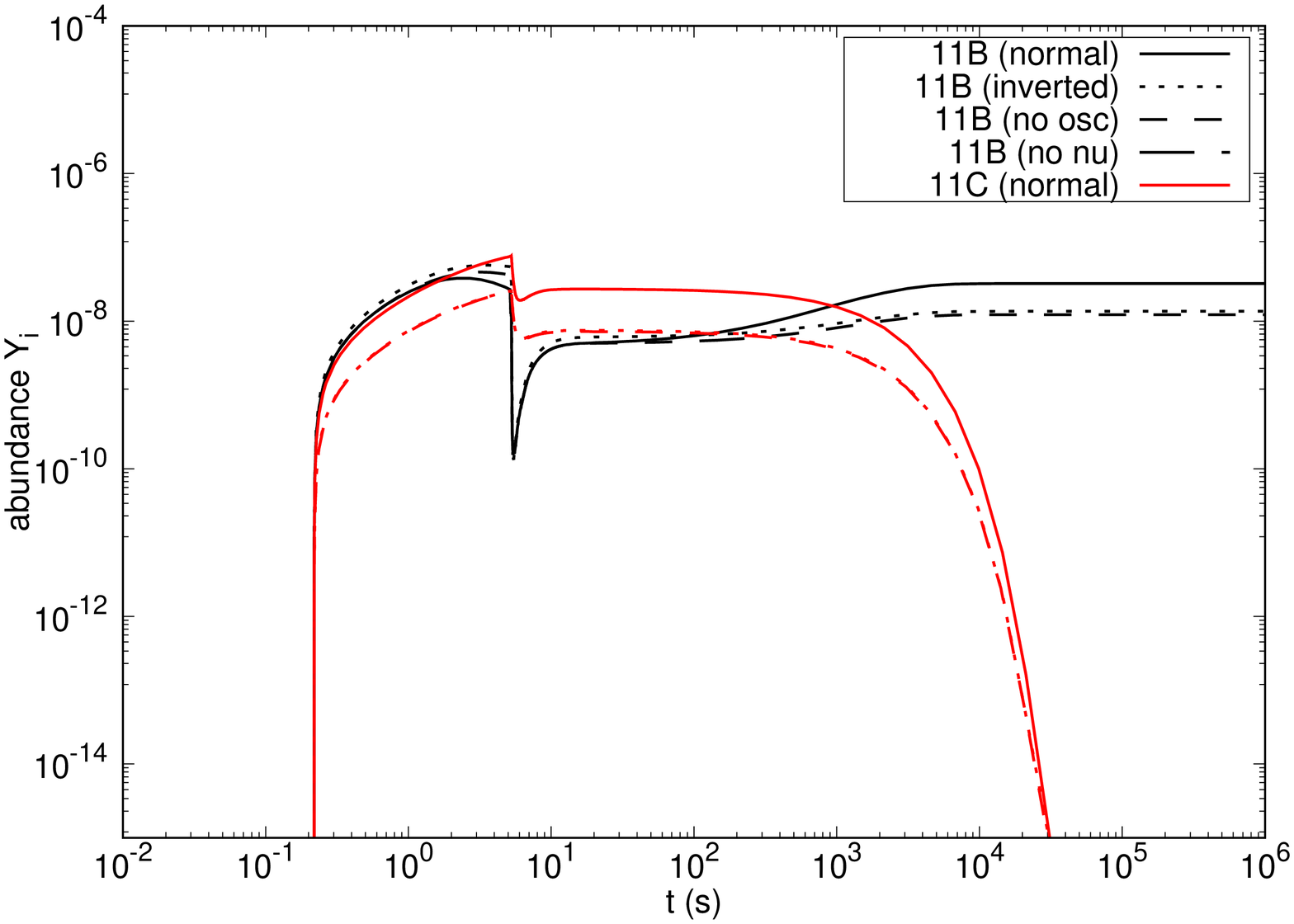}
\caption{Same as Fig. \ref{fig:abun_034} but for shell 2 ($M_r=3.7 M_\sun$).}
\label{fig:abun_173}
\end{figure*}

\begin{figure*}[t!]
\plottwo{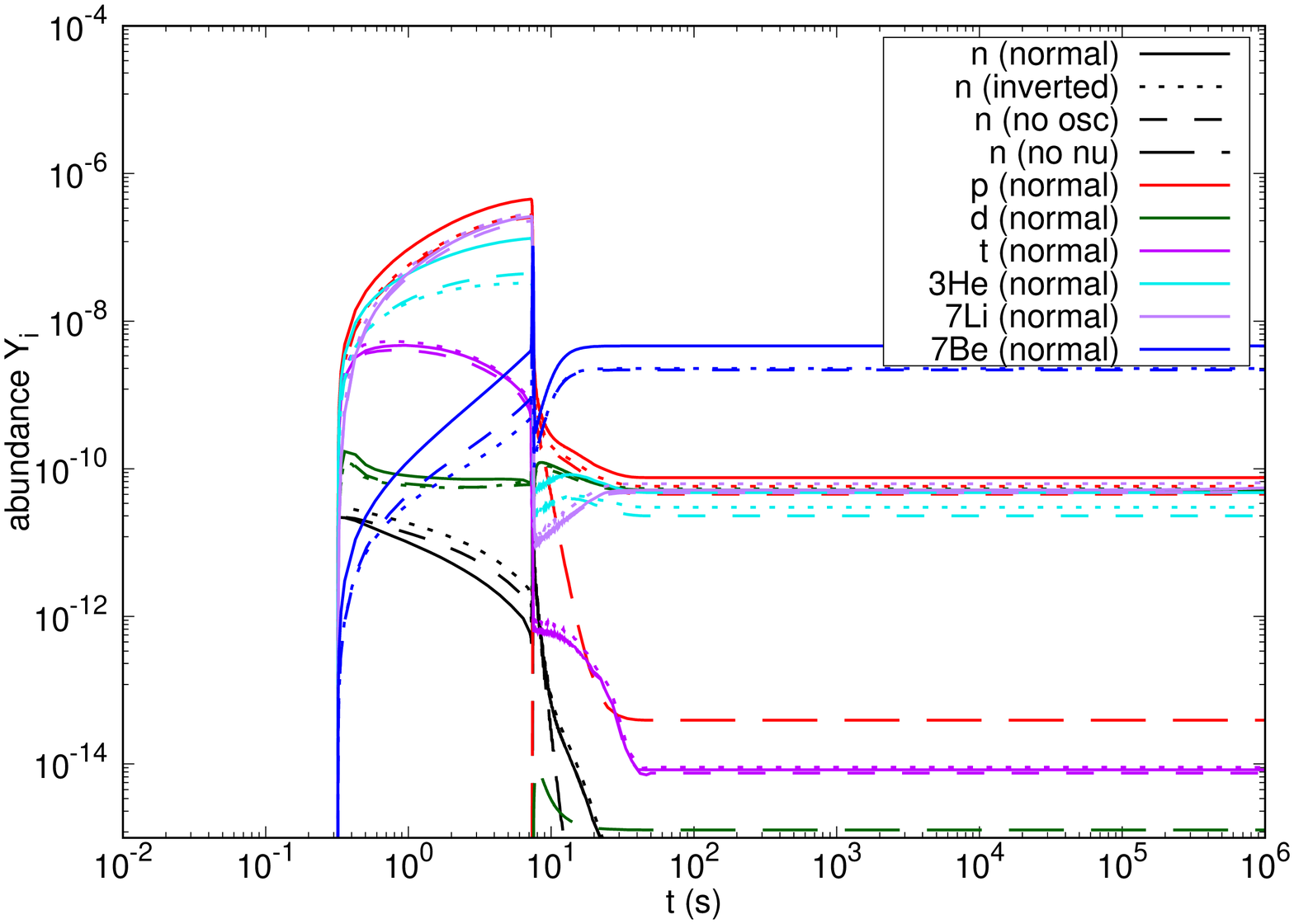}{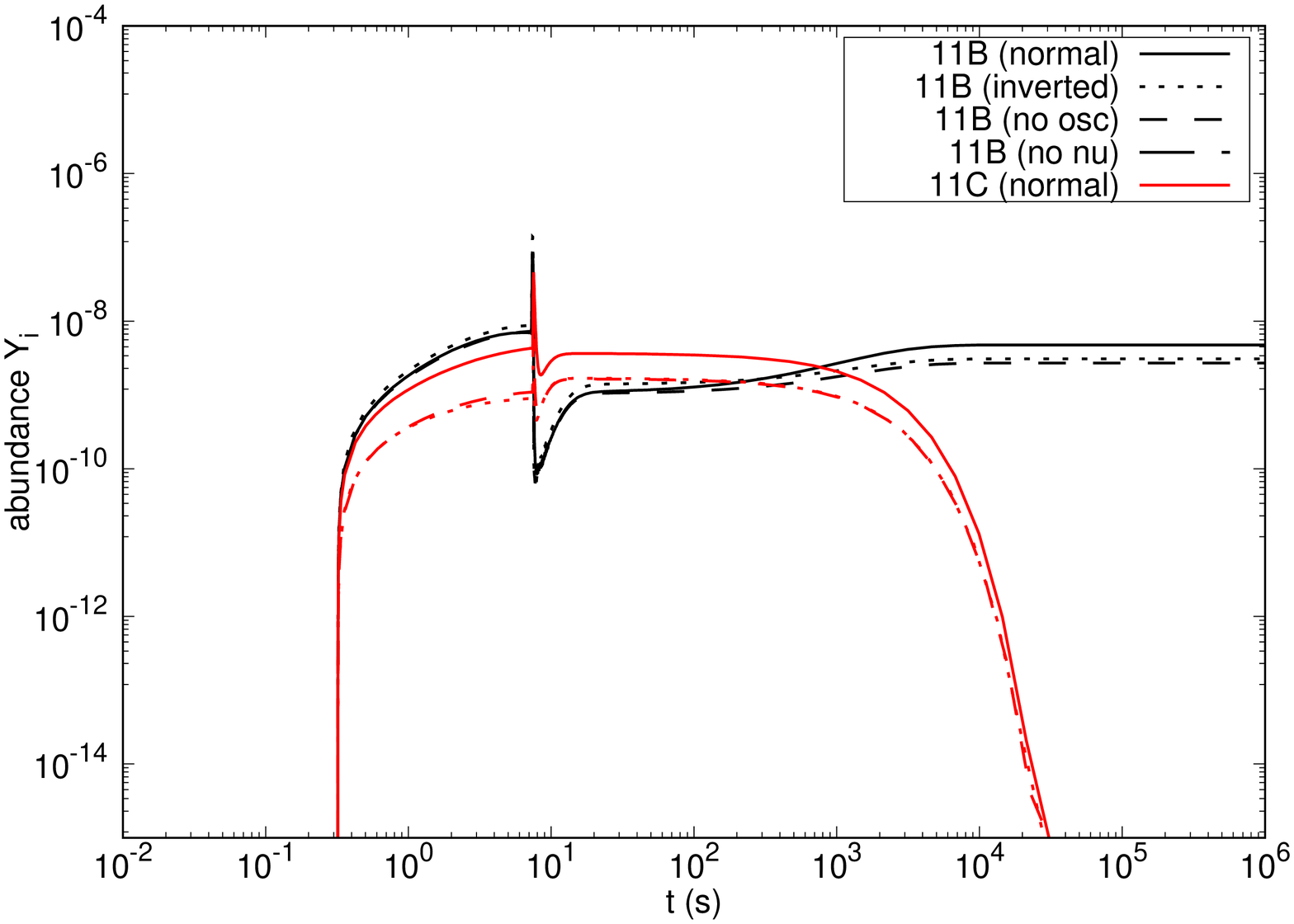}
\caption{Same as Fig. \ref{fig:abun_034} but for shell 3 ($M_r=4 M_\sun$).}
\label{fig:abun_198}
\end{figure*}

\begin{figure*}[t!]
\plottwo{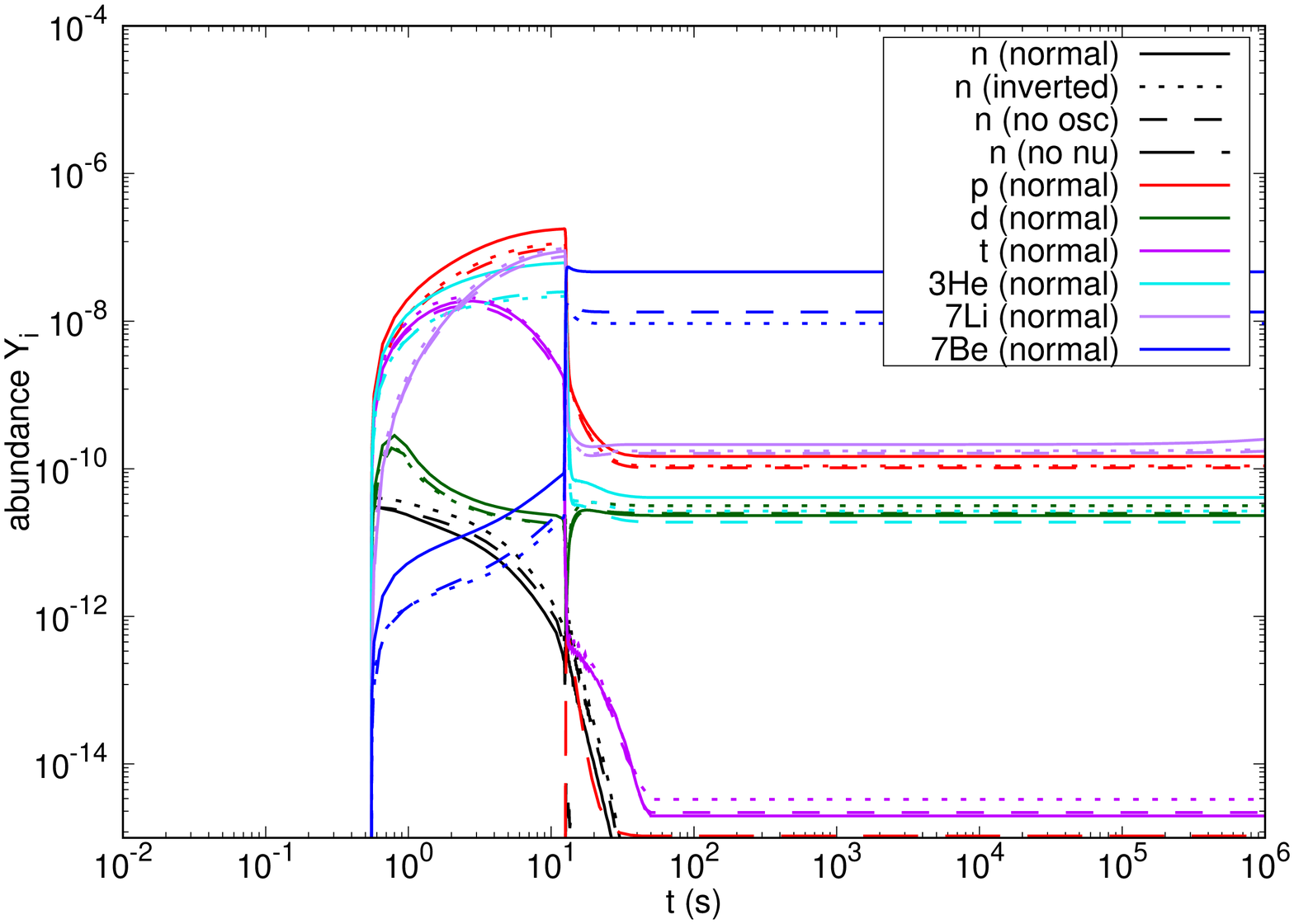}{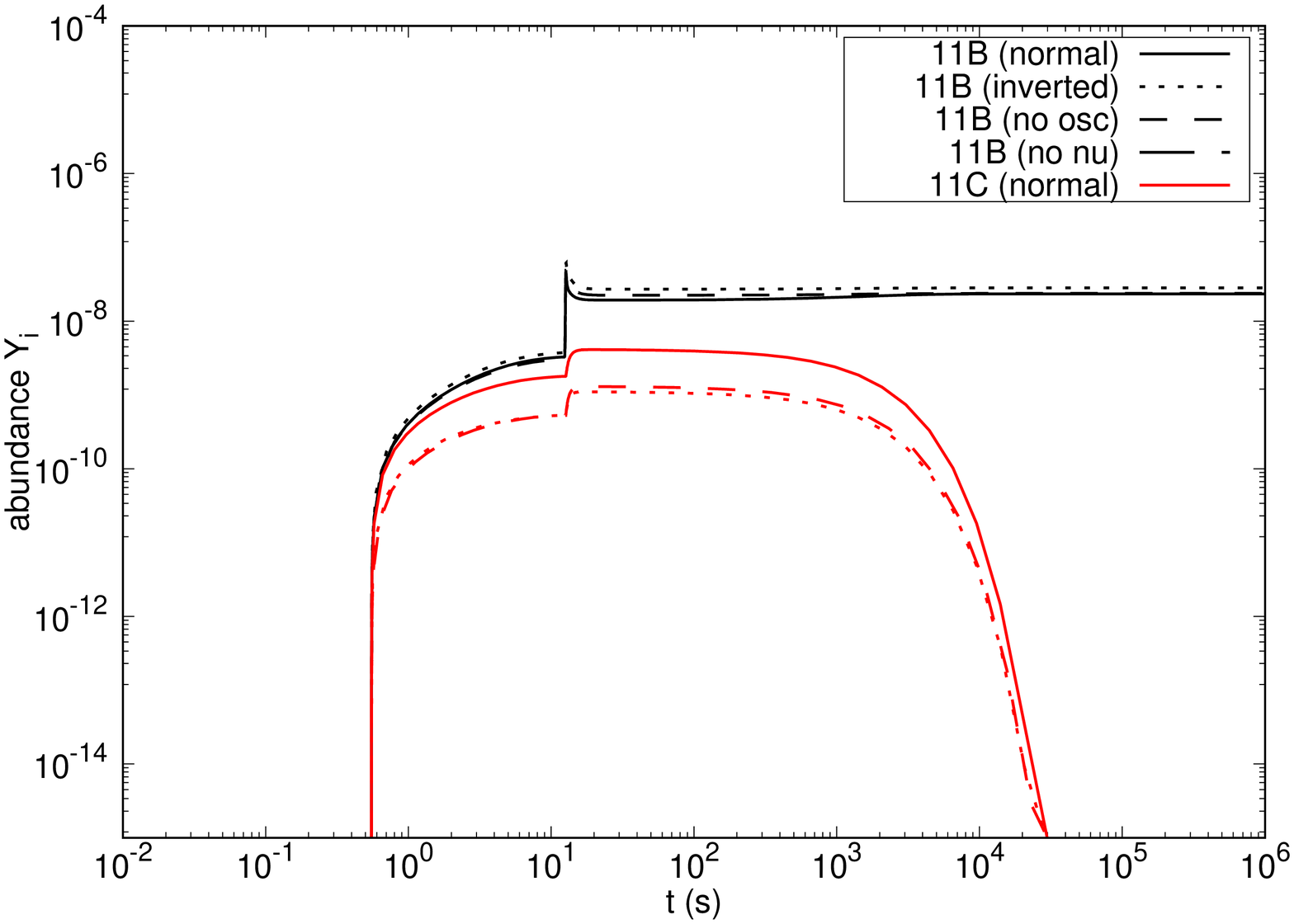}
\caption{Same as Fig. \ref{fig:abun_034} but for shell 4 ($M_r=4.5 M_\sun$).}
\label{fig:abun_238}
\end{figure*}

\begin{figure*}[t!]
\plottwo{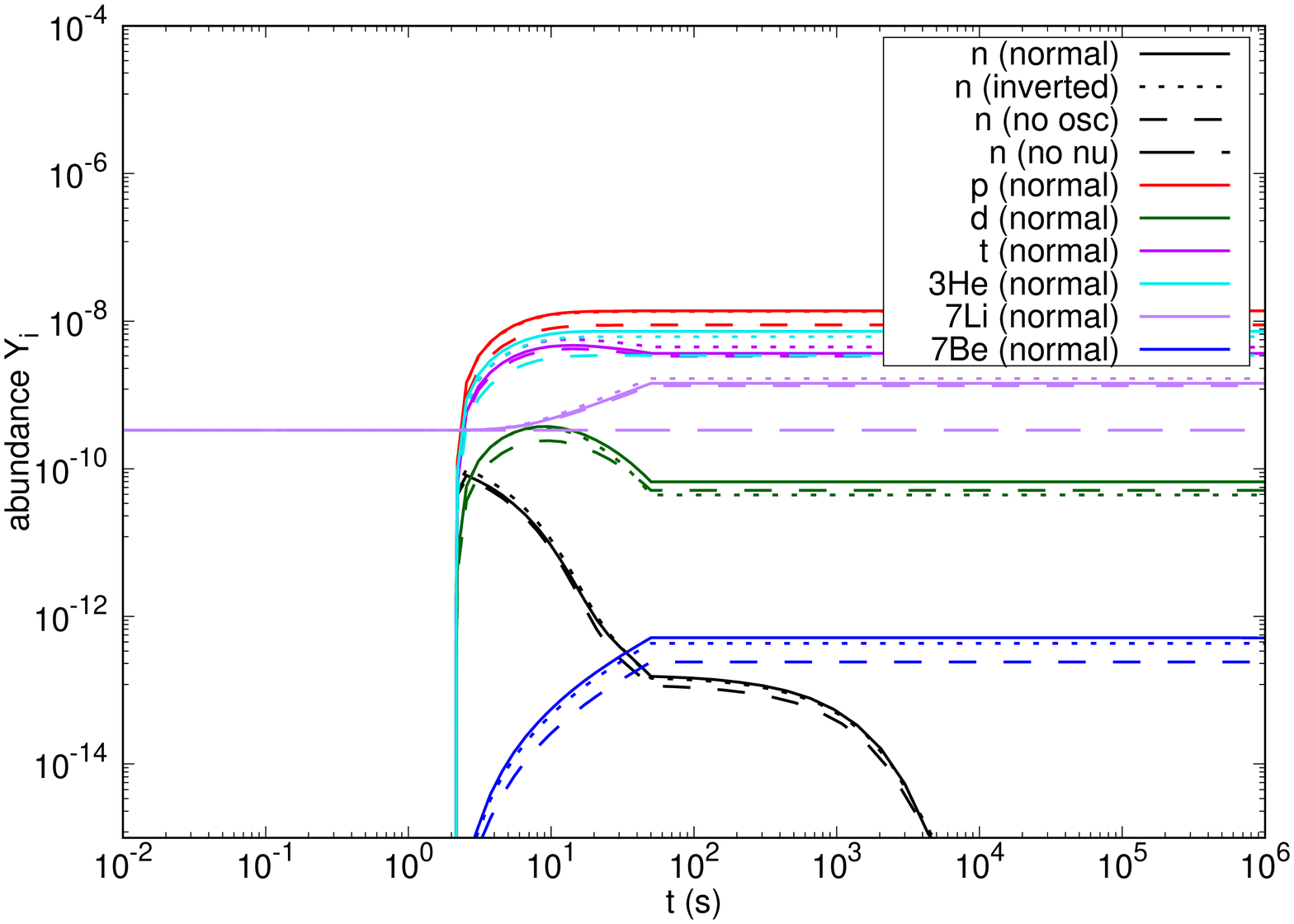}{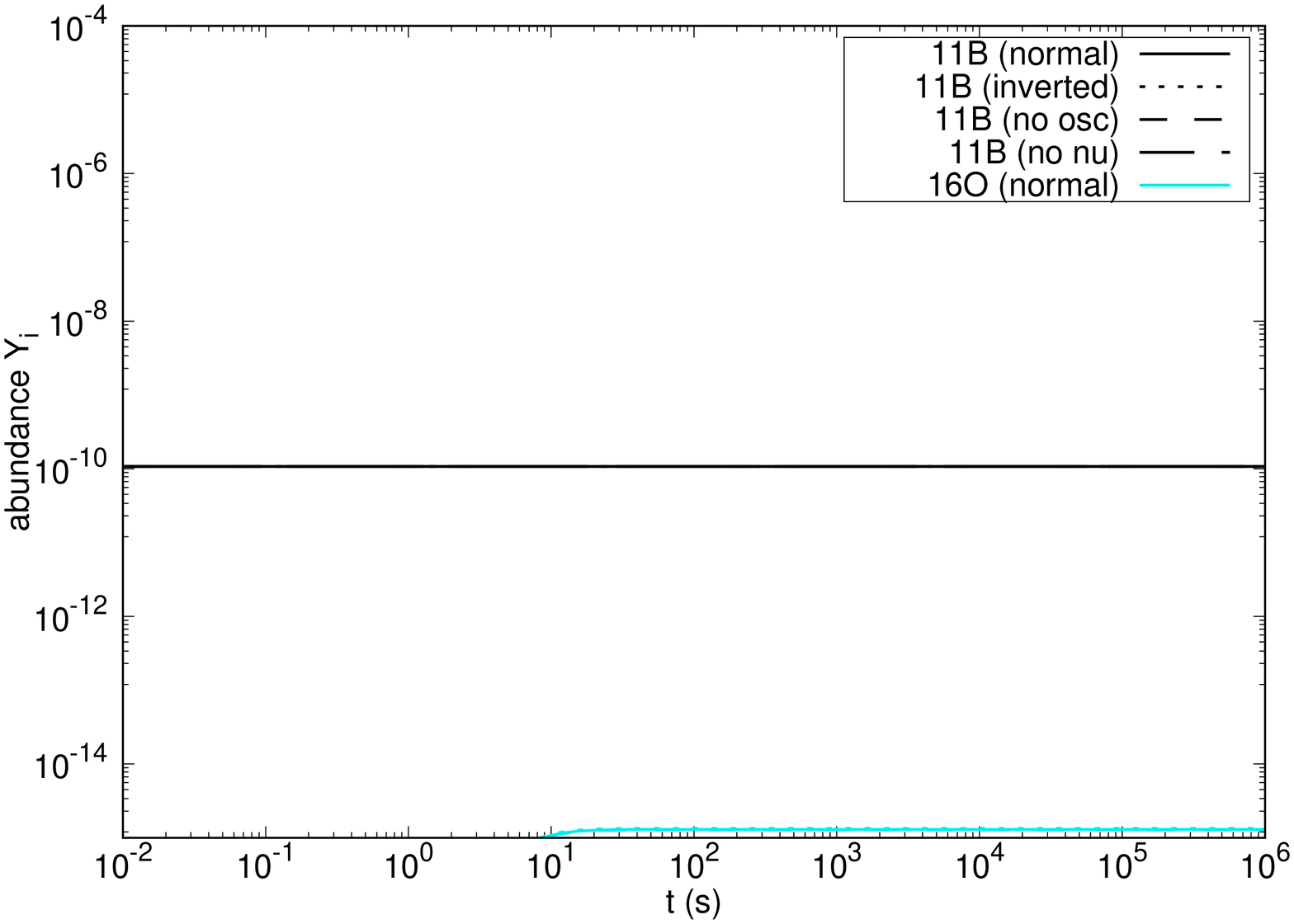}
\caption{Same as Fig. \ref{fig:abun_034} but for shell 5 ($M_r=5.9 M_\sun$).}
\label{fig:abun_353}
\end{figure*}

At $t>10^{-2}$ s, neutrinos from the proto-NS arrive to this layer, and light nuclides are produced via neutrino spallation reactions on $^{16}$O and $^{12}$C. At $t \sim 1$ s, the SN shock arrives and the temperature increases. Then, the destruction of these light nuclides occurs. After the temperature decreases due to the expansion, light nuclides are produced again via neutrino reactions. In this late epoch, differences in the abundances of $^3$H, $^3$He and $^7$Be are seen between the different mass hierarchy cases. The production of $^4$He by the neutrino spallation of $^{12}$C does not occur in the case of no neutrino flux (see long dashed lines in Fig. \ref{fig:abun_034}, right panel).

\subsubsection{Shell 2}
For $M_r =3.7 M_\sun$, neutrinos arrive at $t>10^{-1}$ s, and the shock arrives at $t \lesssim 10$ s.
This region is out of the MSW high resonance region even before the explosion for a typical SN neutrino energy. Therefore, the flavor change effect appears after the neutrino arrival time, and there are differences in the light nuclear abundances.
Before the shock arrives, the temperature in this region is low. Therefore, a significant fraction of protons generated by neutrino spallation survive, and the proton abundance is high. On the other hand, neutrons from the spallation can be easily captured by nuclei because there is no Coulomb penetration factor. The neutron abundance is then kept low.
Large abundances of $^{11}$B and $^{11}$C are produced via the $\nu+^{12}$C reaction.
A smaller but significant abundance of $^7$Be is also produced via $\nu+^{12}$C and survives.
One can see large differences in the $^7$Be and $^{11}$C abundances, while no large differences are seen in the $^7$Li and $^{11}$B abundances.

\subsubsection{Shell 3}
Shell 3 ($M_r=4 M_\sun$) is in the He-rich region. The spallation of $^4$He nuclei produces large abundances of $^3$H and $^3$He. These $^3$H nuclei are then processed into $^7$Li via $^3$H($\alpha$,$\gamma$)$^7$Li. A fraction of the $^3$He nuclei are also processed into $^7$Be via the $^3$He($\alpha$,$\gamma$)$^7$Be reaction. $^7$Be is also produced via $\nu+^{12}$C spallation. $^{11}$B and $^{11}$C nuclei are produced from the neutrino spallation of $^{12}$C.
The shock arrives to this region after the neutrino flux becomes somewhat diminished.
After shock heating, $^7$Be nuclei are produced via the $^3$He($\alpha$,$\gamma$)$^7$Be reaction. Production of $^{11}$B and $^{11}$C via $^7$Li($\alpha$,$\gamma$)$^{11}$B and $^7$Be($\alpha$,$\gamma$)$^{11}$C also operates in this shell.

\subsubsection{Shell 4}
In shell 4 ($M_r=4.5 M_\sun$), the trend of nucleosynthesis is similar to that of the inner shell. In this region, the $^7$Be yields from the $^3$He($\alpha$,$\gamma$)$^7$Be reaction are very large. The temperature is increased sufficiently that the $^7$Li($\alpha$,$\gamma$)$^{11}$B reaction is effective while the $^7$Be($\alpha$,$\gamma$)$^{11}$C reaction is not. Thus, the $^7$Be nuclei survive and the $^{11}$B yields in this layer are large.

\subsubsection{Shell 5}
In shell 5 ($M_r=5.9 M_\sun$), the shock wave does not arrive during the hydrodynamics calculation. 
Similarly to the inner layers, a part of the $^3$H and $^3$He nuclei produced by the neutrino spallation of $^4$He are burned via the $^3$H($\alpha$,$\gamma$)$^7$Li and $^3$He($\alpha$,$\gamma$)$^7$Be reactions, respectively.
However, the $^{11}$B and $^{11}$C yields from the $\nu+^{12}$C spallation reactions are very small due to the small $^{12}$C abundance.

The $^{7}$Be abundances in the normal and inverted hierarchy cases are larger than in the no oscillation case.
The $^7$Li abundance is highest in the inverted hierarchy case. Because of the MSW resonance effect, the $\nueb$ reaction rate is enhanced. Therefore, more $^7$Li is produced.

\section{effects of initial abundances}\label{sec4}
\subsection{Li and B yields}\label{sec4a}

\begin{figure*}[t!]
\plottwo{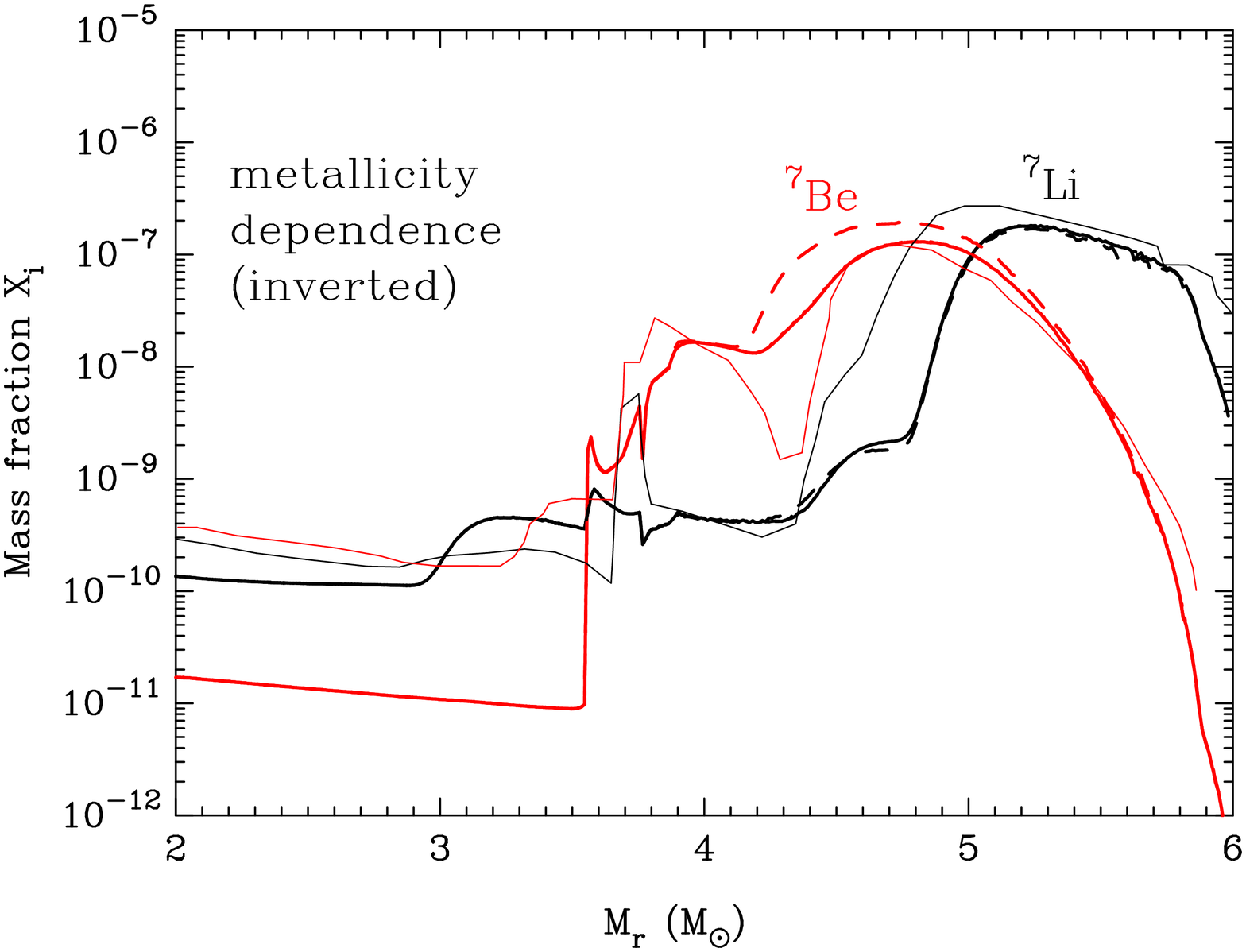}{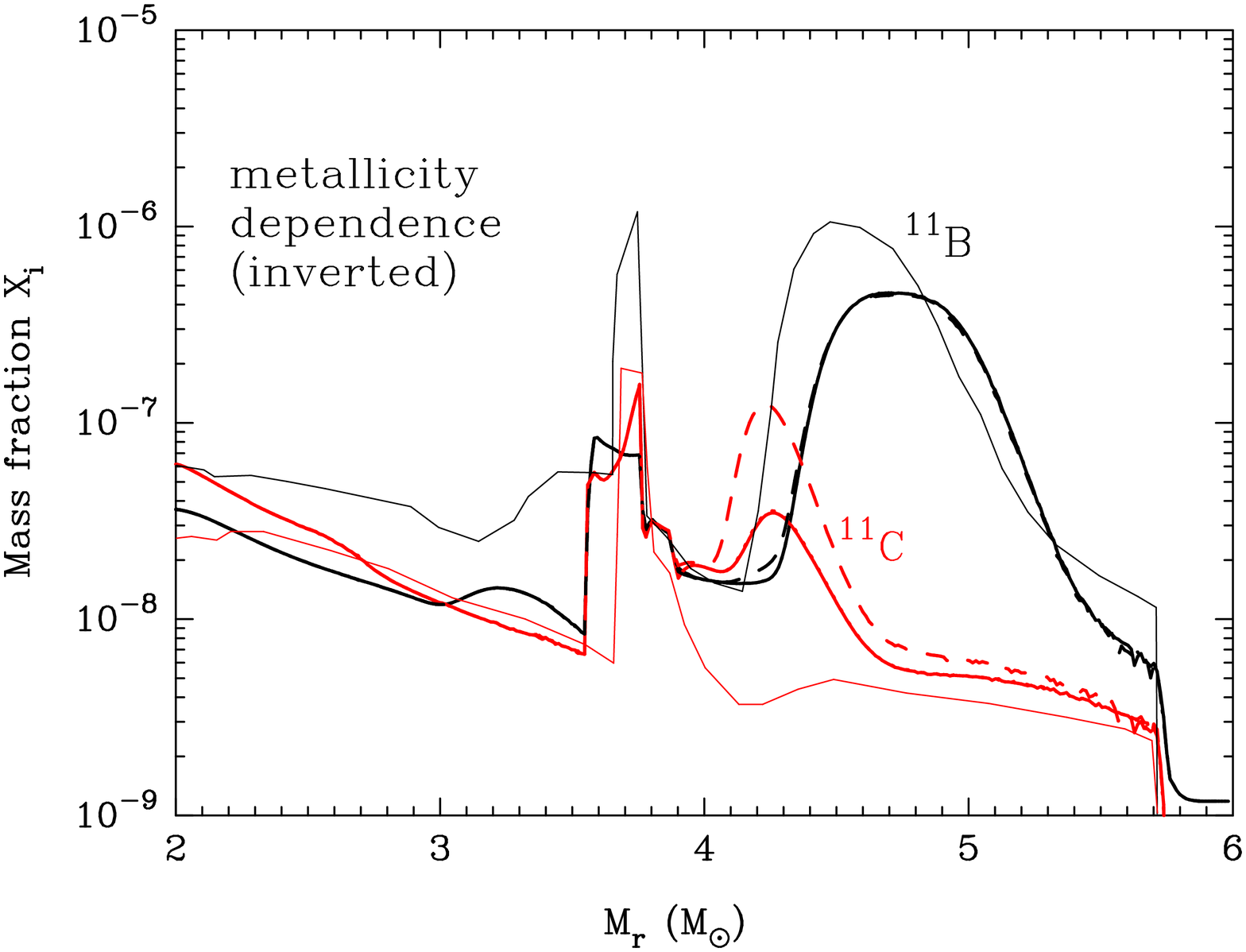}
\caption{Same as Fig. \ref{fig:li_b_nor_inv}, but the solid and dashed lines correspond to Case 2 (the presupernova $s$-abundances for $Z=Z_\sun /4$) and Case 5 (solar abundances), respectively.}
\label{fig:li_b_m_inv}
\end{figure*}

Figure \ref{fig:li_b_m_inv} shows calculated final mass fractions of $^7$Li and $^7$Be (left panel) and $^{11}$B and $^{11}$C (right panel), respectively as a function of the Lagrangian mass coordinate (thick lines). The inverted hierarchy has been assumed for three of the models. Also plotted are the previous result \citep{2006ApJ...649..319Y} for the inverted hierarchy case (thin lines). Solid and dashed lines correspond to Case 2 (the presupernova $s$-abundances for $Z=Z_\sun /4$) and Case 5 (solar abundances), respectively. Although dotted lines are plotted for Case 4 (1/4 of solar abundances), they are overlapping with the solid lines.

The abundances of $^7$Li and $^{11}$B in the He-rich layers ($M_r \gtrsim 3.9 M_\sun$) are almost the same among the three cases, while the difference in the $^7$Be abundance is rather large. At the peak of the yield, the $^7$Be abundance is largest in Case 5.

At the small peak of the $^{11}$C abundance at $M_r \gtrsim 4 M_\sun$, the $^{11}$C abundance is largest in Case 5, similar to the $^7$Be abundance. Differences between Cases 2 and 4 are very small. Thus, the $^7$Li and $^{11}$B yields are not affected by changes in abundances of heavy nuclides during the presupernova evolution that experience neutron capture and photodisintegration reactions during the SN.

\subsection{Abundance evolution}\label{sec4b}

In the O-rich layer (shell 1), there is no difference in abundances of major nuclei. The abundances evolve in all three cases as shown in Figure \ref{fig:abun_034} (dotted lines).

\begin{figure*}[t!]
\plottwo{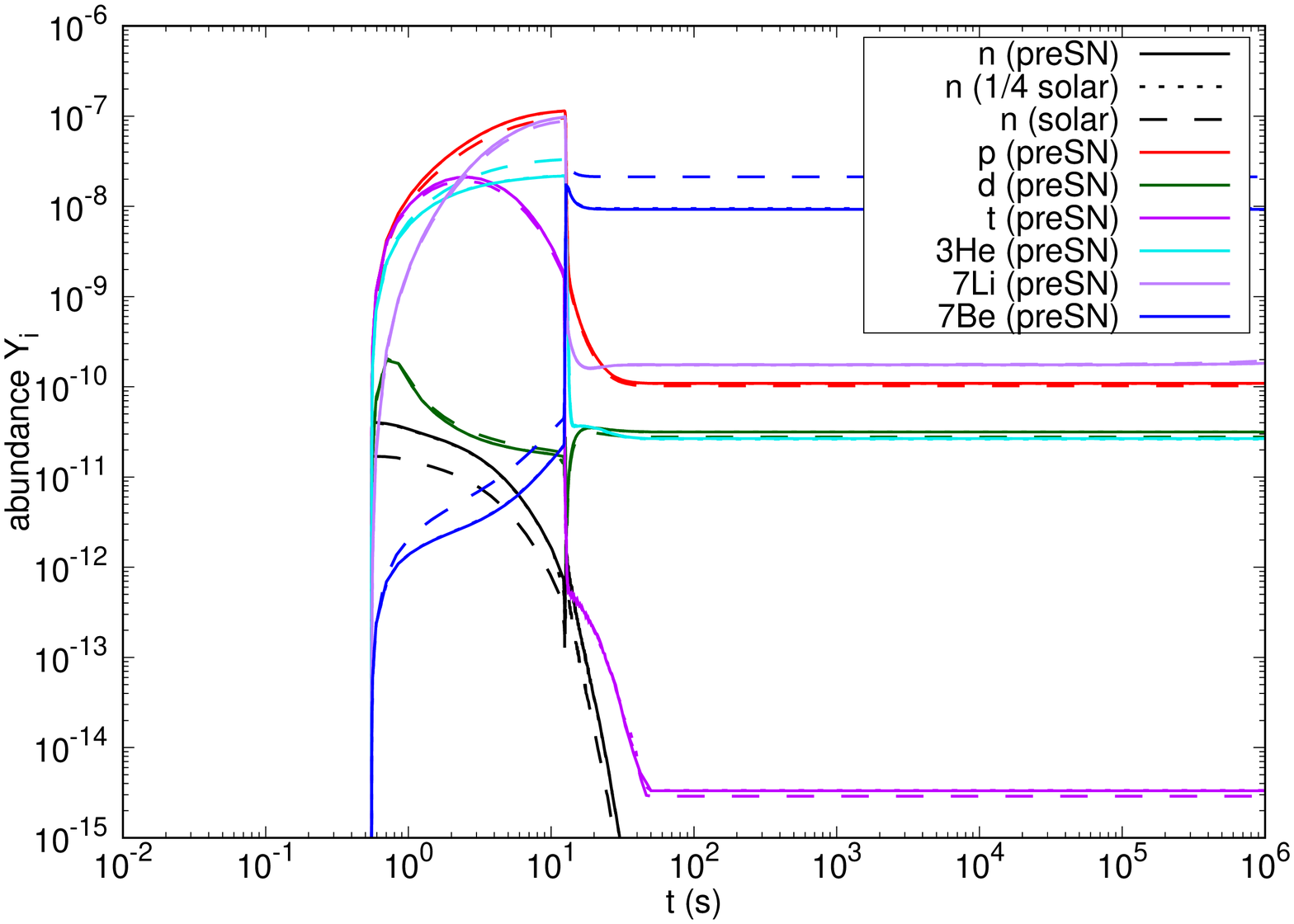}{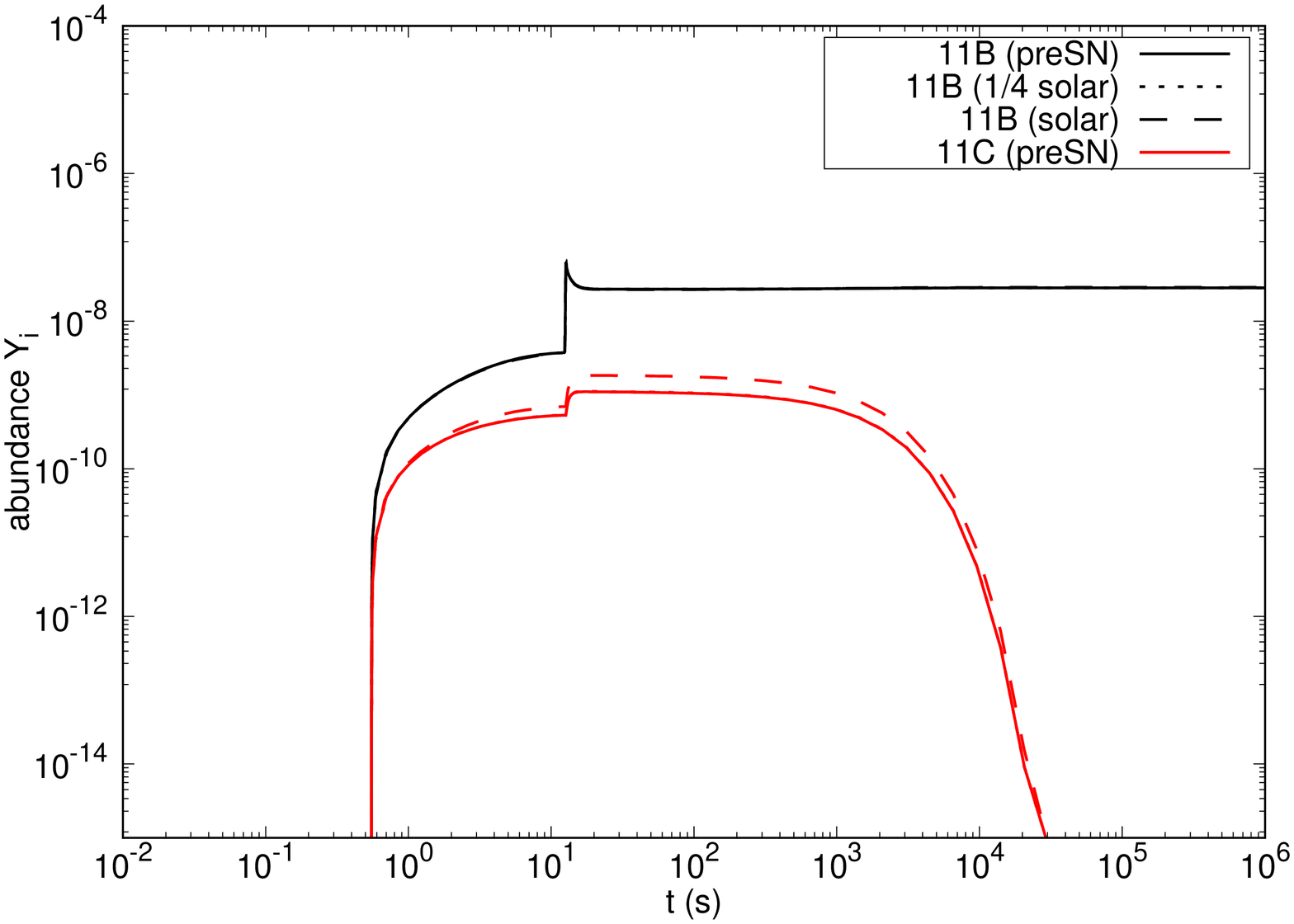}
\caption{Nuclear abundances as a function of time for shell 4. Solid, dotted, and dashed lines correspond to results for heavy elemental abundances taken from the standard $s$-process abundances after stellar evolution of $Z=Z_\sun/4$ (Case 2), the solar abundances divided by 4 (Case 4), and solar abundances (Case 5), respectively. Note, that there is almost no discernable effect from changing the progenitor abundances (solid and dotted lines).}
\label{fig:abun_m_238}
\end{figure*}

Figure \ref{fig:abun_m_238} shows nuclear abundances as a function of time for shell 4. Abundances of $n$, $p$, $d$, $t$, $^3$He, $^7$Li, and $^7$Be (left panel) and $^{11}$B and $^{11}$C (right panel) are plotted. The inverted hierarchy has been assumed. Solid, dotted, and dashed lines correspond to results for heavy elemental abundances taken from the standard $s$-abundances after stellar evolution in the case of $Z=Z_\sun/4$ (Case 2), solar abundances divided by 4 (Case 4), and solar abundances (Case 5), respectively.

Before the shock arrives, the abundances of $^3$He and $^7$Be are largest and those of protons and neutrons are smallest in Case 5.
After shock heating, significant differences in abundances only exist for $^7$Be and $^{11}$C. The $^7$Be and $^{11}$C abundances are largest in Case 5.
The neutron abundance is determined by the abundance of neutron absorbers, i.e., the metallicity. In a metal-rich environment, the neutron abundance is small. This is consistent with the results shown in Fig. \ref{fig:abun_m_238}.
These trends in abundance evolution are observed in shells 2--5 although the magnitude of the differences depends upon location.

\subsection{Neutron absorption}\label{sec4c}

Effects of neutron absorption reactions on the abundances of $^7$Li, $^7$Be, $^{11}$B and $^{11}$C are not large. Our analysis shows that in an inner region with large abundances of C (shells 1--3), the strongest neutron capture reaction is $^{12}$C($n$,$\gamma$)$^{13}$C (cf. Fig. \ref{fig:rate_m_238}). The second largest reaction rate is $^{11}$C($n$,$p$)$^{11}$B, but it is subdominant. Therefore, only a small fraction of neutrons produced via the neutrino reactions or nuclear reactions are available for the conversion of $^{11}$C to $^{11}$B.

In the region of $M_r \sim 4.5 M_\sun$ (shell 4), the $^7$Be($n$,$p$)$^7$Li reaction is the dominant neutron capture reaction, and the $^{12}$C($n$,$\gamma$)$^{13}$C reaction is subdominant. The reason for that is the large $^7$Be abundance and the small $^{12}$C abundance. The operation of the reaction $^7$Be($n$,$p$)$^7$Li has been mentioned in \citet{2006ApJ...649..319Y}. In the outermost region in the He-rich layer (shell 5), the reaction $^3$He($n$,$p$)$^3$H is the strongest neutron capture reaction. Here, only products of the $\nu+^{12}$C reactions participate in nucleosynthesis.

\section{Contribution to Galactic chemical evolution}\label{sec5}

Table \ref{yield1} shows yields of the light nuclides $^7$Li, $^7$Be, $^{11}$B and $^{11}$C for respective models at 50 s after the SN. NH, IH, no, no-$s$, and no-$s$ ($Z_\odot$) correspond to Cases 1 to 5, respectively, and no-$\nu$ corresponds to the case in which the neutrino reactions are switched off. Also shown are yields of $^{16}$O whose solar abundance is believed to be predominantly contributed from SNe II \citep[e.g.,][]{1995MNRAS.277..945T}. The seventh column lists the number ratio $^7$Li/$^{11}$B, and the eighth and ninth columns show the ratios of the overproduction factors ($^7$Li/$^{16}$O) and ($^{11}$B/$^{16}$O), respectively, which are evaluated after the decay of $^7$Be and $^{11}$C into $^7$Li and $^{11}$B, respectively. The ratio is given by
\begin{equation}
  (A/B) =\frac{ M(A) / X_{A,\odot} }{M(B) / X_{B,\odot} },
\label{eq_opf}
\end{equation}
where
$M(i)$ and $X_{i,\odot}$ are the yield from SNe and the solar mass fractions, respectively, of species $i$.

\begin{deluxetable*}{lllllllll}
\caption{\label{yield1}
  Yields of light nuclei (in $M_\odot$), the number ratio $^7$Li/$^{11}$B and normalized overproduction factors}
\tablehead{
  \colhead{model} & \colhead{$M(^7{\rm Li})$} & \colhead{$M(^7{\rm Be})$} & \colhead{$M(^{11}{\rm B})$} & \colhead{$M(^{11}{\rm C})$} & \colhead{$M(^{16}{\rm O})$} & \colhead{$^7$Li/$^{11}$B} & \colhead{($^7$Li/$^{16}$O)} & \colhead{($^{11}$B/$^{16}$O)}
}
\startdata
NH                & $1.0\times 10^{- 8}$ & $2.6\times 10^{- 7}$ & $2.8\times 10^{- 7}$ & $2.1\times 10^{- 7}$ & $1.6$               & $1.2$  & $0.16$              & $0.46$ \\
IH                & $1.2\times 10^{- 7}$ & $9.3\times 10^{- 8}$ & $3.4\times 10^{- 7}$ & $8.1\times 10^{- 8}$ & $1.6$               & $0.80$ & $0.092$             & $0.39$ \\	 
no                & $8.9\times 10^{- 8}$ & $9.4\times 10^{- 8}$ & $2.8\times 10^{- 7}$ & $8.5\times 10^{- 8}$ & $1.6$               & $0.78$ & $0.079$             & $0.34$ \\
no-$\nu$          & $5.7\times 10^{-10}$ & $5.7\times 10^{-19}$ & $3.1\times 10^{-10}$ & $3.7\times 10^{-16}$ & $1.6$               & $2.8$  & $2.5\times 10^{-4}$ & $2.9\times 10^{-4}$ \\
no-$s$            & $1.2\times 10^{- 7}$ & $9.3\times 10^{- 8}$ & $3.4\times 10^{- 7}$ & $8.0\times 10^{- 8}$ & $1.6$               & $0.80$ & $0.093$             & $0.39$ \\
no-$s$ ($Z_\odot$) & $1.1\times 10^{- 7}$ & $1.5\times 10^{- 7}$ & $3.4\times 10^{- 7}$ & $1.0\times 10^{- 7}$ & $1.6$               & $0.93$ & $0.11$             & $0.41$ \\
\enddata
\end{deluxetable*}

It can be seen that the light nuclear yields significantly depend on the neutrino mass hierarchy and the initial nuclear abundances. In the normal mass hierarchy case, both $^7$Li and $^{11}$B overproduction factors are largest, and a significant contribution is expected to the Galactic chemical evolution and the solar abundance. In this case, the $^7$Li/$^{11}$B ratio is also the highest. In the inverted mass hierarchy case, although the $^7$Li and $^{11}$B overproduction factors are larger than those in the no oscillation case, they are significantly smaller than in the normal hierarchy case. The $^7$Li/$^{11}$B ratio is slightly larger than that of the no oscillation case. These enhancements of $^7$Li/$^{11}$B by the neutrino flavor change are consistent with the previous study of \citet{2008ApJ...686..448Y}. However, in this calculation, the ratio in all the three cases are 30--45 \% larger than the previous result \citep{2008ApJ...686..448Y}.

Obviously, if neutrino reactions are nonexistent in SNe, yields of $^7$Li and $^{11}$B are negligibly small (see the row of no-$\nu$). One may consider some exotic situation in which neutrinos energize SN explosions but they change their form or disappear before they travel to the outer region. However, that scenario induces a problem in explaining the Solar $^{10}$B/$^{11}$B isotopic ratio in a Galactic chemical evolution model \citep[see e.g.,][]{2012A&A...542A..67P}.

Effects of the $s$-process on heavy nuclei during the presupernova evolution is small as seen in a comparison of Cases 2 and 4 (fifth row).
An increase of metallicity (no-$s$ $Z_\odot$ case) leads to an enhancement of the $^7$Be yield from that of the no-$s$ case while not affecting other light nuclear yields much. Then, the ($^7$Li/$^{16}$O) ratio is increased by $\sim 22$ \% and the $^7$Li/$^{11}$B ratio increases by $\sim 16$ \%.

Standard theories of grain formation from SN ejecta suggest that mixing of SN ejecta within layers is difficult \citep{2003ApJ...594..312D} \footnote{For effects of the mixing on the dust abundances, see \citet{2003ApJ...598..785N}.} although a convective overturn, i.e. a change of position of materials with different compositions, is possible. Therefore, it is expected that local elemental and isotopic abundances could remain in presolar grains if they originate in SN ejecta, rather than the ratio of total yields from SNe. In this case, as seen in Figs. \ref{fig:li_b_nor_inv} and \ref{fig:li_b_m_inv}, the final abundances of $^7$Li+$^7$Be and $^{11}$B+$^{11}$C significantly depend on location, i.e., $M_r$. Therefore, assuming grain formation in the Li and B production region, we can deduce expected ranges of the $^7$Li/$^{11}$B ratio in SN grains.

\begin{figure}[t!]
  \epsscale{1}
\plotone{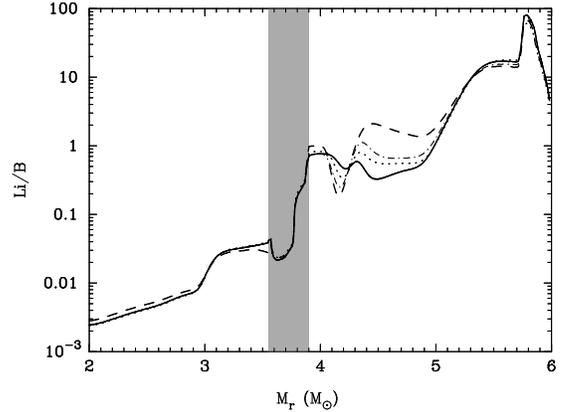}
\caption{The ratio of $^7$Li/$^{11}$B versus Lagrangian mass coordinate. Thick dashed and solid lines correspond to Cases 1 and 2, respectively, while the dotted line shows the result of Case 3. The thin dot-dashed line corresponds to Case 5. The vertical band indicates the C-rich region.}
\label{lib_ratio}
\end{figure}

Figure \ref{lib_ratio} shows the Li/B number ratio as a function of $M_r$. Thick dashed and solid lines correspond to Cases 1 and 2, respectively, while the dotted line shows the result of Case 3. Thin dot-dashed line corresponds to Case 5. The result of Case 4 is indistinguishable from that of Case 2. The Li/B ratio ranges over many orders magnitudes, i.e., Li/B$\sim 10^{-3}$ to $100$. The vertical band indicates the C-rich region. Since the Li and B production is only important in the C-rich and He-rich regions, dust grains containing Li and B from SNe can have a wide range of Li/B ratios, i.e., Li/B$\gtrsim {\mathcal O}$(0.01). If some grains include abundant C as well as Li and B, a possible site of their production is the C-rich layer of SNe. There, the ratio is Li/B$\sim {\mathcal O}$(0.01)--1.

An observational test of our theoretical prediction of the SN $\nu$-process nucleosynthesis is critical. \citet{2011ApJ...730L...7F} reported on enhanced Li and B isotopic abundances in a number of SiC X grains from the Murchison meteorite. Such grains are presumed to arise directly from SN ejecta. Since $^{11}$B and $^{7}$Li are expected to be the main products of $\nu$-process nucleosynthesis \citep[e.g.][]{2005PhRvL..94w1101Y,2008ApJ...686..448Y}, they could be used as a tracer to find evidence for the neutrino oscillations and their associated mass hierarchy.

The possible isotopic anomaly found in their analysis \citep{2011ApJ...730L...7F,2012PhRvD..85j5023M} is still subject to possible laboratory and/or meteoritic contamination that includes other sources of these elements such as Galactic cosmic-ray spallation processes or stellar nucleosynthesis in asymptotic giant branch stars producing light elements after the SiC X grains condensed in the SN ejecta in the early solar system \citep[see e.g.][for the contribution of interstellar matter to meteorites]{2013ApJ...779L...9H}. Another possible test would be the spectroscopic observation of Li and B in SN remnants. A high-dispersion spectrograph equipped on a space telescope like the Hubble Space Telescope might be used to detect $^{11}$B absorption lines at UV wavelengths \citep{1997ApJ...488..338D,1998ApJ...500..241G,1999A&A...343..545P,2000ApJ...530..939C}. It is also possible to detect $^{7}$Li in spectroscopic absorption lines using ground-based large telescopes like SUBARU, EUSO-VLT or KECK. Observational tests probing these light-mass isotopes in SN ejecta or remnants is highly desirable in order to establish this mechanism of Li and B nucleosynthesis and to study the neutrino oscillations.

The main production of $^7$Li and $^{11}$B occurs outside of the MSW H resonance region in core-collapse SNe. This characteristic is different from the production of heavy nuclei by the neutrino process. Therefore, $^7$Li and $^{11}$B abundances are unique tools to constrain the physical processes operating in SNe, such as neutrino spectra and their luminosity evolution, the matter density, and the initial nuclear abundances. In the future, it would be desirable to clarify the dependence of those yields on the initial metallicity, stellar mass, explosion energies, etc. Although nucleosynthesis results \citep[e.g.,][]{2016ApJ...821...38S} are available, the neutrino flavor change in SNe should be taken into account.

The ultimate goal is to understand the contribution of the neutrino process to the Galactic chemical evolution of $^7$Li and $^{11}$B. The dependence of the $^{19}$F yield on stellar parameters has been studied in \citet{2012IAUS..279..339I}. They also considered the Galactic chemical evolution based upon the SN nucleosynthesis result. Observations of the stellar F abundance are consistent with the theoretical result although observational data have a large dispersion \citep{2011ApJ...739L..57K}.

\section{conclusions}\label{sec6}
We have revisited the effects of neutrino oscillations on the $^7$Li and $^{11}$B production in core-collapse SNe. Neutrinos emitted from the proto-NS propagate through stellar regions from the neutrino-sphere to the interstellar region. When they pass through the region with a density of $\rho=10^{2-3}$ g cm$^{-3}$, there is a resonance in the flavor change probability of neutrinos, i.e., the H resonance. Since the neutrino freezeout temperatures are different between the electron neutrinos, antineutrinos, and other types, the resonant flavor change results in changes of the neutrino energy spectra. The neutrino spallation reaction rates are then significantly affected by the flavor change.

In this paper, we have used a new nuclear reaction network code in which neutrino cross sections for $^4$He and $^{12}$C spallation are corrected. Initial nuclear abundances in SNe were calculated with a model for the SN 1987A progenitor, which includes heavy $s$-nuclei. We then analyzed the $^7$Li and $^{11}$B production in SN nucleosynthesis in detail. Effects of neutrino oscillations and the initial nuclear abundances are clearly observed for respective stellar layers. Especially, the effects in the important $^7$Li and $^{11}$B production site at $M_r/M_\odot \sim [3.5, 6]$ are summarized as follows:

\begin{enumerate}

\item oscillations
  
  The neutrino oscillations in the normal hierarchy case increase the CC reaction rates for \nue. Then, yields of proton-rich nuclides, e.g., $^3$He (via the $\nu+^4$He reaction) as well as $^7$Be and $^{11}$C (via the $\nu+^{12}$C reaction) are increased. Since the $^7$Be yield is predominantly contributed to by the reaction $^3$He($\alpha$,$\gamma$)$^7$Be after the shock heating, the enhancement of the $^3$He production rate also results in an enhanced final $^7$Be abundance. For $M_r \lesssim 4.3 M_\odot$, the $^7$Be nuclei are destroyed via the reaction $^7$Be($n$,$p$)$^7$Li followed by the $^7$Li($p$,$\alpha$)$^4$He. The final abundance of $^7$Be is then small. For the outer region, $M_r \gtrsim 5 M_\odot$, the temperature is not high and the neutrino flux is smaller. Therefore, the $^7$Be abundance is small in this region, and a peak of the $^7$Be yield appears inside. There are two peaks in the curve of $^{11}$C abundance versus $M_r$. The inner one is at $M_r \sim 3.55$--$3.75 M_\odot$ from the $\nu+^{12}$C spallation reaction and the outer one is at $M_r \sim 4$--$4.5 M_\odot$ from the $^7$Be($\alpha$,$\gamma$)$^{11}$C reaction. Both peaks are much larger than those of the no oscillation case.

  In the inverted hierarchy case, on the other hand, the CC reaction rate of $\nueb$ is enhanced by the neutrino oscillations. As a result, neutrino spallation products are enriched in neutron-rich nuclei, i.e., $^3$H, $^7$Li, and $^{11}$B. The main reaction for $^7$Li synthesis, $^3$H($\alpha$,$\gamma$)$^7$Li, contributes to larger yields of $^7$Li because of an enhanced $^3$H abundance. In the inner region, $M_r \lesssim 5 M_\odot$, $^7$Li nuclei are produced before the shock arrival. However, the shock heating triggers $^7$Li destruction via the $^7$Li($p$,$\alpha$)$^4$He reaction, and the final yield of $^7$Li is diminished.

\item initial abundances

  The $s$-process during the presupernova stellar evolution changes the abundances of nuclei with $A \gtrsim 56$. Then, the neutron abundances during the SN are affected by the $s$-process since the abundances of neutron poisons are different.
  However, this effect of the seed abundances on the production of $^7$Li and $^{11}$B is found to be unimportant for the metallicity of the Large Magellanic Cloud.
  When the metallicity is increased, the total abundance of nuclei that can capture neutrons is larger. Then, the neutron abundance during SN nucleosynthesis becomes smaller. Therefore, the dependence of yields of $^7$Li and $^{11}$B on the initial stellar metallicity must be considered in future studies of Galactic chemical evolution.
  The initial metallicity does not affect the $^7$Li production much, while they do affect the $^7$Be production.
  When a higher metallicity, e.g., $Z_\odot$ is assumed, the neutron abundance is smaller than in our standard case and the $^7$Be production rate is increased. The final abundance is then significantly higher than that of the standard case in the region with $M_r \sim 4.5$--$5 M_\odot$.

  The $^{11}$C production is significantly affected by the initial metallicity, while $^{11}$B production is not sensitive. In the $^{11}$C production layer of $M_r \sim 4.1$--$4.5 M_\odot$, the $^{11}$C yield is enhanced by increasing the metallicity. A smaller neutron abundance leads to a smaller rate of $^3$He($n$,$p$)$^3$H. Then, the $^3$He abundance is larger and the $^7$Be production rate is also larger via the $^3$He($\alpha$,$\gamma$)$^7$Be reaction. As a result, a larger $^7$Be abundance is obtained. Therefore, the production of $^{11}$C via $^7$Be($\alpha$,$\gamma$)$^{11}$C is more effective.

\end{enumerate}

\acknowledgments

This work was supported in part by the visiting fellow and visiting scholar programs in NAOJ.
Works by M-K. Cheoun were supported by the National Research Foundation of Korea (Grant Nos. NRF-2015K2A9A1A06046598 and NRF-2017R1E1A1A01074023).
K.N. was supported by the World Premier International Research center Initiative (WPI Initiative), MEXT, Japan, and JSPS KAKENHI Grant Numbers JP16H02168 and JP17K05382.
TK was supported by Grants-in-Aid for Scientific Research of JSPS (15H03665, 17K05459).
    Work of GJM supported in part by U.S. DOE nuclear theory grant DE-FG02-95-ER40934.
  We are grateful to Takashi Yoshida for helpful comments regarding the previous calculations of neutrino nucleosynthesis, and Takaya Nozawa for explaining the status of the SN dust formation theory.

%



\software{blcode \citep[\\https://stellarcollapse.org/snec]{ott_blcode},  
          TALYS-1.8 (2015) \citep[http://www.talys.eu/home]{Koning2007}, \\
          MA38 sparse-matrix package \citep[http://www.hsl.rl.ac.uk/catalogue/ma38.html]{MA38}
}



\appendix

\section{partial time derivative of reaction rates}\label{app1}
The rate equation has the form of
\begin{equation}
  \frac{ d \mbox{\boldmath $Y$}}{dt} = \mbox{\boldmath $f$}(t, \mbox{\boldmath $Y$}),
\label{eq_a1}
\end{equation}
where
$\mbox{\boldmath $Y$}=(Y_1, Y_2,~...)$ is the vector of nuclear mole fractions and $\mbox{\boldmath $f$}$ is a function describing the abundance evolution.
When this is solved with a semi-implicit extrapolation method \citep{1992nrfa.book.....P}, we need to evaluate ${\partial \mbox{\boldmath $f$}}/{\partial t}$.  The rate of any reaction is given by
\begin{equation}
  f_{i,a} =\pm \frac{n_i}{\Pi_{k=1}^{N_{\rm nuc}} n_k !}\left[N_{\rm A} \rho_{\rm b} (t) \right]^{\sum_{j=1}^{N_{\rm nuc}} n_j -1} \Pi_{k=1}^{N_{\rm nuc}} Y_k^{n_k} \langle {\rm rate} \rangle(t),
\label{eq_a2}
\end{equation}
where
$i$ and $a$ are indexes for the  nuclide and reaction, respectively.
The plus and minus signs correspond to the production and destruction terms, respectively.
$N_{\rm A}$ is the Avogadro number,
$\rho_{\rm b}(t)$ is the baryonic density,
$N_{\rm nuc}$ is the number of nuclear species involved in the reaction,
$n_j$ is the number of nuclide $j$ reacting in one reaction event, and
$\langle {\rm rate} \rangle(t)$ is the reaction rate in units of [(cm$^3$)$^{\sum_{j=1}^{N_{\rm nuc}} n_j -1}$ s$^{-1}$].  The total reaction rate of species $i$ is then given by summing this term over all reactions $a$:
\begin{equation}
  f_i =\sum_a f_{i,a}.
  \label{eq_a3}
\end{equation}

The time derivative of the abundance change rate $f_{i,a}$ is given by
\begin{equation}
  \frac{\partial f_{i,a}}{\partial t} =f_{i,a} \left[ \left(\sum_{j=1}^{N_{\rm nuc}} n_j -1 \right) \frac{d \ln \rho_{\rm b}}{dt} +
    \frac{\partial \ln \langle {\rm rate} \rangle}{\partial T_9} \frac{d T_9}{dt}
      +\frac{\partial \ln \langle {\rm rate} \rangle}{\partial r} \frac{dr}{dt}
      +\frac{\partial \ln \langle {\rm rate} \rangle}{\partial t}
    \right],
\label{eq_a4}
\end{equation}
where
the derivatives $d \ln \rho_{\rm b}/dt$, $dT_9/dt$, and $dr/dt$ are given as trajectories from the hydrodynamics calculation.

\subsection{Neutrino reactions}
In the region above the mass cut of a SN, the average kinetic energy of nuclides is $E_{\rm nuc}=3T/2 ^<_\sim 1$ MeV.  On the other hand, the average kinetic energy of neutrinos propagating from the surface of the NS is $E_\nu =3.151 {T_\nu} ^>_\sim 10$ MeV.  Since velocities of the nuclides are much smaller than light speed, the kinetic energies of neutrinos in the rest frame of nuclides are unchanged from those in the fluid rest frame, i.e., $E_\nu$.  Therefore, the rates for reactions of neutrinos and nuclei do not depend on the ambient temperature.  Thus, it follows that $\partial \langle {\rm rate} \rangle/\partial T_9 =0$.  In addition, in the SN nucleosynthesis, only two-body neutrino reactions, $\nu+A$, are important.  Then, the sum of the numbers of reacting nuclides is $\sum_{j=1}^{N_{\rm nuc}} n_j =1$.  As a result, we obtain
\begin{eqnarray}
  \frac{\partial f_{i,a}}{\partial t} &\approx& 
  f_{i,a} \left[ \frac{\partial \ln \langle {\rm rate} \rangle}{\partial r} \frac{dr}{dt}
    +\frac{\partial \ln \langle {\rm rate} \rangle}{\partial t}
    \right] \label{eq_a5} \\
  &=&
  \pm n_i Y_k \left[ \frac{\partial \langle {\rm rate} \rangle}{\partial r} \frac{dr}{dt}
    +\frac{\partial \langle {\rm rate} \rangle}{\partial t}
    \right].
  \label{eq_a6}
\end{eqnarray}
The time derivative is then given by this equation with neutrino reaction rates (Equation (\ref{nu3})).

In the present study, we adopted the approximate probability of neutrino flavor change $P_{\beta\alpha}(r; E_\nu)$ evaluated at $t=0$. Then, the probability is independent of $t$, and dependent only on $r$. In addition, the neutrino energy spectra are assumed to be constant, and the luminosity evolution is given by Equation (\ref{nu8}). In this case, the time derivative for a reaction of neutrino flavor $\alpha$ has an analytic form:
\begin{eqnarray}
  \frac{\partial f_{i,a,\alpha}}{\partial t} &=& 
  \pm n_i Y_k
  \sum_{\beta=e,\mu,\tau} \left[
    \frac{L_{\nu_\beta}}{4 \pi r^2} \frac{1}{kT_{\nu_\beta}} \frac{F_2(\eta_{\nu_\beta})}{F_3(\eta_{\nu_\beta})}
    \langle P_{\beta\alpha} \sigma_{\nu_\alpha} \rangle(T_{\nu_\beta}; r)
    \left(
    -\frac{1 -\dot{r}/c}{\tau_\nu} -2 \frac{\dot{r}}{r}
    +\frac{\partial \ln \langle P_{\beta\alpha} \sigma_{\nu_\alpha}  \rangle}{\partial r} \dot{r}
    \right)
    \right]
  \label{eq_a7} \\
  &=& -f_{i,a,\alpha} \left(
    \frac{1 -\dot{r}/c}{\tau_\nu} +2 \frac{\dot{r}}{r} \right)
    \pm n_i Y_k
  \sum_{\beta=e,\mu,\tau} \left[
    \frac{L_{\nu_\beta}}{4 \pi r^2} \frac{1}{kT_{\nu_\beta}} \frac{F_2(\eta_{\nu_\beta})}{F_3(\eta_{\nu_\beta})}
    \frac{\partial \langle P_{\beta\alpha} \sigma_{\nu_\alpha}  \rangle (T_{\nu_\beta}; r)}{\partial r} \dot{r}
    \right].
  \label{eq_a8}
\end{eqnarray}
We note that for the neutral current reactions, the term $\partial \langle P_{\beta\alpha} \sigma_{\nu_\alpha}  \rangle /\partial r=\partial \langle \sigma_{\nu}^{\rm NC}  \rangle /\partial r$ is zero.

\subsection{Nuclear reactions}
Except for reactions triggered by neutrinos, the term $\langle {\rm rate} \rangle$ does not explicitly depend on $r$ and $t$. Therefore, Equation (\ref{eq_a4}) reduces to
\begin{equation}
  \frac{\partial f_{i,a}}{\partial t} =f_{i,a} \left[ \left(\sum_{j=1}^{N_{\rm nuc}} n_j -1 \right) \frac{d \ln \rho_{\rm b}}{dt} +
    \frac{d \ln \langle {\rm rate} \rangle}{dT_9} \frac{d T_9}{dt}
    \right].
\label{eq_a5}
\end{equation}

\subsection{Nuclear reaction rates of JINA REACLIB}
For example, we show the specific form of $d \langle {\rm rate} \rangle/dT_9$ for the JINA REACLIB data adopted in this paper.

\subsubsection{Forward rates}
The forward reaction rates are expressed in the form
\begin{equation}
  \langle {\rm rate} \rangle_{\rm for} =R =\Pi_{i =0}^{5} \exp(a_i T_9^{p_i}) T_9^{a_6},
\label{eq_a6}
\end{equation}
where
$a_i$ ($i =0$, ..., 6) are fitted parameters, and
$p_i= 0,~-1,~-1/3,~1/3,~1$, and $5/3$ (for $i =0$, ..., 5) are fixed power law indexes.

The derivative of the rates with respect to $T_9$ are given by
\begin{equation}
  \frac{d \ln \langle {\rm rate} \rangle_{\rm for}}{d T_9} =
  \sum_{i =0}^{5} \left(a_i p_i T_9^{p_i -1} \right) + \frac{a_6}{T_9}.
\label{eq_a7}
\end{equation}
The specific equation for the JINA REACLIB rates is then given by
\begin{equation}
  \frac{d \ln \langle {\rm rate} \rangle_{\rm for}}{d T_9} =
  -a_1 T_9^{-2} -\frac{1}{3} a_2 T_9^{-4/3} +\frac{1}{3} a_3 T_9^{-2/3} +a_4 +\frac{5}{3} a_5 T_9^{2/3} +a_6 T_9^{-1}.
\label{eq_a8}
\end{equation}

\subsubsection{Reverse rates}
The reverse reaction rates are given by
\begin{equation}
  \langle {\rm rate} \rangle_{\rm rev} =R f_{\rm cor},
\label{eq_a9}
\end{equation}
where
$R$ is given by the function defined in Eq. (\ref{eq_a6}), and
$f_{\rm cor}$ is a multiplication factor for the reverse rates.  
The factor is given for the reaction $a_1 + ... +a_m \rightarrow b_1 + ... +b_n$:
\begin{equation}
  f_{\rm cor} =\frac{\Pi_{i=1}^{n} f_{\rm par}(b_i)}{\Pi_{j=1}^{m} f_{\rm par}(a_j)},
\label{eq_a10}
\end{equation}
where
$f_{\rm par}(k)$ is the nuclear partition function of $k$.
The derivative of this factor is given by
\begin{equation}
  \frac{d \ln f_{\rm cor}}{d T_9} = \sum_{i=1}^n \frac{d \ln f_{\rm par}(b_i)}{d T_9} -\sum_{j=1}^{m} \frac{d \ln f_{\rm par}(a_j)}{d T_9}.
\label{eq_a11}
\end{equation}
The temperature derivative of the reverse reaction rates of the JINA REACLIB database is then given by
\begin{equation}
  \frac{d \ln \langle {\rm rate} \rangle_{\rm rev}}{d T_9} =\frac{ d \ln R}{d T_9} +\frac{d \ln f_{\rm cor}}{d T_9}.
\label{eq_a12}
\end{equation}

\section{effects of neutrino flavor change on abundance change rates}\label{app2} 
Table \ref{tab:2nuc} shows the 13 nuclear reactions relevant to $^7$Li, $^7$Be, $^{11}$B and $^{11}$C synthesis in SNe.
The following reactions occur via $\nu+^4$He spallation: 1) NC reactions
  $^{4}$He($\nu$,$\nu p$)$^3$H,
  $^{4}$He($\nu$,$\nu n$)$^3$He,
  $^{4}$He($\nu$,$\nu d$)$^2$H,
  $^{4}$He($\nu$,$\nu 2np$)$^1$H;
  2) CC reactions 
  $^{4}$He($\nu_e$,$e^-$$p$)$^3$He,
  $^{4}$He($\nu_e$,$e^-$2$p$)$^2$H,
  $^{4}$He($\nueb$,$e^+$$n$)$^3$H, and
  $^{4}$He($\nueb$,$e^+$2$n$)$^2$H.
Also $^{12}$C spallation reactions involve the production of $^3$H, $^3$He, $^7$Li, $^7$Be, $^{11}$B, $^{11}$C, and $^{11}$Be (which are produced only via CC reactions) \citep{2008ApJ...686..448Y}.
    
\begin{deluxetable}{ll}
\caption{\label{tab:2nuc}
Important reactions of $^7$Li, $^7$Be, $^{11}$B, and $^{11}$C in SNe}
\tablehead{
  \colhead{reaction} & \colhead{remark}
}
\startdata
$^{3}$He($n$,$p$)$^{3}$H            & conversion $^3$He$\rightarrow ^3$H \\
$^{3}$H($\alpha$,$\gamma$)$^{7}$Li  & $^7$Li production \\	 
$^{7}$Li($p$,$\alpha$)$^{4}$He      & $^7$Li destruction \\
$^{7}$Li($\alpha$,$\gamma$)$^{11}$B & $^7$Li destruction \& $^{11}$B production \\
$^{3}$He($\alpha$,$\gamma$)$^{7}$Be & $^7$Be production \\
$^{7}$Be($\alpha$,$\gamma$)$^{11}$C & $^7$Be destruction \& $^{11}$C production \\
$^{7}$Be($n$,$p$)$^{7}$Li           & conversion $^7$Be$\rightarrow ^7$Li \\
$^{11}$B($p$,2$\alpha$)$^{4}$He     & $^{11}$B destruction \\
$^{11}$B($\alpha$,$p$)$^{14}$C      & $^{11}$B destruction \\
$^{11}$B($\alpha$,$n$)$^{14}$N      & $^{11}$B destruction \\
$^{11}$C($n$,$p$)$^{11}$B           & conversion $^{11}$C$\rightarrow ^{11}$B \\
$^{12}$C($p$,$\gamma$)$^{13}$N      & $p$ capture \\
$^{12}$C($n$,$\gamma$)$^{13}$C      & $n$ capture \\
\enddata
\end{deluxetable}

\subsection{Trends of $\nu+^4$He reactions}
Because of the MSW resonance effect, CC reaction rates are enhanced in the normal and inverted hierarchy cases.
In the normal hierarchy case, the $\nue$ rate is enhanced so that $^3$He and $p$ are produced in a greater abundance.
In the inverted hierarchy case, the $\nueb$ rate is enhanced and the production rates of $^3$H and $n$ are larger. 
In this case there is a tendency for $^7$Li and $^{11}$B to be produced via the $^3$H($\alpha$,$\gamma$)$^7$Li($\alpha$,$\gamma$)$^{11}$B reaction sequence.

\subsection{Trends of $\nu+^{12}$C reactions}
\subsubsection{$^7$Li}
The cross section for the CC $^7$Li production via the reaction $^{12}$C+$\nueb$ is much larger than that of $^{12}$C+$\nu_e$ \citep{2008ApJ...686..448Y}.
Therefore, the neutrino oscillations increase the $^7$Li yield in the inverted hierarchy case (shells 2, 3, and 5; Sec. \ref{sec3b}).
The $^7$Li production cross section of the NC $^{12}$C+$\nu$ reaction is close to that of the CC $^{12}$C+$\nueb$ reaction.
Therefore, the change in the CC rate of $^{12}$C+$\nueb$ does not much affect the $^7$Li yield.

\subsubsection{$^7$Be}
The effective energy threshold for the CC $^7$Be production is lower for the $\nu_e$ reactions than for the $\nueb$ reactions \citep[cf. Tables 4 and 5 in][]{2008ApJ...686..448Y}. Therefore, when the flux of energetic $\nu_e$ is increased by the flavor change, the $^7$Be production is more effective.
In the normal hierarchy case, a complete $\nux \leftrightarrow \nue$ transition occurs \citep[See Fig. 1 of][]{2006ApJ...649..319Y}.
Therefore, the $^7$Be yield is larger.
In the inverted hierarchy case, the $\nux \leftrightarrow \nue$ transition is incomplete \citep[See Fig. 2 of][]{2006ApJ...649..319Y}.
Thus, the $^7$Be yield is not changed much from the no oscillation case.
The $^7$Be production cross section of the NC $^{12}$C+$\nu$ reaction is smaller than that of the CC $^{12}$C+$\nu_e$ reaction.
Therefore, the effect of flavor change $\nu_x \leftrightarrow \nu_e$ is relatively large.

\subsubsection{$^{11}$B and $^{11}$C}
The MSW effect changes the $\nu_e$ spectrum (for shells 2--5).
The $^{11}$B nuclei are produced by the CC $^{12}$C+$\nueb$ reaction only, while $^{11}$C nuclei are produced by the CC $^{12}$C+$\nu_e$ reaction only. Then, the $^{11}$C abundance is affected. The CC cross section is somewhat larger than the NC cross section. An enhancement of the $^{11}$C abundance is then apparent.

\subsection{Analysis}

\begin{figure*}[t!]
\gridline{\fig{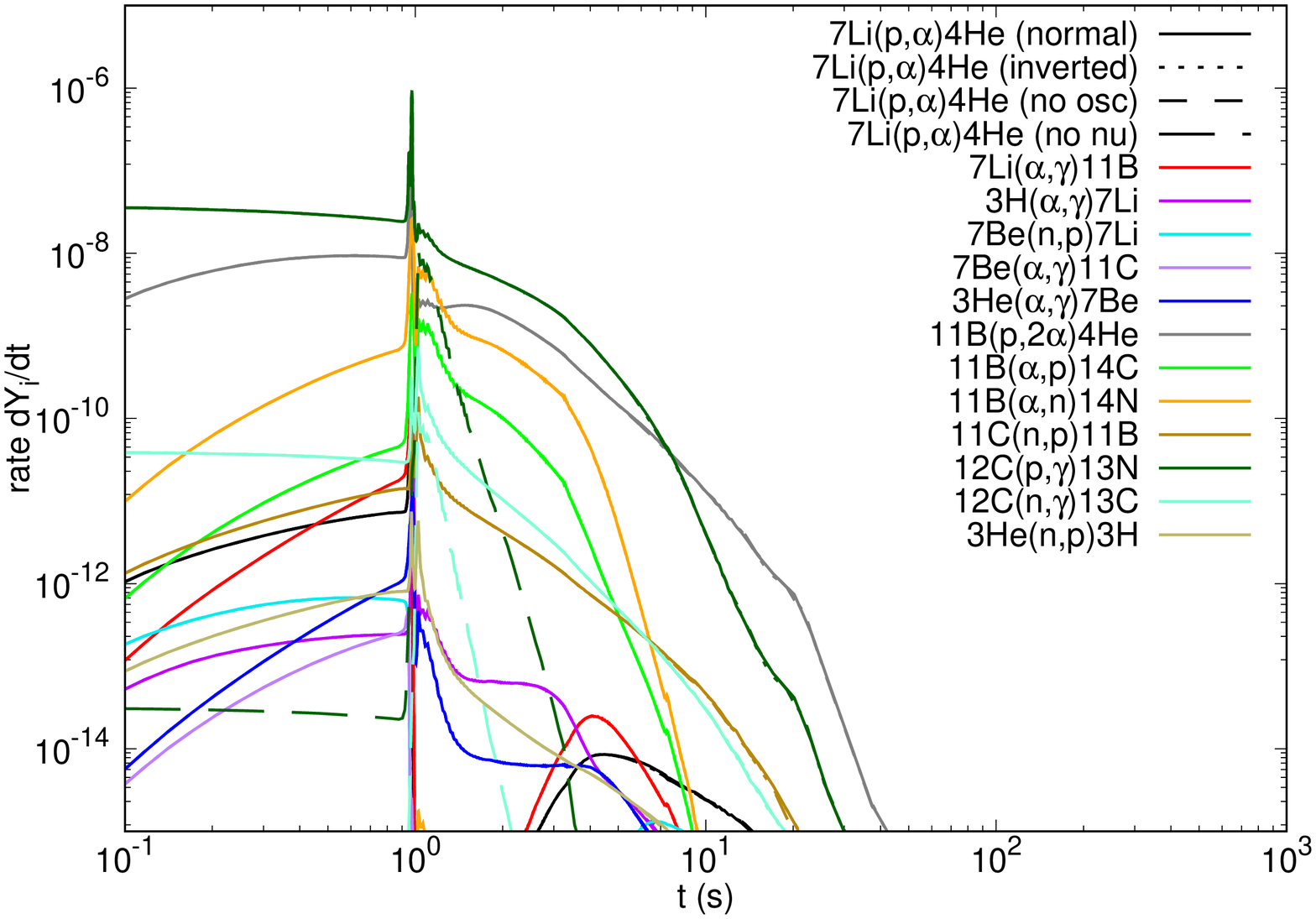}{0.3\textwidth}{(a)}
\fig{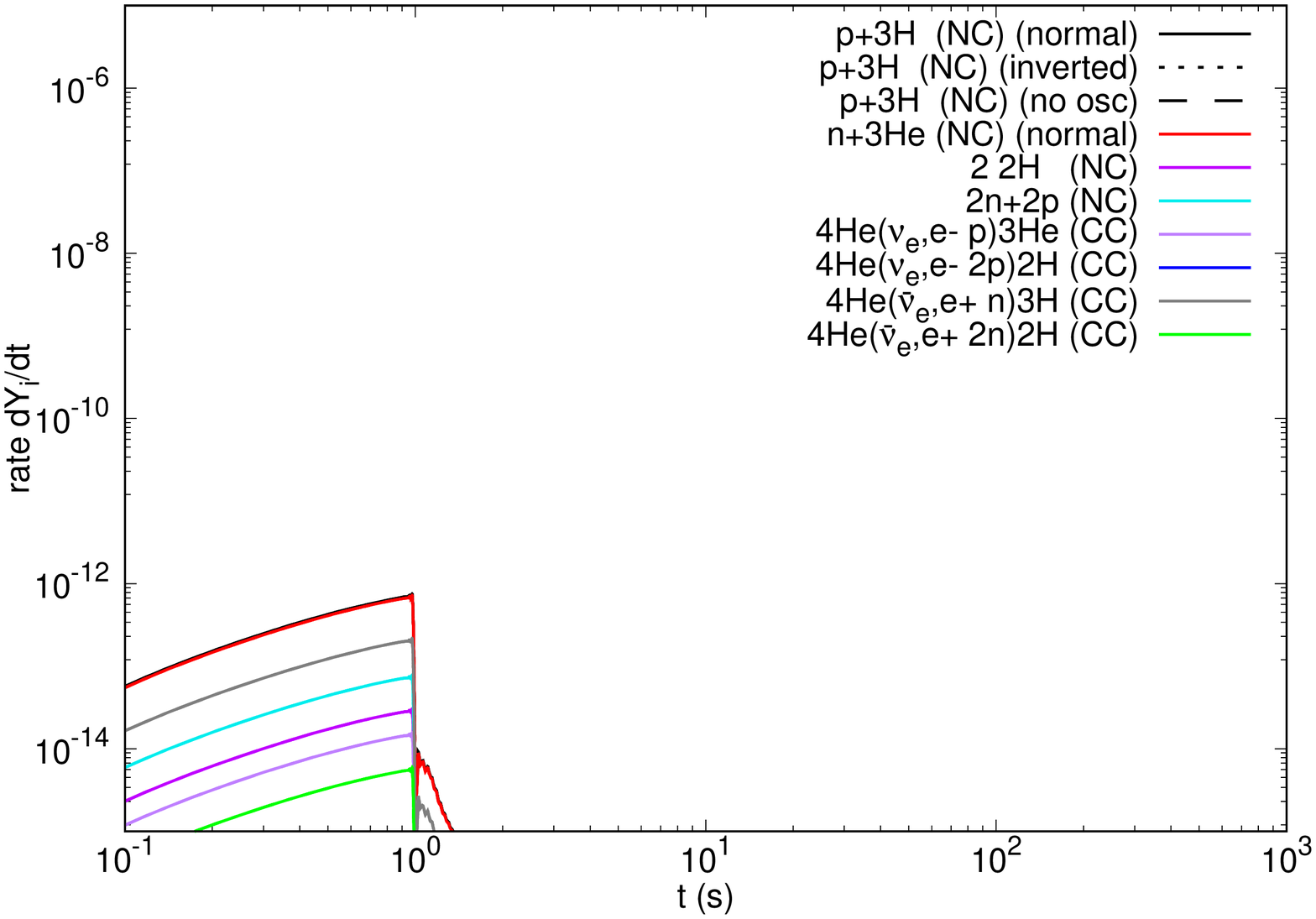}{0.3\textwidth}{(b)}
\fig{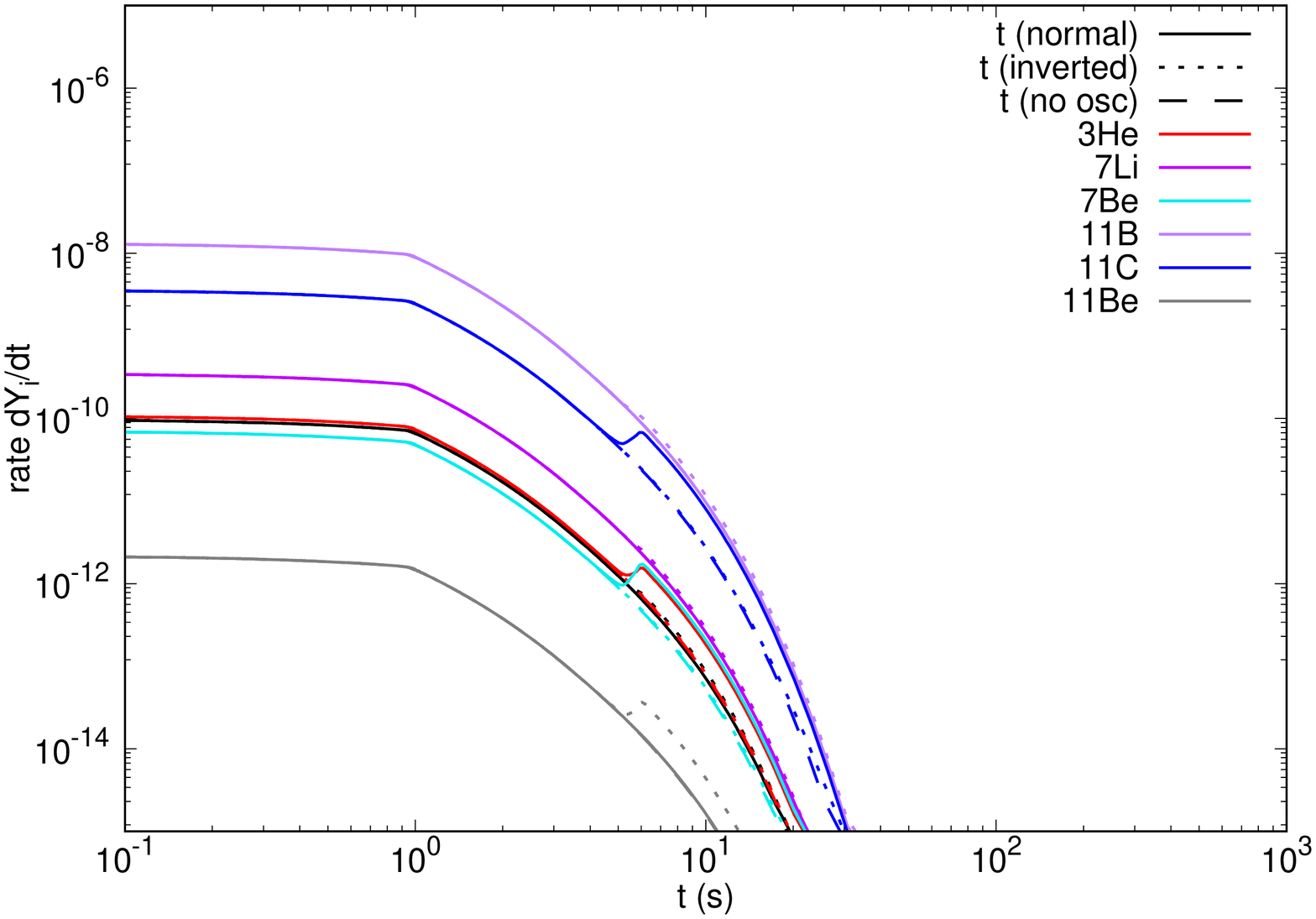}{0.3\textwidth}{(c)}}
\caption{Rates of abundance change $dY_i/dt$ via respective reactions versus time for shell 1. Solid, dotted, and dashed lines show results of the normal, inverted hierarchy, and no neutrino oscillation cases, respectively. Long dashed lines (not always visible in this figure) correspond to results for the no neutrino flux case. Panels correspond to: (a) nuclei in the initial states of two-body nuclear reactions; (b) $^4$He destruction via $\nu+^4$He reactions; (c) products of $\nu+^{12}$C reactions.}
\label{fig:rate_034}
\end{figure*}

\begin{figure*}[t!]
\gridline{\fig{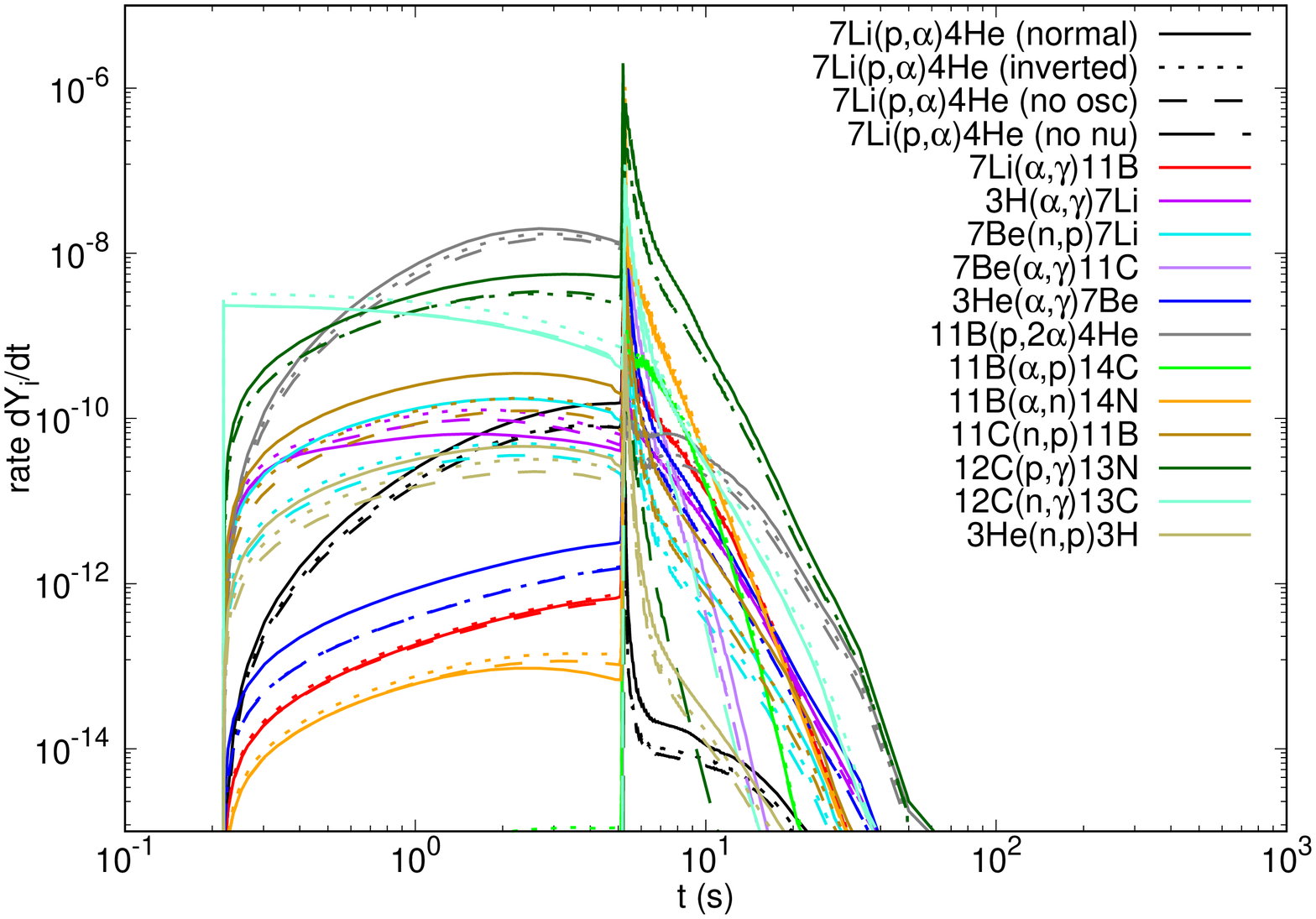}{0.3\textwidth}{(a)}
\fig{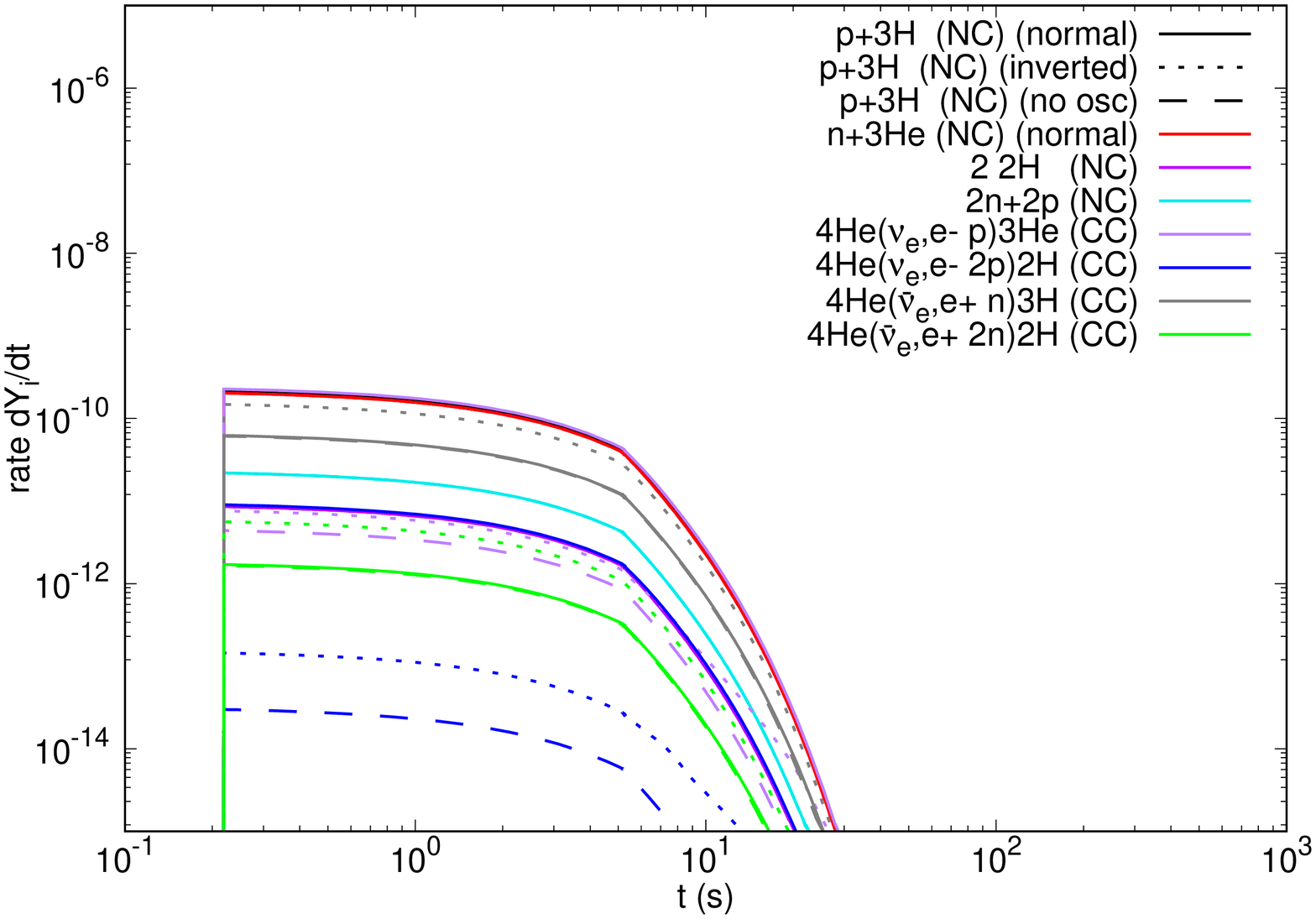}{0.3\textwidth}{(b)}
\fig{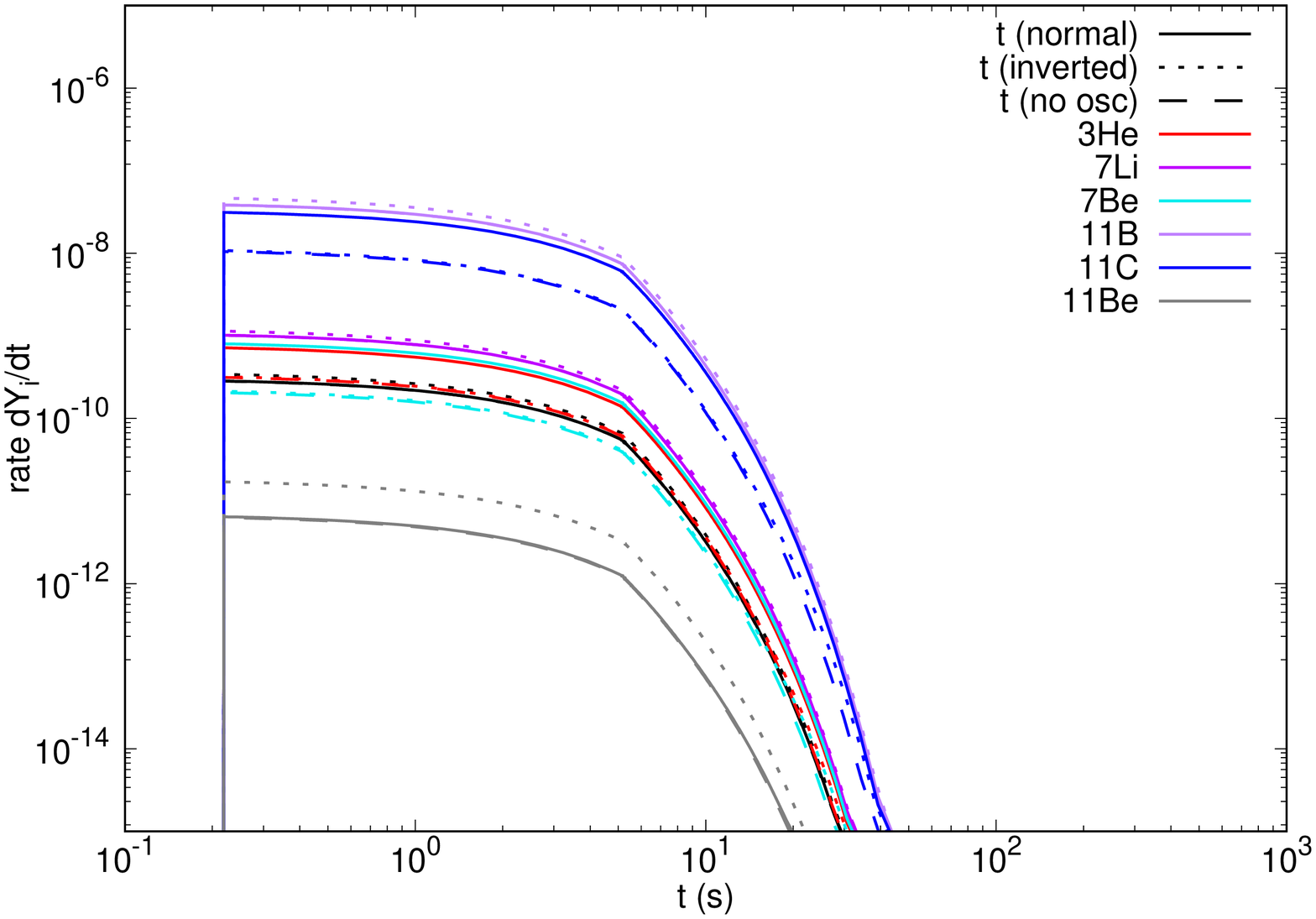}{0.3\textwidth}{(c)}}
\caption{Same as Fig. \ref{fig:rate_034} but for shell 2.}
\label{fig:rate_173}
\end{figure*}

\begin{figure*}[t!]
\gridline{\fig{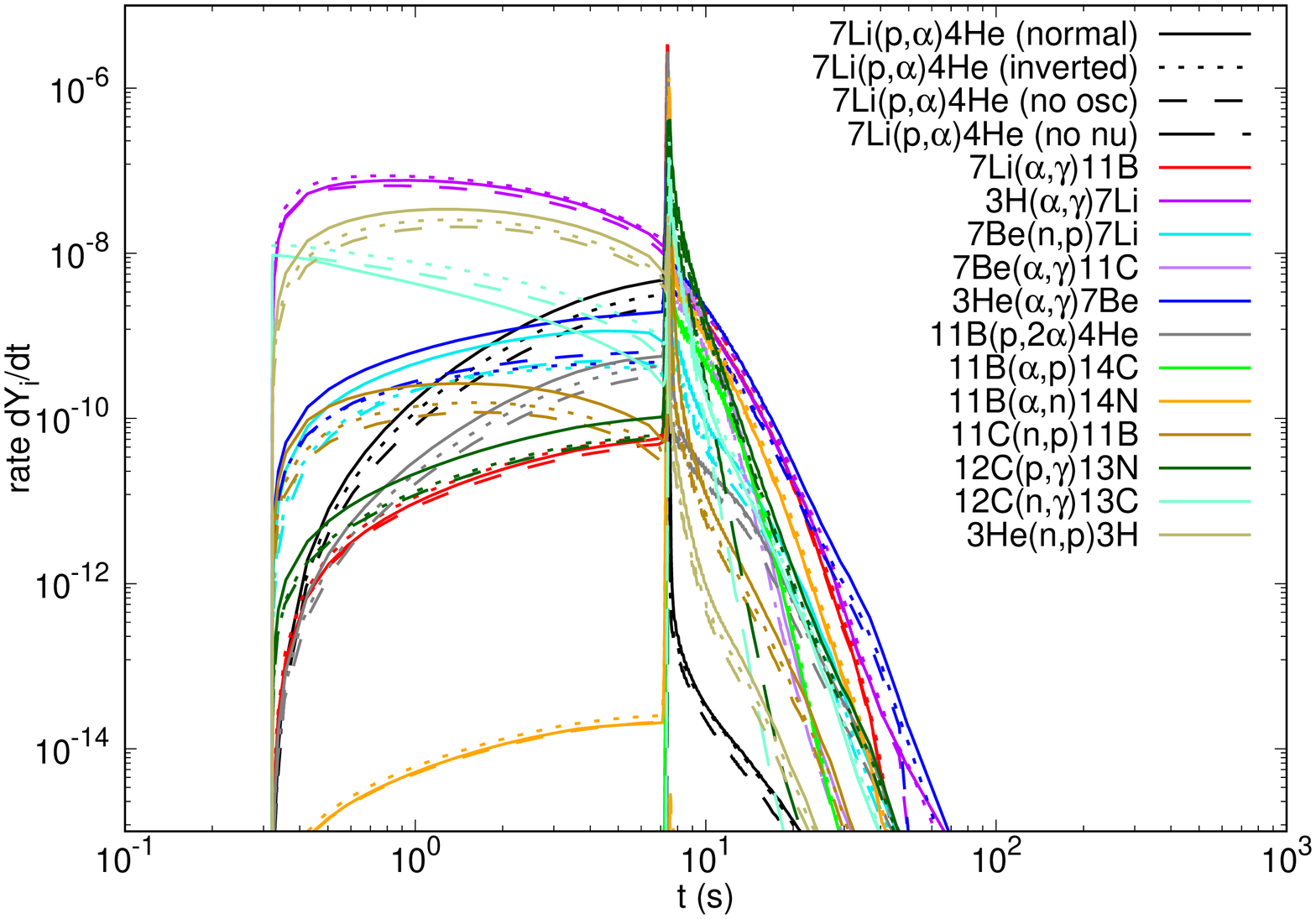}{0.3\textwidth}{(a)}
\fig{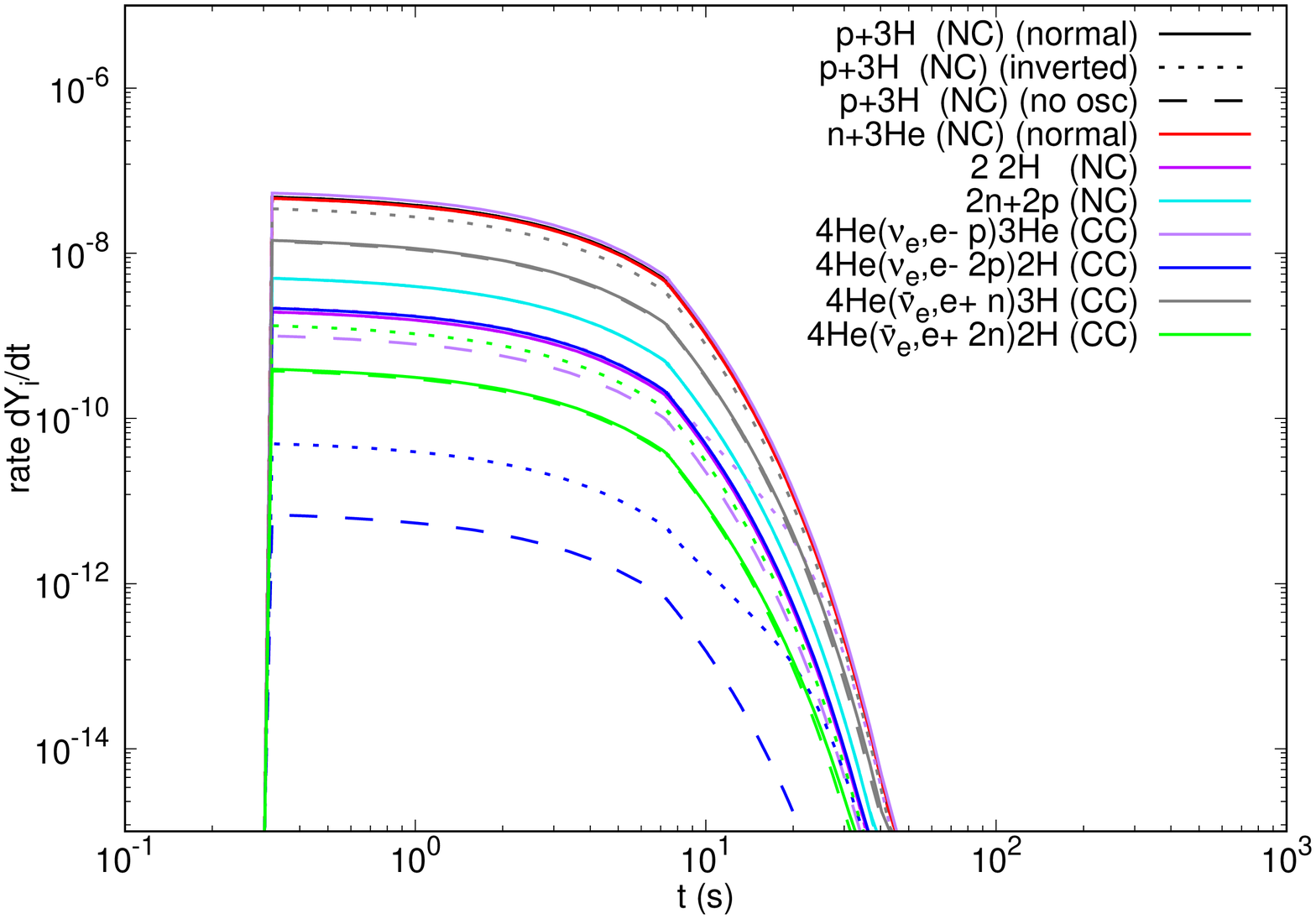}{0.3\textwidth}{(b)}
\fig{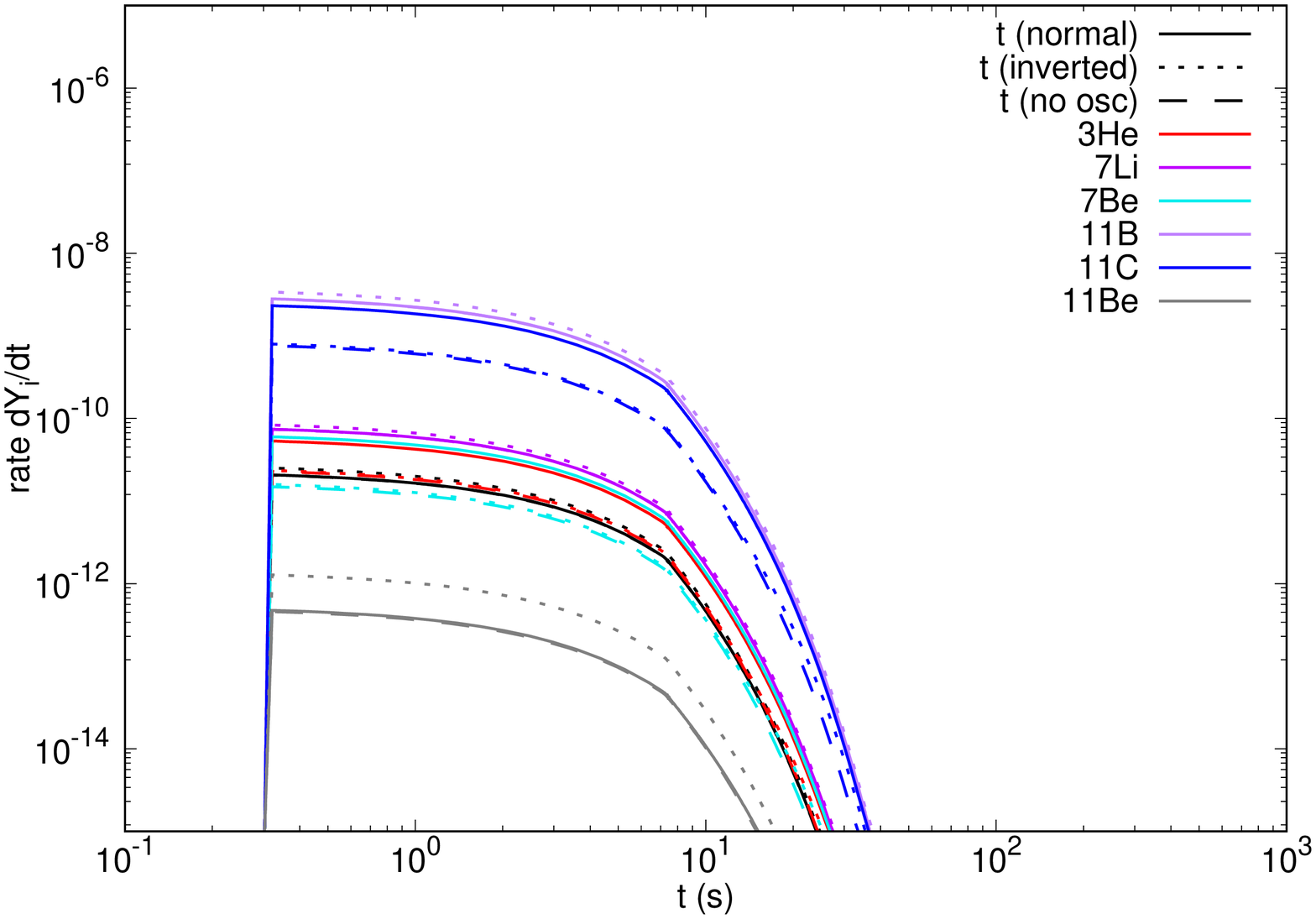}{0.3\textwidth}{(c)}}
\caption{Same as Fig. \ref{fig:rate_034} but for shell 3.}
\label{fig:rate_198}
\end{figure*}

\begin{figure*}[t!]
\gridline{\fig{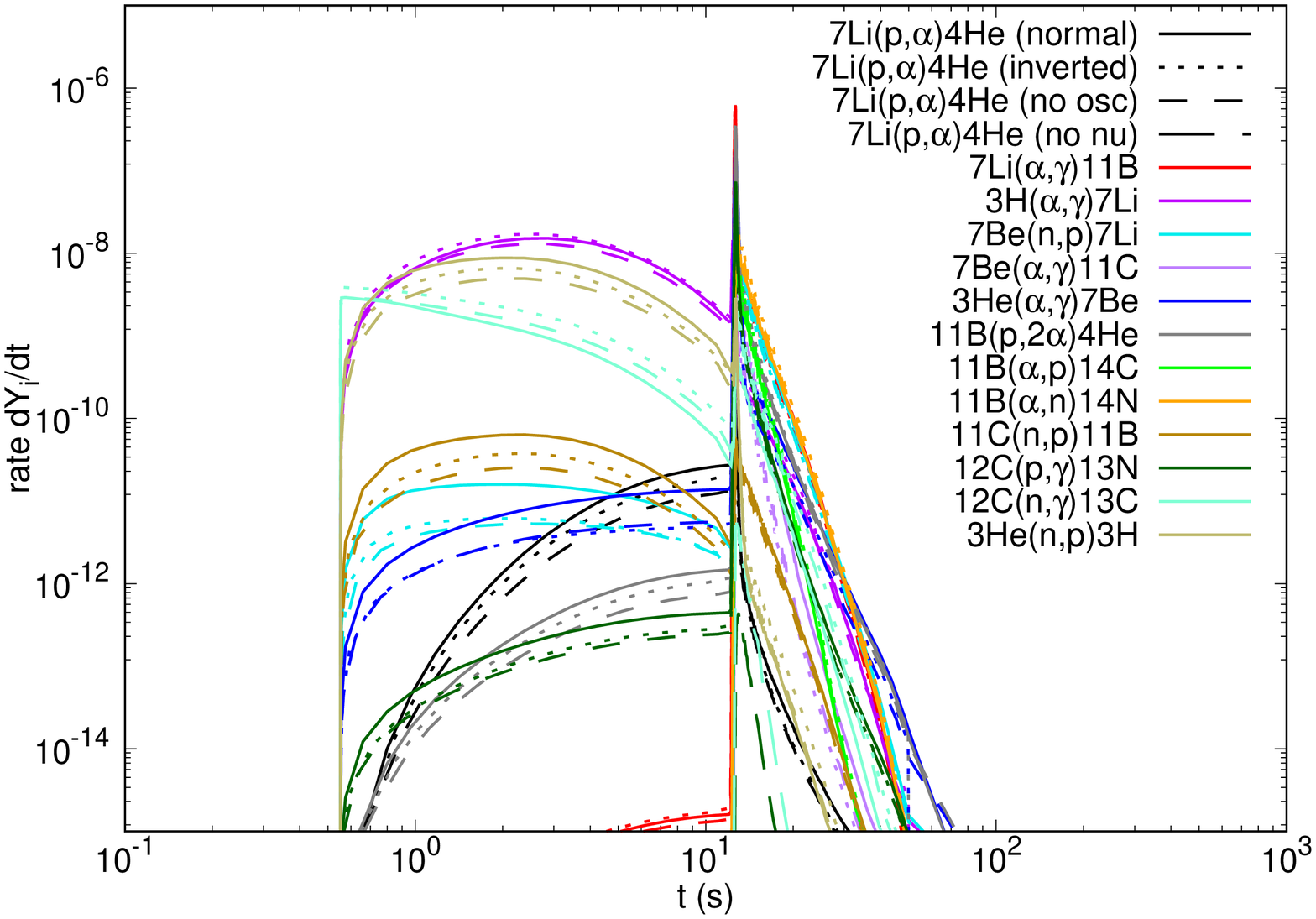}{0.3\textwidth}{(a)}
\fig{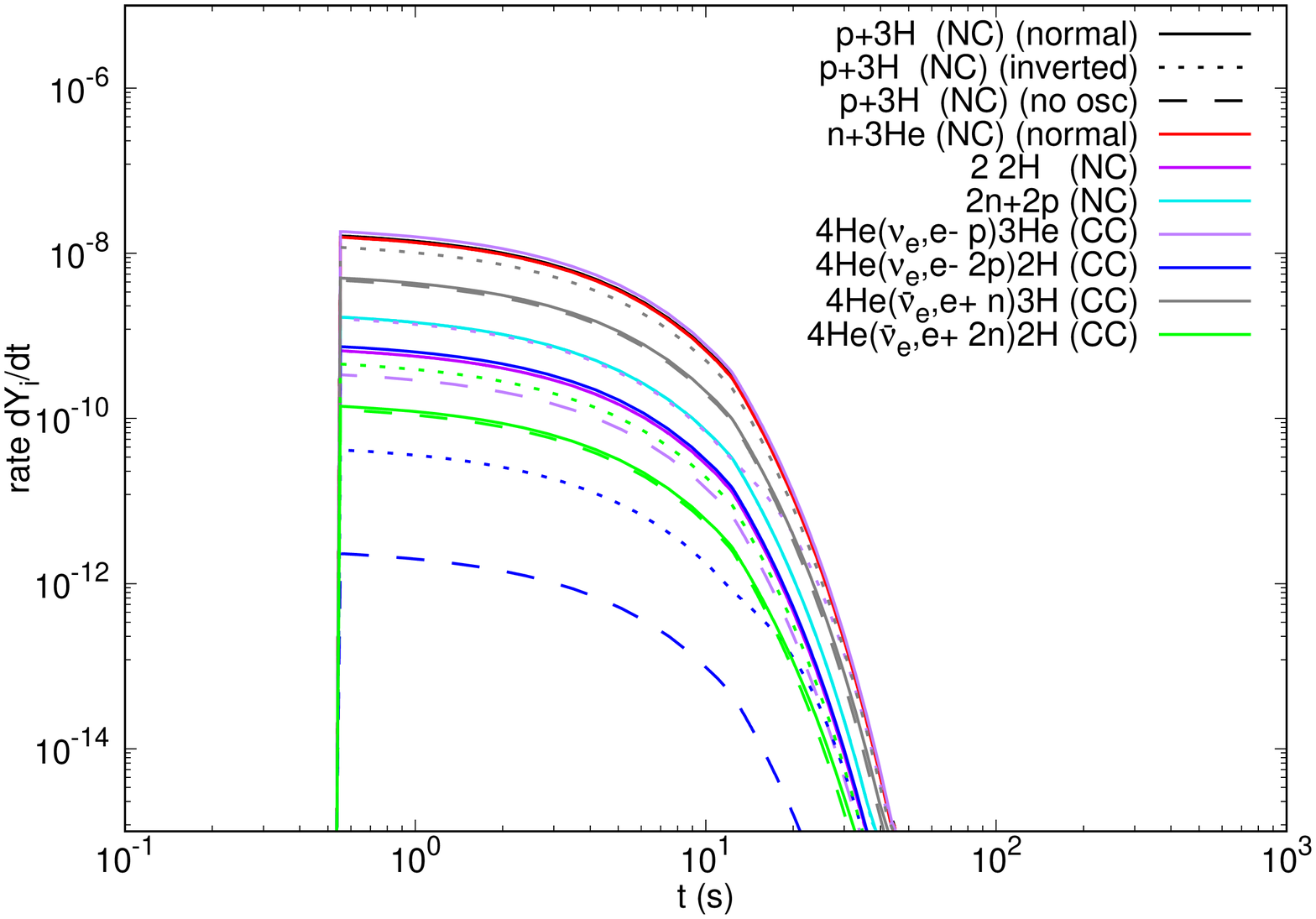}{0.3\textwidth}{(b)}
\fig{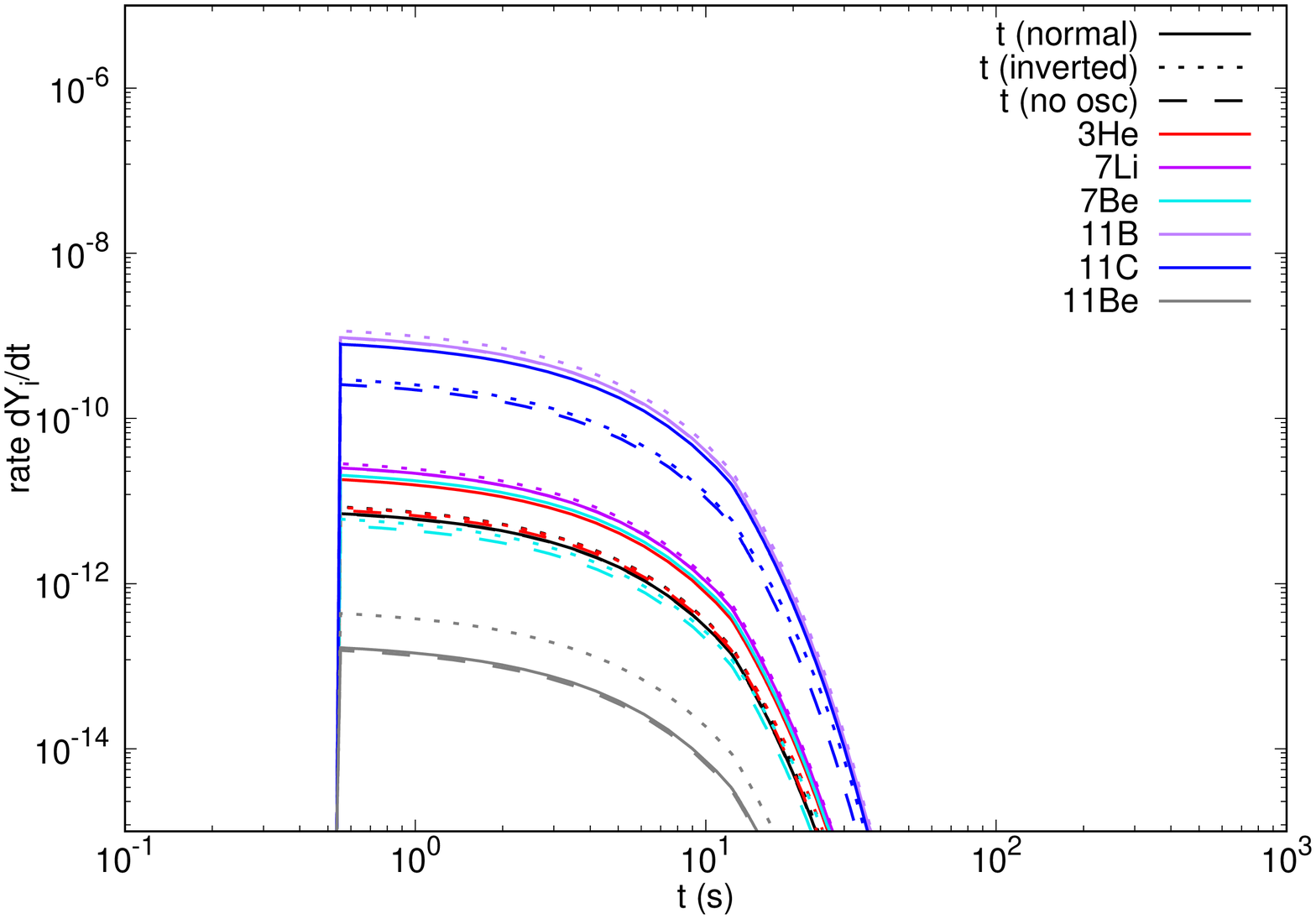}{0.3\textwidth}{(c)}}
\caption{Same as Fig. \ref{fig:rate_034} but for shell 4.}
\label{fig:rate_238}
\end{figure*}

\begin{figure*}[t!]
\gridline{\fig{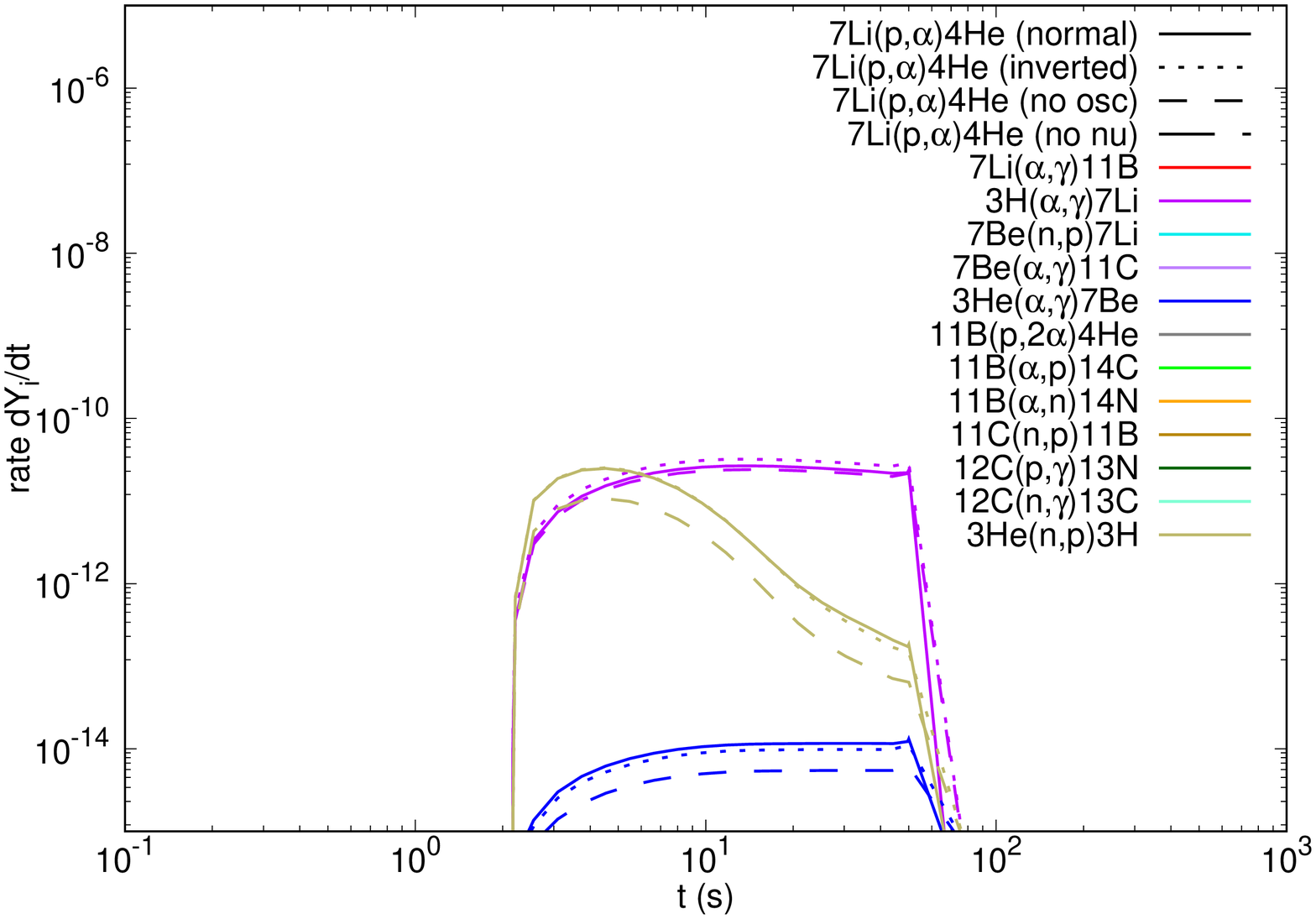}{0.3\textwidth}{(a)}
\fig{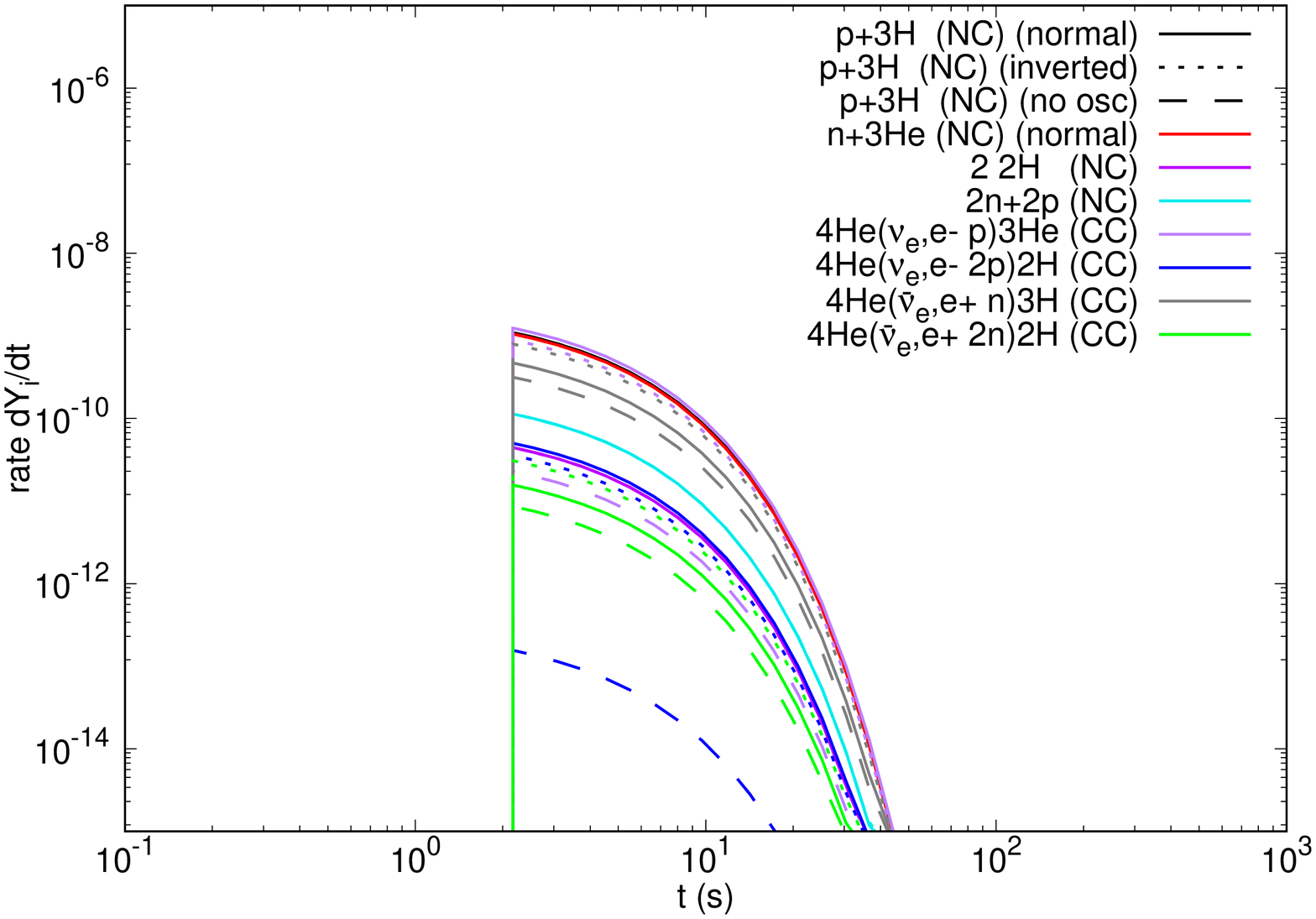}{0.3\textwidth}{(b)}}
\caption{Same as Fig. \ref{fig:rate_034} but for shell 5.}
\label{fig:rate_353}
\end{figure*}

Figures \ref{fig:rate_034}--\ref{fig:rate_353} show the rates of the abundance change $dY_i/dt$ versus time for shells 1--5, respectively. Solid, dotted, and dashed lines correspond to the normal hierarchy, inverted hierarchy, and no neutrino oscillation cases, respectively. Long dashed lines (not always visible in these figures) correspond to results of the no neutrino flux case. Respective panels show values for (a) nuclei in the initial states of the two-body nuclear reactions; (b) $^4$He via the $\nu+^4$He reactions; (c) products of $\nu+^{12}$C reactions.

\subsubsection{Shell 1}

\paragraph{2-body nuclear reactions:}
Before the shock passage at $t\sim 1$ s, the $^{12}$C($p$,$\gamma$)$^{13}$N and $^{11}$B($p$,2$\alpha$)$^4$He reactions are operative. After the shock arrives, the $^{11}$B($\alpha$,$n$)$^{14}$N and $^{11}$B($\alpha$,$p$)$^{14}$C reactions are also activated. It is found that the $^{12}$C($p$,$\gamma$)$^{13}$N reaction is much weaker in the no neutrino flux case except during the shock passage epoch. This reaction is driven by protons generated from $\nu$-spallation reactions.

\paragraph{$\nu$+$^4$He}
Although the $^4$He spallation proceeds, its effect is very small because of the small $^4$He abundance (Fig. \ref{fig:abun_034}).

\paragraph{$\nu$+$^{12}$C}
All products ($^{11}$B, $^{11}$C, $^7$Li, $^7$Be, ...) are produced. At the late time of $t \gtrsim 5$ s, $^{11}$C and $^7$Be production is more efficient in the normal case, while the $^{11}$Be production is more efficient in the inverted case. This is because this shell passes through the MSW resonance density at about that time.

\subsubsection{Shell 2}

\paragraph{2-body nuclear reactions:}
Similar to the shell 1, before the shock passage, the $^{11}$B($p$,2$\alpha$)$^4$He and $^{12}$C($p$,$\gamma$)$^{13}$N reactions are operative. After the shock arrival, the $^{12}$C($p$,$\gamma$)$^{13}$N and $^{11}$B($\alpha$,$n$)$^{14}$N reactions are strong. The $^{12}$C($p$,$\gamma$)$^{13}$N reaction is much weaker in the no neutrino flux case except during the shock passage epoch. This reaction is the most efficient in the normal hierarchy case. This is related to a larger proton production by $\nu_e$ after the neutrino flavor change.

\paragraph{$\nu$+$^4$He}
The $^4$He spallation is unimportant in this shell also.

\paragraph{$\nu$+$^{12}$C}
In the normal hierarchy case, the yields of $^{11}$C and $^7$Be are enhanced via the change of the $\nu_e$ spectrum. In the inverted hierarchy case, the yields of $^{11}$B and $^7$Li are slightly enhanced via the increased $\nueb$ reaction rates. Although the $^{11}$Be yield is also enhanced, it is much smaller than those of $^{11}$B and $^{11}$C.

\subsubsection{Shell 3}

\paragraph{2-body nuclear reactions:}
Before the shock passage at $t \sim 7$ s, the $^7$Li production reaction $^3$H($\alpha$,$\gamma$)$^7$Li and the charge exchange reaction $^3$He($n$,$p$)$^3$H are strong.
During the shock heating, the $^7$Li($\alpha$,$\gamma$)$^{11}$B, $^{12}$C($p$,$\gamma$)$^{13}$N, $^3$H($\alpha$,$\gamma$)$^7$Li, $^3$He($\alpha$,$\gamma$)$^7$Be, $^{11}$B($\alpha$,$n$)$^{14}$N, and $^{12}$C($n$,$\gamma$)$^{13}$C reactions effectively operate.
In the normal hierarchy, the rates of the $^3$He($\alpha$,$\gamma$)$^7$Be and $^{12}$C($p$,$\gamma$)$^{13}$N reactions are the largest due to the enhanced proton richness in the neutrino spallation products.
In the inverted hierarchy, the rates of the $^3$H($\alpha$,$\gamma$)$^7$Li and $^7$Li($\alpha$,$\gamma$)$^{11}$B reactions are enhanced due to the neutron richness.

\paragraph{$\nu$+$^4$He}
The NC reactions are effective independently of the neutrino oscillations.
Rates of the CC reactions $^{4}$He($\nu_e$,$e^-$$p$)$^3$He and $^{4}$He($\nu_e$,$e^-$2$p$)$^2$H are extremely spectrum-dependent. The decreasing order of rate is the normal, inverted hierarchy, and no oscillation cases.
On the other hand, rates for $^{4}$He($\nueb$,$e^+$$n$)$^3$H and $^{4}$He($\nueb$,$e^+$2$n$)$^2$H are enhanced only in the inverted hierarchy case.

\paragraph{$\nu$+$^{12}$C}
The trends of reaction rates are the same as those of shell 2.

\subsubsection{Shell 4}

\paragraph{2-body nuclear reactions:}
Before the shock passage, rates of $^3$H($\alpha$,$\gamma$)$^7$Li and $^3$He($n$,$p$)$^3$H are large.
During the shock heating, the $^{11}$B($\alpha$,$n$)$^{14}$N, $^7$Li($\alpha$,$\gamma$)$^{11}$B, $^7$Be($n$,$p$)$^7$Li, $^{11}$B($p$,2$\alpha$)$^4$He, $^{11}$B($\alpha$,$p$)$^{14}$C, $^7$Be($\alpha$,$\gamma$)$^{11}$C, $^{12}$C($n$,$\gamma$)$^{13}$C, $^{12}$C($p$,$\gamma$)$^{13}$N, $^3$H($\alpha$,$\gamma$)$^7$Li, and $^3$He($\alpha$,$\gamma$)$^7$Be reactions become effective.
In the normal hierarchy, the rates of $^{12}$C($p$,$\gamma$)$^{13}$N and $^3$He($\alpha$,$\gamma$)$^7$Be are largest, while
in the inverted hierarchy, the rates of $^{11}$B($\alpha$,$n$)$^{14}$N, $^3$H($\alpha$,$\gamma$)$^7$Li, and $^{12}$C($n$,$\gamma$)$^{13}$C are enhanced for the same reason explained for shell 3.

\paragraph{$\nu$+$^4$He}
Reaction rates behave similarly to those of shell 3.

\paragraph{$\nu$+$^{12}$C}
Reaction rates behave similarly to those of shell 2.

\subsubsection{Shell 5}

\paragraph{2-body nuclear reactions:}
The shock does not reach to this shell. Before the shock passage, only the rates of the $^3$H($\alpha$,$\gamma$)$^7$Li, $^3$He($\alpha$,$\gamma$)$^7$Be and $^3$He($n$,$p$)$^3$H reactions are significant.
The rate of $^3$H($\alpha$,$\gamma$)$^7$Li is enhanced in the inverted hierarchy because of the enhanced $^3$H yield or neutron richness.
The rate of $^3$He($n$,$p$)$^3$H is enhanced in the normal and inverted hierarchy cases because of the enhanced $^3$He yields.

\paragraph{$\nu$+$^4$He}
The trends for the CC rates of $^{4}$He($\nu_e$,$e^-$$p$)$^3$He and $^{4}$He($\nu_e$,$e^-$2$p$)$^2$H are similar to those of shell 3. However, in this shell, the differences of the rates between the normal and inverted hierarchy cases are small. We can interpret that this is because the flavor change at the low MSW resonance is complete, and the $\nue$ spectrum has changed in the inverted hierarchy case also \citep[cf. Fig. 2 (left panel) in][]{2006ApJ...649..319Y}.
Enhancements of the rate for the $^{4}$He($\nueb$,$e^+$$n$)$^3$H and $^{4}$He($\nueb$,$e^+$2$n$)$^2$H reactions in the inverted hierarchy case are, on the other hand, larger than those in the normal hierarchy case.

\paragraph{$\nu$+$^{12}$C}
Reaction rates are too small to affect light element yields.

\section{effects of initial abundances on abundance change rates}\label{app3}

In the inner region (shell 1), there is no difference in production rates of light nuclides, and obtained rates of Cases 4 and 5 are the same as those of Case 2 (Fig. \ref{fig:rate_034}).
No effects of different $s$-nuclear abundances are expected to be seen since light nuclides including $^{16}$O, $^{20}$Ne, $^{24,25}$Mg, and $^{28}$Si as the main components are predominant in this inner region of the presupernova stars. Their compositions do not depend on the metallicity or the $s$-nuclear abundances, and determine the neutron abundances.

\begin{figure*}[t!]
\gridline{\fig{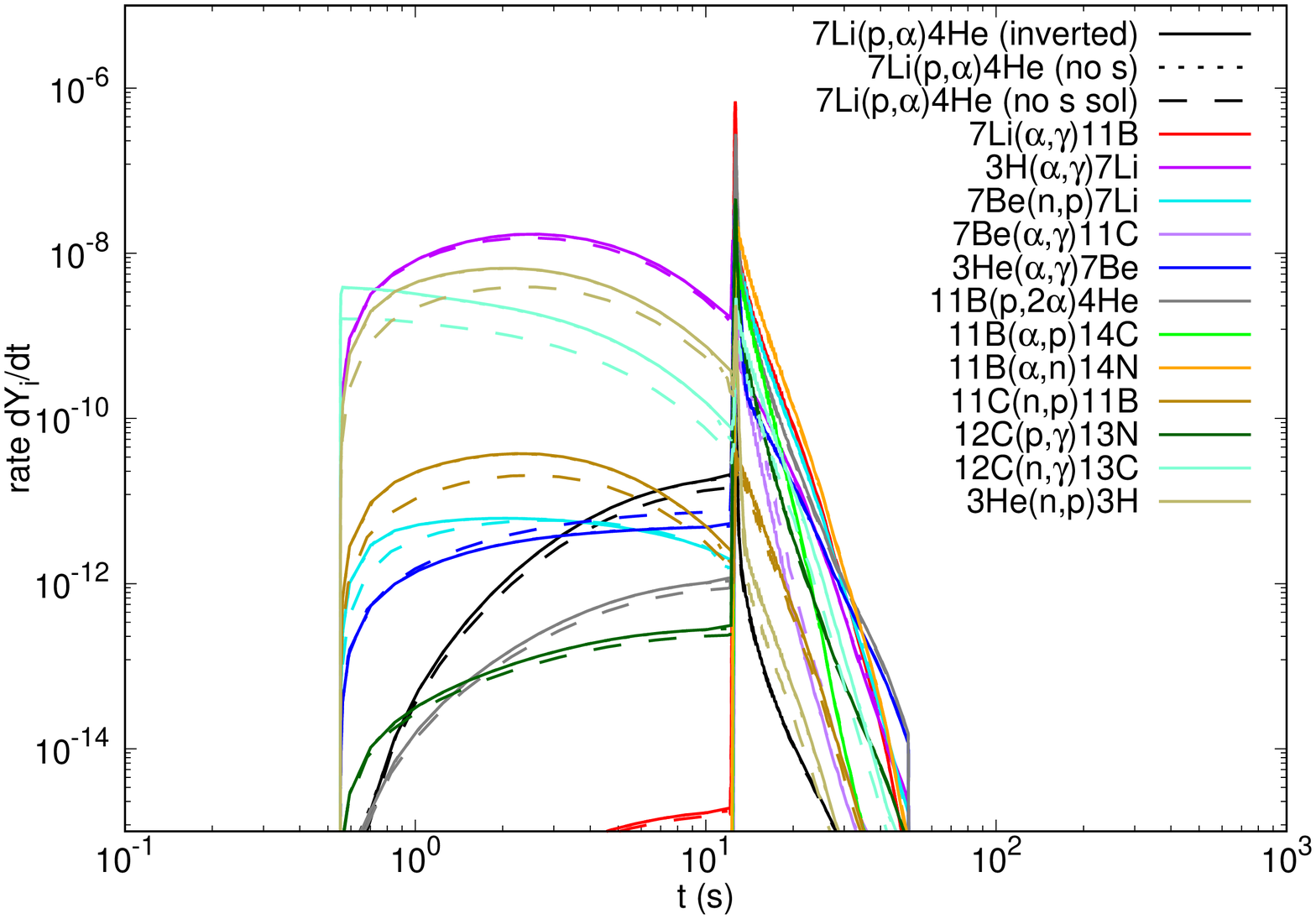}{0.3\textwidth}{(a)}
\fig{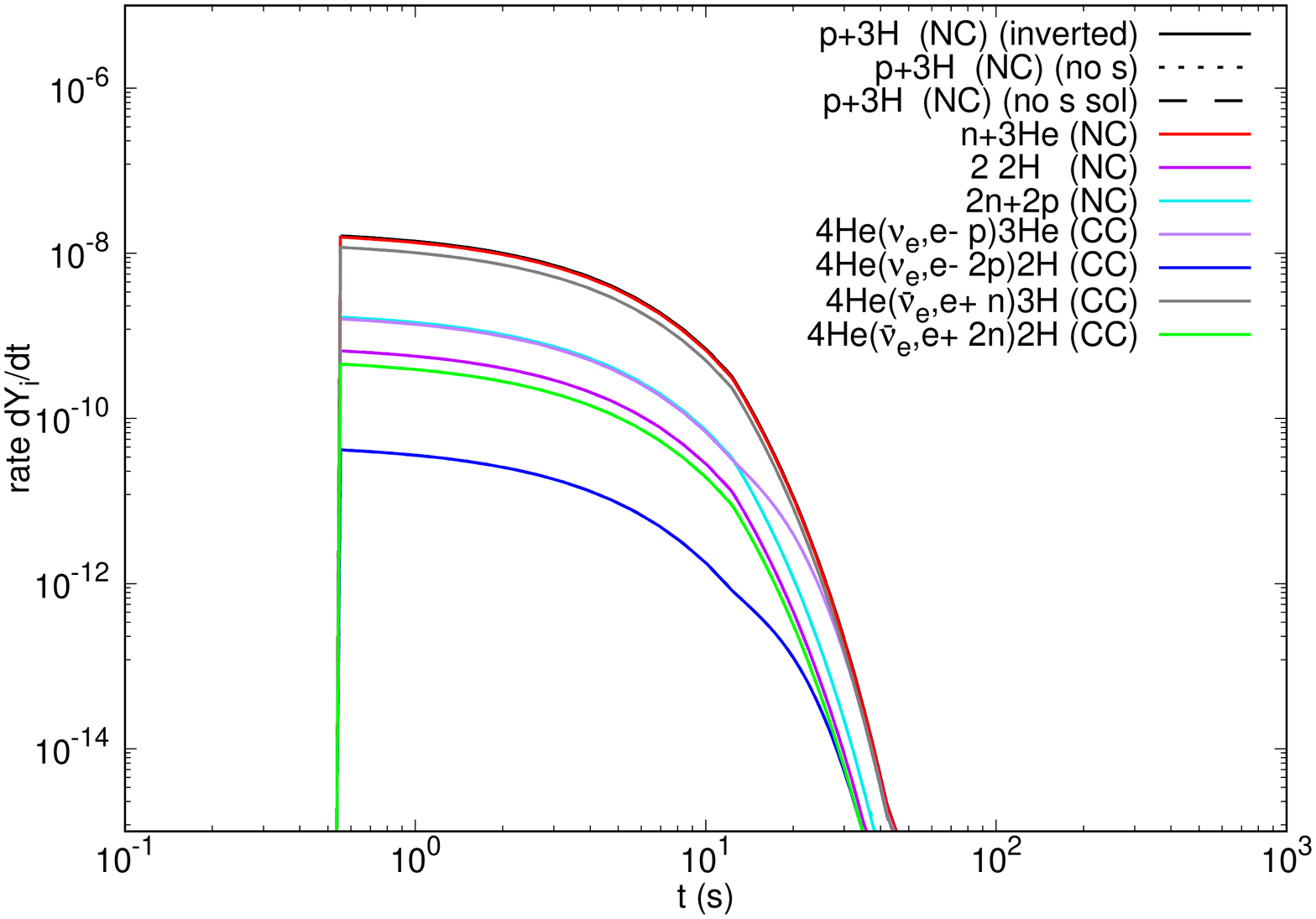}{0.3\textwidth}{(b)}
\fig{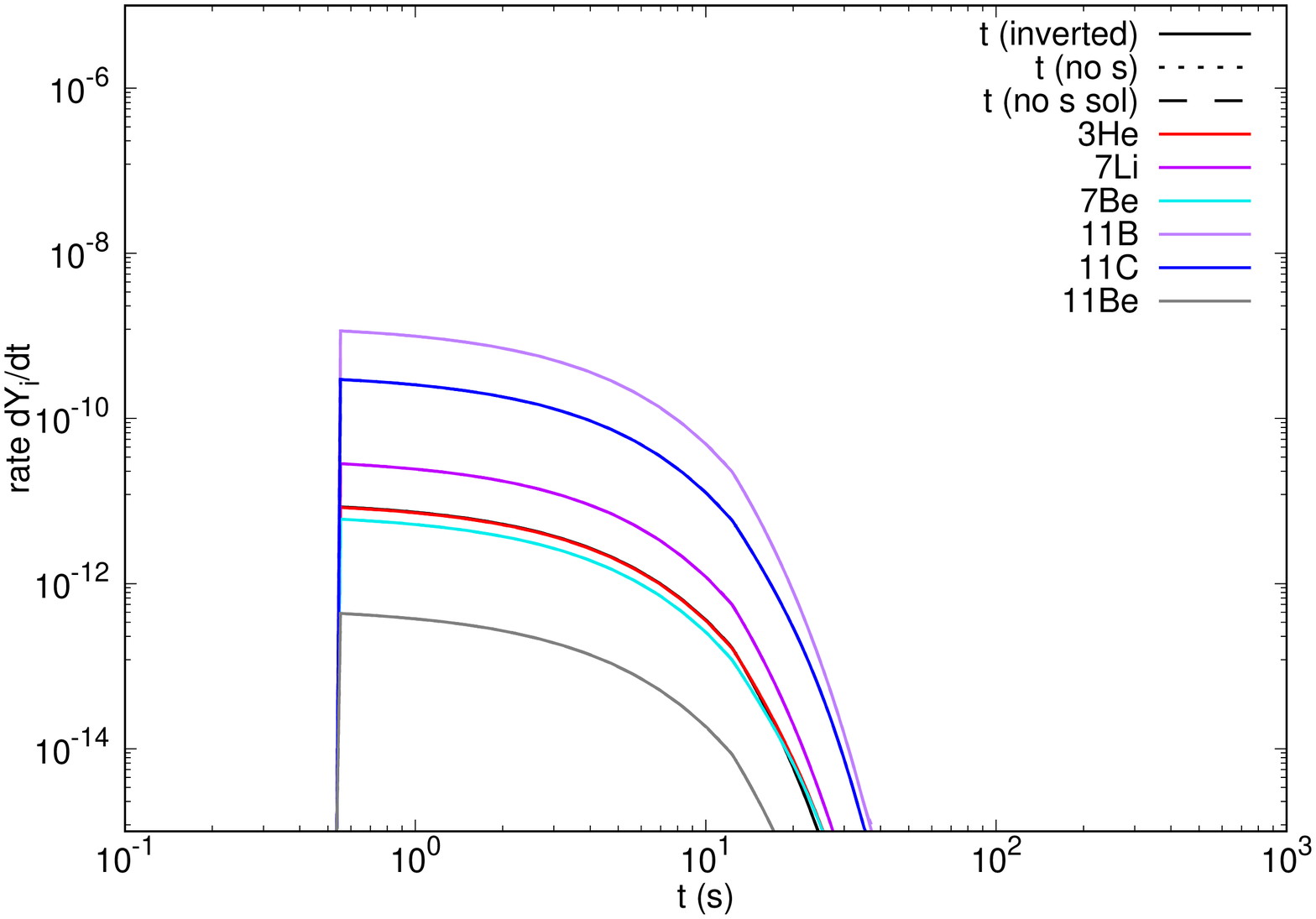}{0.3\textwidth}{(c)}}
\caption{Same as Fig. \ref{fig:rate_238}, but solid and dashed lines correspond to results for heavy elemental abundances taken from the standard $s$-process abundances after stellar evolution of $Z=Z_\sun/4$ (Case 2) and solar abundances (Case 5), respectively. Dotted lines for heavy elemental abundances from the solar abundances divided by 4 (Case 4) overlap with solid lines and are not seen.}
\label{fig:rate_m_238}
\end{figure*}

Figure \ref{fig:rate_m_238} shows the rates of abundance changes $dY_i/dt$ similar to Fig. \ref{fig:rate_238} for shell 4. Solid and dashed lines correspond to results for Cases 2 and 5, respectively. The result for Case 4 is almost the same as that for Case 4 in the whole stellar region.

Immediately after the arrival of the neutrinos at $t \gtrsim 0.5$ s, all neutron capture reaction rates shown in Fig. \ref{fig:rate_m_238} are smaller in Case 5. In Case 5, the abundance of neutron poisons including heavy $s$-nuclei are larger, and the neuron abundance is kept lower.

The rate of the $^{3}$He($\alpha$,$\gamma$)$^{7}$Be reaction is larger in Case 5 because of a difference in the $^3$He abundance (Fig. \ref{fig:abun_m_238}), that is controlled by the reaction $^{3}$He($n$,$p$)$^{3}$H.

All proton capture reactions are faster in Case 2. This reflects differences in the proton abundances. Once the differences in the $^7$Be abundance becomes significant, the $^{7}$Be($n$,$p$)$^{7}$Li rate follows that of the $^{3}$He($\alpha$,$\gamma$)$^{7}$Be reaction as well as the neutron abundance.

After the shock passage, all neutron capture reaction rates are larger in Case 2 reflecting the different neutron abundances. All proton capture reaction rates are larger in Case 2 because of the larger proton abundances. 

We note that the $^7$Li nuclei are violently burned during the shock passage and their abundance decreases significantly. The rate of $^{7}$Be($n$,$p$)$^{7}$Li differs within $\mathcal{O}(10)$\% between the models. Since this rate is marginally effective and near the freeze-out, this small difference results in a significant difference in the survival probability of $^{7}$Be after the shock. Once $^{7}$Be nuclei are converted to $^{7}$Li, they are quickly burned. The more effective conversion in Case 5 leads to a small final $^{7}$Be abundance.

Since the $^4$He abundance does not change between the three models, the differences in the $\alpha$-capture reaction rates reflect the number abundances of target nuclei.

  In shell 2--4, trends of reaction rates are similar although the sizes of differences depend on the location.

In the outermost region of the He-layer (shell 5), the main nuclear components are $^4$He that has been produced from the initial hydrogen and $^{14}$N from the initial CNO elements. Therefore, this region does not experience an effective $s$-process before the SN. Then, the heavy nuclear abundances are almost the same in Cases 2 and 4. The neutron abundances in these two models are the same, while those in Case 5 are smaller because of the larger abundance of neutron absorbers.

The only significant difference is in the rate of $^{3}$He($n$,$p$)$^{3}$H, which is proportional to the neutron abundance. This rate is the smallest in Case 5.


\begin{thebibliography}{}

\bibitem[Aoki et al.(2009)]{2009ApJ...698.1803A} Aoki, W., Barklem, P.~S., Beers, T.~C., et al.\ 2009, \apj, 698, 1803
  
\bibitem[Austin et al.(2011)]{2011PhRvL.106o2501A} Austin, S.~M., Heger, A., \& Tur, C.\ 2011, Physical Review Letters, 106, 152501 

\bibitem[Austin et al.(2014)]{2014PhRvL.112k1101A} Austin, S.~M., West, C., \& Heger, A.\ 2014, Physical Review Letters, 112, 111101 

\bibitem[Bader \& Deuflhard(1983)]{Bader1983} 
  Bader, G., \&  Deuflhard, P.\
  1983, Numerische Mathematik, 41, 373

\bibitem[Balantekin \& Y{\"u}ksel(2005)]{2005NJPh....7...51B} Balantekin, A.~B., \& Y{\"u}ksel, H.\ 2005, New Journal of Physics, 7, 51

\bibitem[Banerjee et al.(2016a)]{2016EPJWC.10906001B} Banerjee, P., Qian, Y.-Z., Heger, A., \& Haxton, W.\ 2016, European Physical Journal Web of Conferences, 109, 06001 

\bibitem[Banerjee et al.(2016b)]{2016NatCo...713639B} Banerjee, P., Qian, Y.-Z., Heger, A., \& Haxton, W.~C.\ 2016, Nature Communications, 7, 13639
  
\bibitem[Bird et al.(2008)]{2008PhRvD..78h3010B} Bird, C., Koopmans, K., \& Pospelov, M.\ 2008, \prd, 78, 083010 

\bibitem[Cameron(1955)]{1955ApJ...121..144C} Cameron, A.~G.~W.\ 1955, \apj, 121, 144
  
\bibitem[Cameron \& Fowler(1971)]{1971ApJ...164..111C} Cameron, A.~G.~W., \& Fowler, W.~A.\ 1971, \apj, 164, 111
  
\bibitem[Cheoun et al.(2012)]{2012PhRvC..85f5807C} Cheoun, M.-K., Ha, E., Hayakawa, T., et al.\ 2012, \prc, 85, 065807 

\bibitem[Cunha et al.(2000)]{2000ApJ...530..939C} Cunha, K., Smith, V.~V., Boesgaard, A.~M., \& Lambert, D.~L.\ 2000, \apj, 530, 939 
  
\bibitem[Cyburt et al.(2010)]{2010ApJS..189..240C} Cyburt, R.~H., Amthor, A.~M., Ferguson, R., et al.\ 2010, \apjs, 189, 240 

\bibitem[Davis \& Duff(1995)]{MA38} Davis, T. A., \& Duff, I. S.\ 1995, HSL. A collection of Fortran codes for large scale scientific computation.
    
\bibitem[Deneault et al.(2003)]{2003ApJ...594..312D} Deneault, E.~A.-N., Clayton, D.~D., \& Heger, A.\ 2003, \apj, 594, 312

\bibitem[Domogatskii et al.(1978)]{1978Ap&SS..58..273D} Domogatskii, G.~V., Eramzhian, R.~A., \& Nadezhin, D.~K.\ 1978, \apss, 58, 273

\bibitem[Duncan et al.(1997)]{1997ApJ...488..338D} Duncan, D.~K., Primas, F., Rebull, L.~M., et al.\ 1997, \apj, 488, 338

\bibitem[Fujiya et al.(2011)]{2011ApJ...730L...7F} Fujiya, W., Hoppe, P., \& Ott, U.\ 2011, \apjl, 730, L7

\bibitem[Garcia Lopez et al.(1998)]{1998ApJ...500..241G} Garcia Lopez, R.~J., Lambert, D.~L., Edvardsson, B., et al.\ 1998, \apj, 500, 241 

\bibitem[Hayakawa et al.(2018)]{Hayakawa2017} Hayakawa, T., Ko, H., Cheoun, M.~K., Kusakabe, M., Kajino, T., Usang, M. D., Chiba, S., Nakamura, K., Tolstov, A., Nomoto, K., Hashimoto, M.-a., Ono, M., Mathews, G. J.\ 2018, \prl, 121, 102701.

\bibitem[Hayakawa et al.(2013)]{2013ApJ...779L...9H} Hayakawa, T., Nakamura, K., Kajino, T., et al.\ 2013, \apjl, 779, L9 

\bibitem[Heger et al.(2005)]{2005PhLB..606..258H} Heger, A., Kolbe, E., Haxton, W.~C., et al.\ 2005, Physics Letters B, 606, 258

\bibitem[Hernanz et al.(1996)]{1996ApJ...465L..27H} Hernanz, M., Jose, J., Coc, A., \& Isern, J.\ 1996, \apjl, 465, L27

\bibitem[Izutani et al.(2012)]{2012IAUS..279..339I} Izutani, N., Umeda, H., \& Yoshida, T.\ 2012, Death of Massive Stars: Supernovae and Gamma-Ray Bursts, 279, 339 

\bibitem[Kawasaki \& Kusakabe(2011)]{2011PhRvD..83e5011K} Kawasaki, M., \& Kusakabe, M.\ 2011, \prd, 83, 055011 

\bibitem[Kikuchi et al.(2015)]{2015PTEP.2015f3E01K} Kikuchi, Y., Hashimoto, M.-a., Ono, M., \& Fukuda, R.\ 2015, Progress of Theoretical and Experimental Physics, 2015, 063E01 

\bibitem[Ko et al.(2019a)]{Ko2019} Ko, H., Cheoun, M.-K., Ha, E., et al.\ 2019a, arXiv:1903.02086

\bibitem[Ko et al.(2019b)]{Ko2019b} Ko, H., Cheoun, M.-K., et al., \ 2019b, in preparation.
  
\bibitem[Kobayashi et al.(2011)]{2011ApJ...739L..57K} Kobayashi, C., Izutani, N., Karakas, A.~I., et al.\ 2011, \apjl, 739, L57 

\bibitem[Koning et al.(2008)]{Koning2007} 
  Koning, A.~J., Hilaire, S., \& Duijvestijn, M.~C.,
  Proceedings of the International Conference on Nuclear Data for Science and Technology, April 22-27, 2007, Nice, France,
  editors O.~Bersillon, F.~Gunsing, E.~Bauge, R.~Jacqmin, and S.~Leray, EDP Sciences, 2008, p. 211-214.


\bibitem[Kuo \& Pantaleone(1987)]{1987PhLB..198..406K} Kuo, T.~K., \& Pantaleone, J.\ 1987, Physics Letters B, 198, 406

\bibitem[Kusakabe et al.(2009)]{2009PhRvD..80j3501K} Kusakabe, M., Kajino, T., Yoshida, T., \& Mathews, G.~J.\ 2009, \prd, 80, 103501
  
\bibitem[Kusakabe et al.(2014)]{2014ApJS..214....5K} Kusakabe, M., Kim, K.~S., Cheoun, M.-K., et al.\ 2014, \apjs, 214, 5


\bibitem[Lodders et al.(2009)]{2009LanB...4B..712L} Lodders, K., Palme, H., \& Gail, H.-P.\ 2009, Landolt B{\"o}rnstein, 712

\bibitem[Mathews et al.(2012)]{2012PhRvD..85j5023M} Mathews, G.~J., Kajino, T., Aoki, W., Fujiya, W., \& Pitts, J.~B.\ 2012, \prd, 85, 105023 

\bibitem[Meneguzzi et al.(1971)]{1971A&A....15..337M} Meneguzzi, M., Audouze, J., \& Reeves, H.\ 1971, \aap, 15, 337 

\bibitem[Nadyozhin \& Panov(2014)]{2014MNRAS.441..733N} Nadyozhin, D.~K., \& Panov, I.~V.\ 2014, \mnras, 441, 733 

\bibitem[Nozawa et al.(2003)]{2003ApJ...598..785N} Nozawa, T., Kozasa, T., Umeda, H., Maeda, K., \& Nomoto, K.\ 2003, \apj, 598, 785 

\bibitem[Olive \& Particle Data Group(2014)]{2014ChPhC..38i0001O} Olive, K.~A., \& Particle Data Group 2014, Chinese Physics C, 38, 090001

\bibitem[Ott et al.(2014)]{ott_blcode} Ott, C. Morozova, V., \& Piro, A. L.\ 2014

\bibitem[Pagel(1997)]{1997nceg.book.....P} Pagel, B.~E.~J.\ 1997, Nucleosynthesis and Chemical Evolution of Galaxies, by Bernard E.~J.~Pagel, pp.~392.~ISBN 0521550610.~Cambridge, UK: Cambridge University Press, October 1997., 392


\bibitem[Prantzos(2012)]{2012A&A...542A..67P} Prantzos, N.\ 2012, \aap, 542, A67 

\bibitem[Press et al.(1992)]{1992nrfa.book.....P} Press, W.~H., Teukolsky, S.~A., Vetterling, W.~T., \& Flannery, B.~P.\ 1992, Numerical recipes in FORTRAN. The art of scientific computing (2nd ed.; Cambridge: Cambridge Univ. Press)

\bibitem[Primas et al.(1999)]{1999A&A...343..545P} Primas, F., Duncan, D.~K., Peterson, R.~C., \& Thorburn, J.~A.\ 1999, \aap, 343, 545 
  
\bibitem[Reeves(1970)]{1970Natur.226..727R} Reeves, H.\ 1970, \nat, 226, 727

\bibitem[Ryan et al.(2000)]{2000ApJ...530L..57R} Ryan, S.~G., Beers, T.~C., Olive, K.~A., Fields, B.~D., \& Norris, J.~E.\ 2000, \apjl, 530, L57 

\bibitem[Sackmann \& Boothroyd(1999)]{1999ApJ...510..217S} Sackmann, I.-J., \& Boothroyd, A.~I.\ 1999, \apj, 510, 217 

\bibitem[Sbordone et al.(2010)]{2010A&A...522A..26S} Sbordone, L., Bonifacio, P., Caffau, E., et al.\ 2010, \aap, 522, A26 

\bibitem[Shigeyama \& Nomoto(1990)]{1990ApJ...360..242S} Shigeyama, T., \& Nomoto, K.\ 1990, \apj, 360, 242

\bibitem[Sieverding et al.(2015)]{2015arXiv150501082S} Sieverding, A., Huther, L., Langanke, K., Mart{\'{\i}}nez-Pinedo, G., \& Heger, A.\ 2015, arXiv:1505.01082
  
\bibitem[Sieverding et al.(2016)]{2016EPJWC.10906004S} Sieverding, A., Huther, L., Mart{\'{\i}}nez-Pinedo, G., Langanke, K., \& Heger, A.\ 2016, European Physical Journal Web of Conferences, 109, 06004 

\bibitem[Sieverding et al.(2018a)]{2018JPhCS.940a2054S} Sieverding, A., Huther, L., Mart{\'{\i}}nez-Pinedo, G., Langanke, K., \& Heger, A.\ 2018, Journal of Physics Conference Series, 940, 012054
  
\bibitem[Sieverding et al.(2018b)]{2018ApJ...865..143S} Sieverding, A., Mart{\'{\i}}nez-Pinedo, G., Huther, L., Langanke, K., \& Heger, A.\ 2018, \apj, 865, 143

\bibitem[Sieverding et al.(2018c)]{2018EPJWC.16501045S} Sieverding, A., Mart{\'{\i}}nez Pinedo, G., Langanke, K., Harris, J.~A., \& Hix, W.~R.\ 2018, European Physical Journal Web of Conferences, 165, 01045 
  
\bibitem[Spite \& Spite(1982)]{1982A&A...115..357S} Spite, F., \& Spite, M.\ 1982, \aap, 115, 357 

  
\bibitem[Sukhbold et al.(2016)]{2016ApJ...821...38S} Sukhbold, T., Ertl, T., Woosley, S.~E., Brown, J.~M., \& Janka, H.-T.\ 2016, \apj, 821, 38 

\bibitem[Timmes(1999)]{1999ApJS..124..241T} Timmes, F.~X.\ 1999, \apjs, 124, 241
  
\bibitem[Tsujimoto et al.(1995)]{1995MNRAS.277..945T} Tsujimoto, T., Nomoto, K., Yoshii, Y., et al.\ 1995, \mnras, 277, 945
  
\bibitem[Ventura \& D'Antona(2010)]{2010MNRAS.402L..72V} Ventura, P., \& D'Antona, F.\ 2010, \mnras, 402, L72

\bibitem[Woosley et al.(1990)]{1990ApJ...356..272W} Woosley, S.~E., Hartmann, D.~H., Hoffman, R.~D., \& Haxton, W.~C.\ 1990, \apj, 356, 272

\bibitem[Wu et al.(2015)]{2015PhRvD..91f5016W} Wu, M.-R., Qian, Y.-Z., Mart{\'{\i}}nez-Pinedo, G., Fischer, T., \& Huther, L.\ 2015, \prd, 91, 065016 

\bibitem[Yokomakura et al.(2002)]{2002PhLB..544..286Y} Yokomakura, H., Kimura, K., \& Takamura, A.\ 2002, Physics Letters B, 544, 286

\bibitem[Yoshida et al.(2005)]{2005PhRvL..94w1101Y} Yoshida, T., Kajino, T., \& Hartmann, D.~H.\ 2005, Physical Review Letters, 94, 231101

\bibitem[Yoshida et al.(2006a)]{2006PhRvL..96i1101Y} Yoshida, T., Kajino, T., Yokomakura, H., et al.\ 2006a, Physical Review Letters, 96, 091101

\bibitem[Yoshida et al.(2006b)]{2006ApJ...649..319Y} Yoshida, T., Kajino, T., Yokomakura, H., et al.\ 2006b, \apj, 649, 319 

\bibitem[Yoshida et al.(2008)]{2008ApJ...686..448Y} Yoshida, T., Suzuki, T., Chiba, S., et al.\ 2008, \apj, 686, 448

\bibitem[Yoshida et al.(2004)]{2004ApJ...600..204Y} Yoshida, T., Terasawa, M., Kajino, T., \& Sumiyoshi, K.\ 2004, \apj, 600, 204

\end{thebibliography}
\end{document}